\documentclass[graybox,envcountchap,sectrefs]{svmono}

\usepackage{mathptmx}
\usepackage{helvet}
\usepackage{courier}
\usepackage{type1cm}         
\usepackage{makeidx}         
\usepackage{graphicx}        
\usepackage{multicol}        

\def\R{{\bf R}}

\begin{document}

\author{\center Ricardo L\'opez-Ruiz, Jaime Sa\~nudo, \\ Elvira Romera, Xavier Calbet}
\title{\center Statistical Complexity and \\ Fisher-Shannon Information. \\ Applications}
\subtitle{\center -- Book Chapter --}
\maketitle

\tableofcontents

%
%
%


\chapter{Statistical Complexity and Fisher-Shannon Information. Applications}
\label{intro} 


\abstract{\emph{In this chapter, a statistical measure of complexity and the Fisher-Shannon information product
are introduced and their properties are discussed. These measures are based on the interplay between 
the Shannon information, or a function of it,
and the separation of the set of accessible states to a system from the equiprobability distribution, 
i.e. the disequilibrium or the Fisher information, respectively. 
Different applications in discrete and continuous systems are shown. Some of them are concerned with quantum systems,
from prototypical systems such as the H-atom, the harmonic oscillator and the square well to other ones such as
He-like ions, Hooke's atoms or just the periodic table. In all of them, these statistical indicators show an
interesting behavior able to discern and highlight some conformational properties of those systems.}}

\newpage
\section{A Statistical Measure of Complexity. Some Applications}

This century has been told to be the century of {\it complexity} \cite{hawking00}.
Nowadays the question {\it ``what is complexity?"} is circulating over 
the scientific crossroads of physics, biology, mathematics and computer science, 
although under the present understanding of the world could be no urgent
to answer this question. However, many different points of view have been developed 
to this respect and hence a lot of different answers can be found in the literature. 
Here we explain in detail one of these options.

On the most basic grounds, an object, a
procedure, or system is  said to be ``complex" when it does not match 
patterns regarded as simple. This sounds rather like an oxymoron 
but common knowledge tells us what is simple and complex: 
simplified systems or idealizations are always a starting point to solve
scientific problems. The notion of ``complexity"  
in physics  \cite{anderson91,parisi93}  starts
 by considering the perfect crystal and the 
isolated ideal gas as examples of simple models and therefore as systems 
with zero ``complexity".
Let us briefly recall their main characteristics with
``order", ``information" and ``equilibrium". 

A perfect crystal is completely ordered and the  
atoms are arranged  following stringent rules of symmetry. 
The probability distribution for the states accessible to the perfect
crystal is centered around a prevailing state of
perfect symmetry. A small piece of ``information" is enough to describe
the perfect crystal: the distances and 
the symmetries that define the elementary cell.
The ``information"  stored in this system can be considered minimal.
On the other hand, the isolated ideal gas is completely
disordered. The system  can be found in any of its accessible states with 
the same probability. All of them contribute in equal measure to
the ``information" stored in the ideal gas.
It has therefore a maximum ``information". These two  simple systems
are extrema in the scale of ``order" and ``information". It follows that
the definition of ``complexity" must not be made in terms of just ``order"
or ``information". 

It might seem reasonable to propose a measure of ``complexity"
by adopting some kind of distance from the equiprobable distribution of the
accessible states of the system. Defined in this way,
``disequilibrium"  would give an idea of
the probabilistic hierarchy of the system. 
``Disequilibrium" would be different from zero if there are privileged,
or more probable, states among those accessible. But this would not work.
Going back to the two examples we began with, it is readily seen that a perfect
crystal is far from an equidistribution among the accessible states
because one of them is totally prevailing, and so ``disequilibrium"
would be maximum. For the ideal gas, ``disequilibrium" would be zero by
construction. Therefore such a distance or ``disequilibrium" (a measure
of a probabilistic hierarchy) cannot be directly associated 
with ``complexity". 

\begin{figure}[h]  
\centerline{\includegraphics[width=12cm]{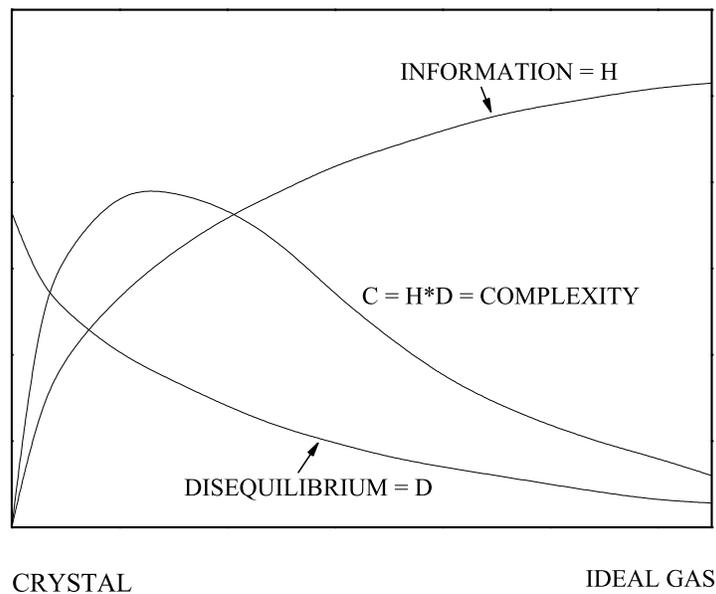}}  
\caption{Sketch of the intuitive notion of the magnitudes of 
``information" (H) and ``disequilibrium" (D) for the physical systems and
the behavior intuitively required for the magnitude 
``complexity". The quantity $C=H\cdot D$ is proposed to measure
such a magnitude.}  
\label{figIntro0}  
\end{figure}

In Figure \ref{figIntro0} we sketch an intuitive qualitative behavior for ``information"
$H$ and ``disequilibrium" $D$ for systems ranging from 
the perfect crystal to the ideal gas. This graph suggests that
the product of these two quantities  could be used 
as a measure of ``complexity": $C = H \cdot D$.
The function $C$ has indeed the features and asyntotical properties
that one would expect intuitively: it vanishes for the 
perfect crystal and for the
isolated ideal gas, and it is different from zero for the rest of the
systems of particles. We will follow these guidelines to establish a quantitative
measure of ``complexity". 

Before attempting any further progress, however, we must recall that 
``complexity" cannot be measured univocally, because it depends on the
nature of the description (which always involves a reductionist process)
and  on the scale of observation. Let us take an example to illustrate
this point. A computer chip can look very different at different scales.
It is an entangled array of electronic elements at microscopic scale but 
only an ordered set of pins attached to a black box at a macroscopic scale.

We shall now discuss a measure of ``complexity" based on the 
statistical description of systems.
Let us assume that the system has $N$ accessible states $\{x_1,x_2,...,x_N\}$ 
when observed at a given scale. We will call this an $N$-system. 
Our understanding of the behavior of this system determines the 
corresponding probabilities $\{p_1,p_2,...,p_N\}$ (with the 
condition $\sum_{i=1}^{N}p_i =1$) of each state ($p_i>0$ for all $i$). 
Then the knowledge of the underlying
physical laws at this scale is incorporated into a probability distribution 
for the accessible states. It is possible to find a quantity
measuring the amount of ``information".
Under to the most elementary conditions of 
consistency, Shannon \cite{shannon49} determined the unique function 
$H(p_1,p_2,...,p_N)$ that accounts for the ``information" stored
in a system:
\begin{equation}
H = -K \sum_{i=1}^{N} p_i\log p_i \, ,
\end{equation}
where $K$ is a positive constant. The  quantity $H$ is called 
{\it information}. The redefinition of information $H$ as some type of monotone function 
of the Shannon entropy can be also useful in many contexts as we shall show in the next sections. 
In the case of a crystal, a state $x_c$ would be the most probable
$p_c\sim 1$, and all others $x_i$ 
would be very improbable, $p_i\sim 0$ $i\neq c$. Then $H_c \sim 0$. 
On the other side, equiprobability
characterizes an isolated ideal gas, $p_i\sim 1/N$ so $H_g\sim K\log N$,
i.e., the maximum of information for a N-system.
(Notice that if one assumes equiprobability and $K=\kappa\equiv Boltzmann$ 
$constant$, $H$ is identified  with the thermodinamic
entropy, $S=\kappa\log N$). Any other N-system will have an amount of
information between those two extrema. 

Let us propose a definition of {\it disequilibrium} $D$ 
in a $N$-system \cite{prigo77}. The intuitive notion suggests that some kind of 
distance from an equiprobable distribution should be adopted.
Two requirements are imposed on the magnitude of $D$: $D>0$ in order to have a
positive measure of ``complexity" and $D=0$ on the limit of equiprobability.
The straightforward solution is to add the quadratic distances of 
each state to the equiprobability  as follows: 
\begin{equation}
D = \sum_{i=1}^{N}\left(p_i - \frac{1}{N}\right)^2\,.
\label{def-D}
\end{equation}
According to this definition, a crystal has maximum disequilibrium
(for the dominant state,
$p_c\sim 1$, and $D_c\rightarrow 1$ for $N\rightarrow \infty$)
while the disequilibrium for an 
ideal gas vanishes ($D_g\sim 0$) by construction. For any other system 
$D$ will have a value between these two extrema.

We now introduce the definition of {\it complexity} $C$ of 
a $N$-system \cite{lopezruiz94,lopezruiz95}.
This is simply the interplay between the information stored in 
the system and its disequilibrium:
\begin{equation}
C = H \cdot D = -\left ( K\sum_{i=1}^{N} p_i\log p_i \right ) \cdot
\left (\sum_{i=1}^{N}\left(p_i - \frac{1}{N}\right)^2 \right )\,.
\label{def-C}
\end{equation} 
This definition fits the intuitive arguments.
For a crystal, disequilibrium is large but the information stored
is vanishingly small, so $C\sim 0$.
On the other hand, $H$ is large for an ideal gas, but $D$ is small, 
so $C\sim 0$ as well. Any other system will have an 
intermediate behavior and therefore $C>0$.

As was intuitively suggested, the definition of complexity (\ref{def-C})
also depends on the {\it scale}.
At each scale of observation a new set of accessible states appears
with its corresponding probability distribution so that
complexity changes. Physical laws at each level
of observation allow us to infer the probability
distribution of the new set of
accessible states, and therefore different values for $H$, $D$ and $C$ 
will be obtained. 
The straightforward passage to the case of a continuum number of states,
$x$, can be easily inferred. Thus we must treat with probability distributions
with a continuum support, $p(x)$, and normalization condition
$\int_{-\infty}^{+\infty}p(x)dx=1$. Disequilibrium has the limit
$D=\int_{-\infty}^{+\infty}p^2(x)dx$ and the complexity could be defined by:
\begin{equation} 
C=H\cdot D=-\left(K\int_{-\infty}^{+\infty}p(x)\log p(x)dx \right)
\cdot\left(\int_{-\infty}^{+\infty}p^2(x)dx \right)\,.
\label{def-C-continuo}
\end{equation} 
As we shall see, other possibilities for the continuous extension of $C$
are also possible.

\begin{figure}[h]  
\centerline{\includegraphics[width=8.5cm,angle=-90]{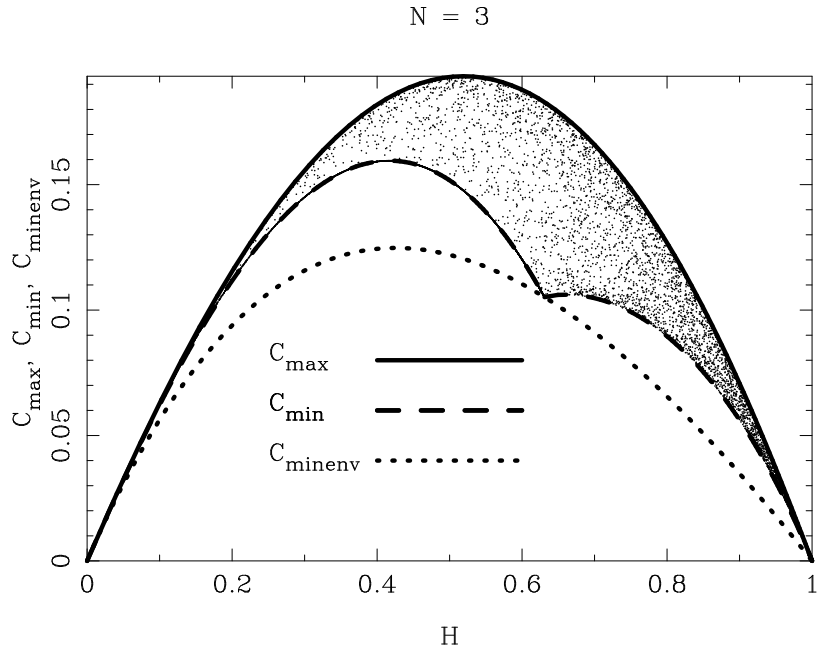}}  
\caption{In general, dependence of complexity ($C$)
on normalized information ($H$) is not univocal: many distributions $\{p_i\}$
can present the same value of $H$ but different $C$.
This is shown in the case $N=3$.}  
\label{figIntro1}  
\end{figure}  

\begin{figure}[h]  
\centerline{\includegraphics[width=7.5cm,angle=-90]{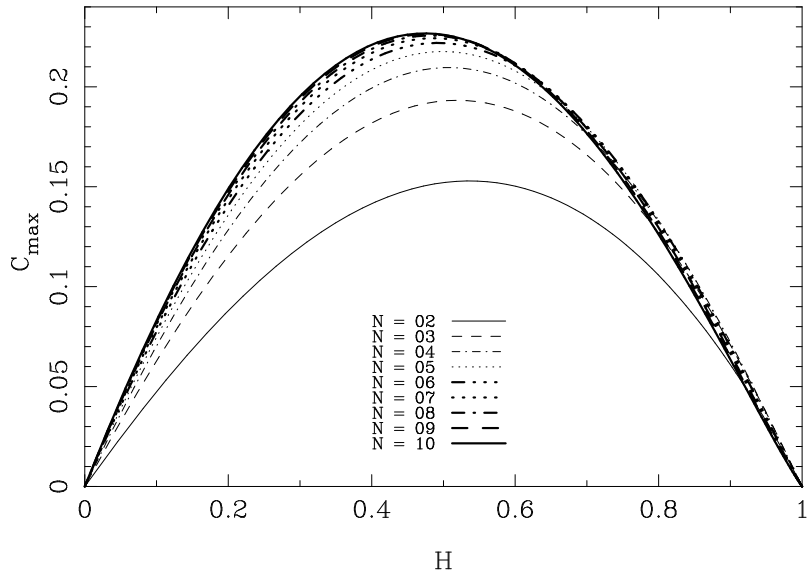}}  
\caption{Complexity ($C=H\cdot D$) as a function
of the normalized information ($H$) for a system with two accessible
states ($N=2$). Also curves of maximum complexity ($C_{max}$) 
are shown for the cases: $N=3,\ldots,10$.}  
\label{figIntro2}  
\end{figure}

Direct simulations of the definition give the values
of $C$ for general $N$-systems.
The set of all the possible distributions $\{p_1,p_2,...,p_N\}$ 
where an $N$-system could be found 
is sampled. For the sake of simplicity $H$ is normalized to the interval
$[0,1]$. Thus $H=\sum_{i=1}^{N} p_i\log p_i/\log N$.
 For each distribution $\{p_i\}$ the normalized information 
 $H(\{p_i\})$, and the disequilibrium 
$D(\{p_i\})$ (eq. \ref{def-D}) are calculated.
In each case the normalized complexity
$C=H\cdot D$ is obtained and the
pair $(H,C)$ stored.
These two magnitudes are plotted on a diagram $(H,C(H))$ in order
to verify the qualitative behavior predicted in Figure \ref{figIntro0}.
For $N=2$ an analytical expression for the curve 
$C(H)$ is obtained.
If the probability of one state is $p_1 =x$, that 
of the second one is simply $p_2 =1-x$. 
The complexity of the system will be:
\begin{equation}
C(x)= H(x)\cdot D(x)=-\frac{1}{\log 2}
[x\log\left(\frac{x}{1-x}\right)+ \log(1-x)]\cdot 2\left(x-\frac{1}{2}\right)^2\,.
\end{equation}
Complexity vanishes for the two simplest $2$-systems:
the crystal ($H=0$; $p_1 =1$, $p_2 =0$) and the ideal gas 
($H=1$; $p_1 =1/2$, $p_2 =1/2$). 
Let us notice that this curve is the simplest one that fulfills all
the conditions discussed in the introduction. 
The largest complexity is reached for $H\sim 1/2$ and its value
is: $C(x\sim 0.11)\sim 0.151$. 
For $N>2$ the relationship between 
$H$ and $C$ is not
univocal anymore. Many different distributions $\{p_i\}$ store
the same information $H$ but have different complexity 
$C$. Figure \ref{figIntro1} displays such a behavior for $N=3$. 
If we take the maximum complexity $C_{max}(H)$ associated with each 
$H$ a curve similar to the one for a $2$-system is recovered.
Every $3$-system will have a complexity below this line and upper the line
of $C_{min}(H)$ and also upper the minimum envelope complexity $C_{\rm minenv}$.
These lines will be analytically found in 
a next section. In Figure \ref{figIntro2} 
curves $C_{max}(H)$ for the cases 
$N=3,\ldots,10$ are also shown. 
Let us observe the shift of the complexity-curve peak 
to smaller values of entropy for rising $N$. This fact 
agrees with the intuition telling us that the biggest complexity
(number of possibilities of `complexification') be reached for lesser
entropies for the systems with bigger number of states.

Let us return to the point at which we started this discussion. 
Any notion of complexity in physics \cite{anderson91,parisi93} 
should only be made on the basis of a well defined or operational 
magnitude \cite{lopezruiz94,lopezruiz95}.
But two additional requirements are needed in order to obtain a 
good definition of complexity in physics: ($1$) the new magnitude
must be measurable in many different physical systems and ($2$) 
a comparative relationship and a physical interpretation 
between any two measurements should be possible.

Many different definitions of complexity have been proposed to date, mainly
in the realm of physical and computational sciences. Among these, several can be cited: 
algorithmic complexity (Kolmogorov-Chaitin) \cite{kolmogorov65,chaitin66},
the Lempel-Ziv complexity \cite{lempel76}, the logical depth of Bennett 
\cite{bennett85}, the effective measure complexity of
Grassberger \cite{grassberger86}, the complexity of a system based in its 
diversity \cite{huberman86}, the thermodynamical depth \cite{lloyd88},
the $\epsilon$-machine complexity \cite{crutchfield89} ,
the physical complexity of genomes \cite{adami00},
complexities of formal grammars, etc.
The definition of complexity (\ref{def-C}) proposed in this section offers 
a new point of view, based on a statistical
description of systems at a given {\it scale}. In this 
scheme, the knowledge of the physical laws governing the dynamic evolution 
in that scale is used to find its accessible states and its
probability distribution. This process would immediately indicate the
value of complexity. In essence this is nothing but an interplay
between the information stored by the system and the 
{\it distance from equipartition} (measure of a probabilistic hierarchy
between the observed parts) of the probability distribution of
its accessible states. 
Besides giving the main features of a ``intuitive" notion
of complexity, we will show in this chapter that we can go one step further 
and to compute this quantity in other relevant physical
situations and in continuum systems. The most important point is that the new definition 
successfully enables us to discern situations regarded as complex.
For example, we show here two of these applications in complex systems with
some type of discretization: 
one of them is the study of this magnitude in 
a phase transition in a coupled map lattice \cite{sanchez05}
and the other one is its calculation for 
the time evolution of a discrete gas out of equilibrium \cite{calbet01}.
Other applications to more realistic systems can also be found
in the literature \cite{escalona09}.


\subsection{Complexity in a phase transition: coupled map lattices}

If by complexity it is to be understood that property present in all systems 
attached under the epigraph of `complex systems', this property should be reasonably quantified 
by the measures proposed in the different branches of knowledge.
As discussed above, this kind of indicators is found in those fields where 
the concept of information is crucial, 
from physics \cite{grassberger86,lloyd88} to computational 
sciences \cite{kolmogorov65, chaitin66,lempel76,crutchfield89}.

In particular, taking into account the statistical properties of a system,
the indicator called the {\it LMC (L\'{o}pezRuiz-Mancini-Calbet) complexity}
has been introduced  \cite{lopezruiz94,lopezruiz95} in the former section. 
This magnitude identifies the entropy or information $H$ stored in a system 
and its disequilibrium $D$, i.e. the distance from its actual state to the 
probability distribution of equilibrium, as the two basic ingredients for calculating
its complexity. Hence, the LMC complexity $C$ is given by the formula (\ref{def-C}),
$C(\bar p)  =  H(\bar p)\cdot D(\bar p)$,
where $\bar p=\{p_i\}$, with $p_i>0$ and $i=1,\cdots,N$, represents the
distribution of the $N$ accessible states to the system, and $k$ is a constant
taken as $1/\log N$. 

As well as the Euclidean distance $D$ is present in the original LMC complexity,
other kinds of disequilibrium measures have been proposed in order to
remedy some statistical characteristics considered troublesome for
some authors \cite{feldman98}.
In particular, some attention has been focused \cite{lin91,martin03} on 
the Jensen-Shannon divergence $D_{JS}$ as a measure for evaluating 
the distance between two different distributions $(\bar p_1,\bar p_2)$. 
This distance reads:
\begin{equation} 
D_{JS}(\bar p_1,\bar p_2) = H(\pi_1\bar p_1+\pi_2\bar p_2)-\pi_1H(\bar p_1)-\pi_2H(\bar p_2)\,,
\label{eq:d-js}
\end{equation}
with $\pi_1,\pi_2$ the weights of the two probability distributions
$(\bar p_1,\bar p_2)$ verifying $\pi_1,\pi_2\geq 0$ and $\pi_1+\pi_2=1$.
The ensuing statistical complexity 
\begin{equation} 
C_{JS}=H\cdot D_{JS}
\label{eq:c-js}
\end{equation} 
becomes intensive and also keeps the property of distinguishing 
among distinct degrees of periodicity \cite{lamberti04}. In this section,
we consider $\bar p_2$ the equiprobability distribution and $\pi_1=\pi_2=0.5$.

As it can be straightforwardly seen, all these LMC-like complexities vanish 
both for completely ordered and for completely 
random systems as it is required for the correct asymptotic properties 
of a such well-behaved measure. Recently, they have been successfully used to
discern situations regarded as complex in discrete systems out of
equilibrium \cite{calbet01,guozhang98,shiner99,zuguo00,rosso03,rosso05,lovallo05,lopezruiz05}. 

Here, the local transition to chaos via
intermittency \cite{pomeau80} in the logistic map, $x_{n+1}=\lambda x_n(1-x_n)$
presents a sharp transition when $C$ is plotted
versus the parameter $\lambda$ in the region around
the instability for $\lambda\sim \lambda_t=3.8284$. 
When $\lambda<\lambda_t$ the system approaches the laminar regime and 
the bursts become more unpredictable. The complexity increases. When the point
$\lambda=\lambda_t$ is reached a drop to zero occurs for the magnitude $C$.
The system is now periodic and it has lost its complexity.
The dynamical behavior of the system is finally well reflected in the magnitude $C$
as it has been studied in \cite{lopezruiz95}).

When a one-dimensional array of such maps is put together a more complex behavior
can be obtained depending on the coupling among the units. Ergo the phenomenon
called {\it spatio-temporal intermittency} can emerge \cite{chate87,jensen90,jensen98}. 
This dynamical regime corresponds 
with a situation where each unit is weakly oscillating around a laminar state
that is aperiodically and strongly perturbed for a traveling burst. 
In this case, the plot of the one-dimensional lattice evolving in time 
gives rise to complex patterns on the plane. If the coupling among units
is modified the system can settle down in an absorbing phase where its dynamics 
is trivial \cite{coullet97,toral00} and then homogeneous patterns are obtained.
Therefore an abrupt transition to spatio-temporal intermittency can be depicted by
the system \cite{pomeau86,sinha03} when modifying the coupling parameter.   

Now we are concerned with measuring $C$ and $C_{JS}$ in a such
transition for a coupled map lattice of logistic type.
Our system will be a line of sites, $i=1,\ldots,L$, with periodic
boundary conditions. In each site $i$ a local variable $x_i^{n}$ evolves 
in time ($n$) according to a discrete logistic equation. The interaction 
with the nearest neighbors takes place via a multiplicative coupling:
\begin{equation}
x_i^{n+1} = (4-3pX_i^{n})x_i^{n}(1-x_i^{n})\,,
\label{eq:xn}
\end{equation}  
where $p$ is the parameter of the system measuring the strength of the coupling ($0<p<1$).
The variable $X_i^{n}$ is the digitalized local mean field,
\begin{equation}
X_i^{n} = nint \left[\frac{1}{2}\: ({x_{i+1}^{n}+x_{i-1}^{n}}) \right] \, ,
\end{equation}
with {\it $nint(.)$} the integer function rounding its argument to the nearest integer.
Hence $X_i^{n}=0$ or $1$.

\begin{figure}[h]
  \centering
  {\includegraphics[angle=0, width=10cm]{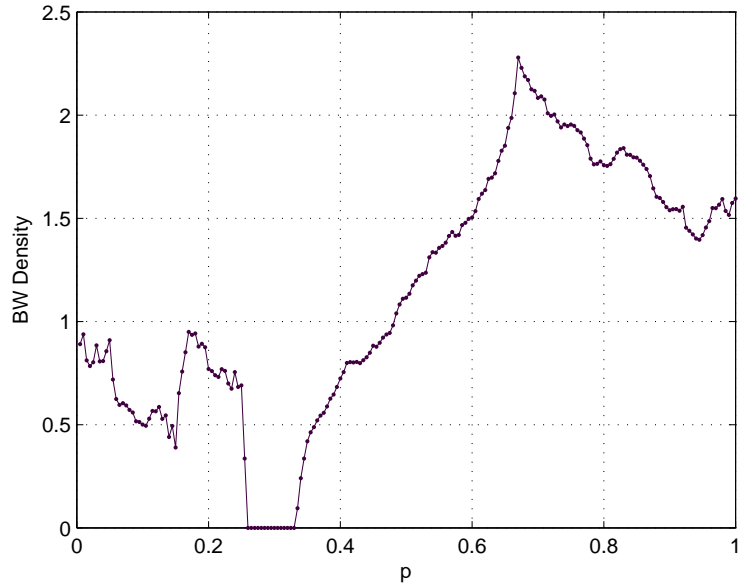}}
  \caption{$\beta$ versus $p$. The $\beta$-statistics (or BW density) 
  for each $p$ is the rate between the number of {\it black} and {\it white} cells 
  depicted by the system in the two-dimensional representation of its after-transient time evolution.
  (Computations have been performed with $\Delta p=0.005$ for a lattice of $10000$ sites after a transient 
  of $5000$ iterations and a running of other $2000$ iterations).}
  \label{figCML1}
\end{figure}

There is a biological motivation behind this kind of systems \cite{lopezruiz04,lopezruiz05-1}.
It could represent a {\it colony of interacting competitive individuals}.
They evolve randomly when they are independent ($p=0$). If some competitive interaction 
($p>0$) among them takes place the local dynamics loses its erratic component and becomes
chaotic or periodic in time depending on how populated the vicinity is.
Hence, for bigger $X_i^n$ more populated is the neighborhood of the individual $i$ and 
more constrained is its free action. At a first sight, it would seem that some particular 
values of $p$ could stabilize the system. In fact, this is the case. 
Let us choose a number of individuals for the colony ($L=500$ for instance),
let us initialize it randomly in the range $0<x_i<1$  and 
let it evolve until the asymptotic regime is attained.
Then the {\it black/white} statistics of the system is performed. That is,
the state of the variable $x_i$ is compared with the critical level $0.5$ for $i=1,\ldots,L$:
if $x_i>0.5$ the site $i$ is considered {\it white} (high density cell) and a counter $N_w$ is 
increased by one, or if $x_i<0.5$ the site $i$ is considered {\it black} (low density cell) and 
a counter $N_b$ is increased by one. This process is executed in the stationary regime
for a set of iterations. The {\it black/white} statistics is then the rate $\beta=N_b/N_w$.
If $\beta$ is plotted versus the coupling parameter $p$ the Figure \ref{figCML1} is obtained.

\begin{figure}[h]
  \centering
  {\includegraphics[angle=0, width=10cm]{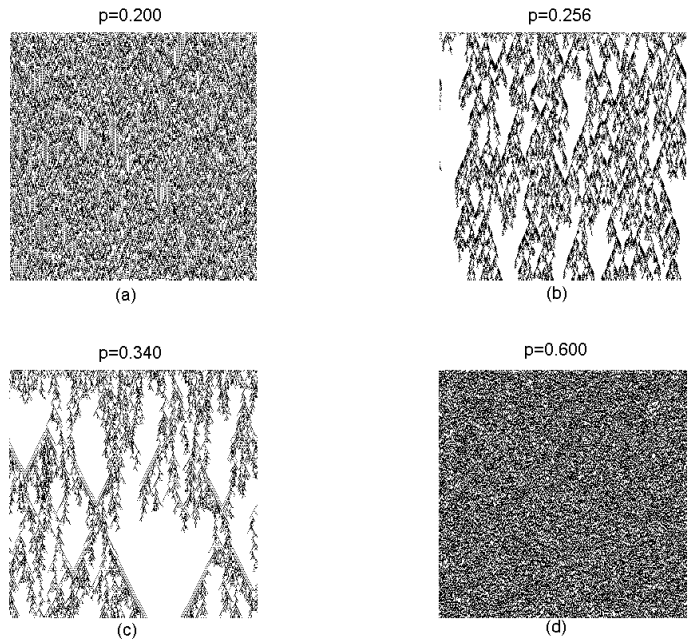}}
  \caption{Digitalized plot of the one-dimensional coupled map lattice (axe OX) evolving in time (axe OY)
  according to Eq. (\ref{eq:xn}): if $x_i^n>0.5$ the $(i,n)$-cell is put in white color 
  and if $x_i^n<0.5$ the $(i,n)$-cell is put in black color. The discrete time
  $n$ is reset to zero after the transitory. (Lattices of $300\times 300$ sites, 
  i.e., $0<i<300$ and $0<n<300$).}
  \label{figCML2}
\end{figure}

The region $0.258<p<0.335$ where $\beta$ vanishes is remarkable. 
As stated above, $\beta$ represents the rate between the number of black cells and the number of white 
cells appearing in the two-dimensional digitalized representation of the colony evolution.
A whole white pattern is obtained for this range
of $p$. The phenomenon of spatio-temporal intermittency is displayed by the system 
in the two borders of this parameter region (Fig. \ref{figCML2}). 
Bursts of low density (black color) travel in an
irregular way through the high density regions (white color). 
In this case two-dimensional complex patterns 
are shown by the time evolution of the system (Fig. \ref{figCML2}b-c). 
If the coupling $p$ is far enough from this region, i.e., $p<0.25$ or $p>0.4$,
the absorbent region loses its influence on the global dynamics and less structured 
and more random patterns than before are obtained (Fig.  \ref{figCML2}a-d).   
For $p = 0$ we have no coupling of the maps, and each map generates so called fully developed
chaos, where the invariant measure is well-known to be symmetric around $0.5$. 
From this we conclude that $\beta(p = 0) = 1$. Let us observe that this symmetrical behavior
of the invariant measure is broken for small $p$, and $\beta$ decreases slightly 
in the vicinity of $p=0$.     

\begin{figure}[h]
  \centering
  {\includegraphics[angle=0, width=10cm]{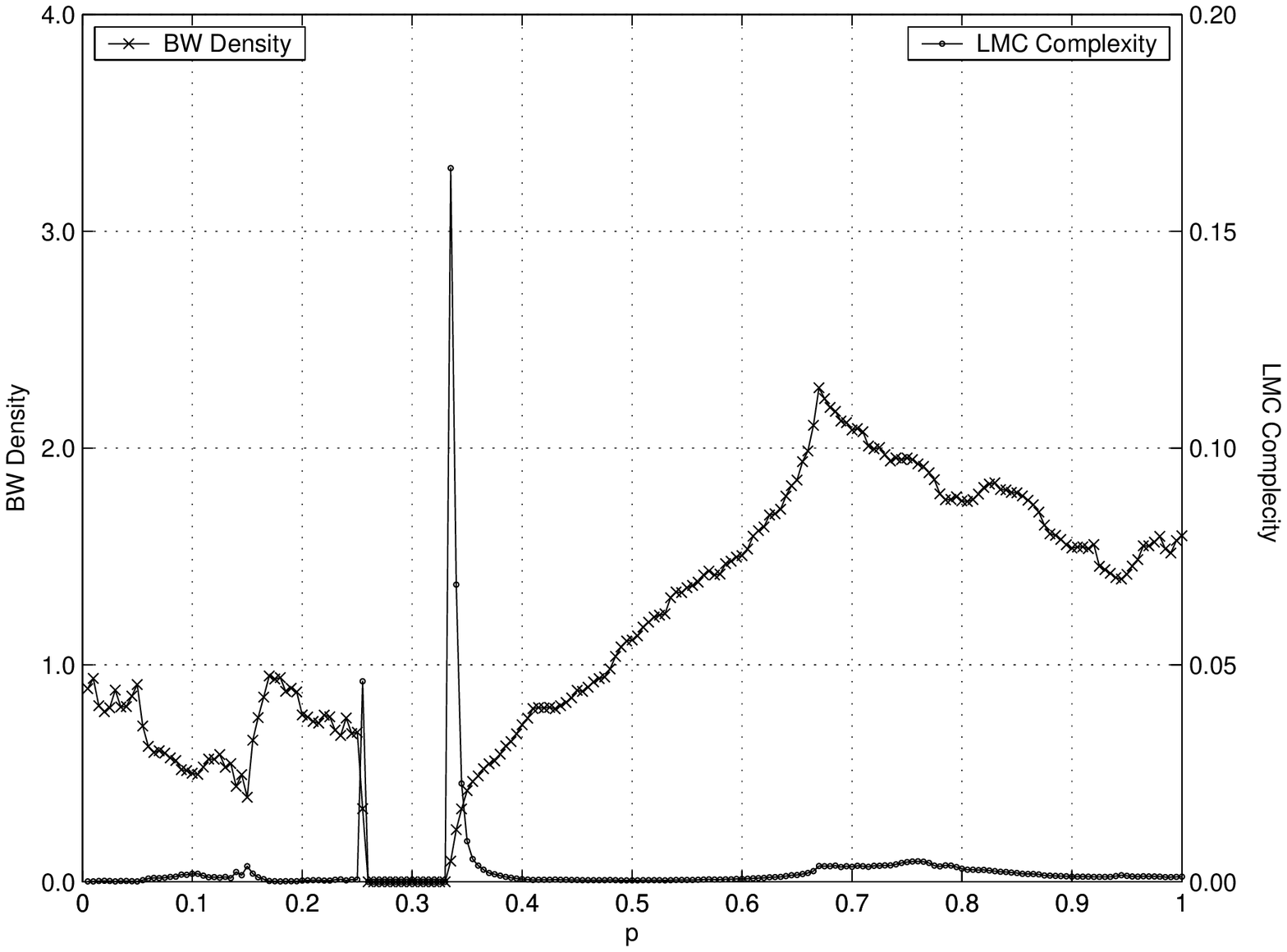}}
  \caption{($\bullet$) $C$ versus $p$. Observe the peaks of the LMC complexity located
  just on the borders of the absorbent region $0.258<p<0.335$,
  where $\beta=0$ ($\times$). 
  (Computations have been performed with $\Delta p=0.005$ for a lattice of $10000$ sites 
  after a transient of $5000$ iterations and a running of other $2000$ iterations).}
  \label{figCML3}
\end{figure}

\begin{figure}[h]
  \centering
  {\includegraphics[angle=0, width=10cm]{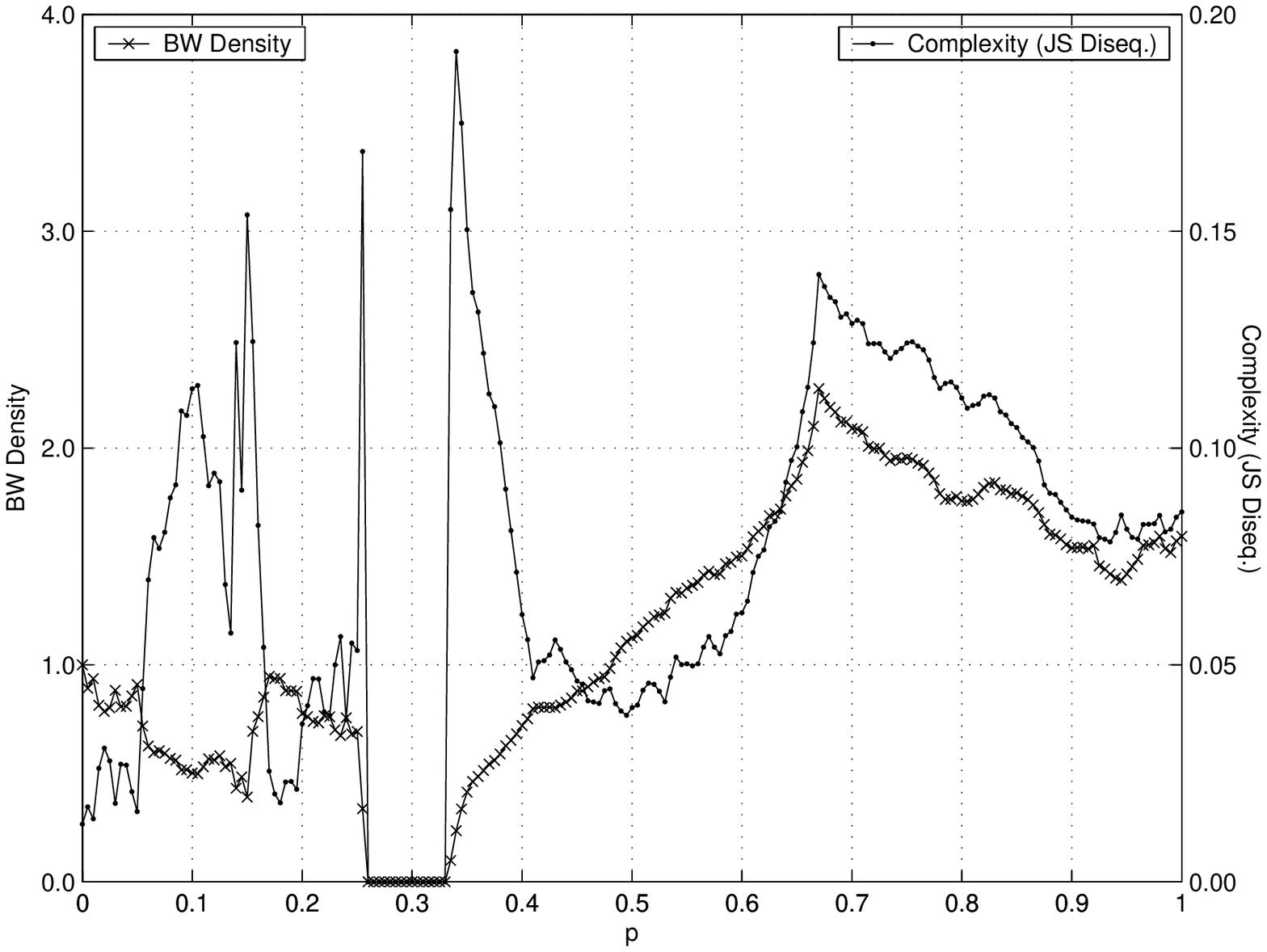}}
  \caption{($\cdot$) $C_{JS}$ versus $p$. The peaks of this modified LMC complexity
  are also evident just on the borders of the absorbent region $0.258<p<0.335$,
  where $\beta=0$ ($\times$). 
  (Computations have been performed with $\Delta p=0.005$ for a lattice of $10000$ sites 
  after a transient of $5000$ iterations and a running of other $2000$ iterations).}
  \label{figCML4}
\end{figure}

If the LMC complexities are quantified as function of $p$, 
our {\it intuition} is confirmed.
The method proposed in Ref. \cite{lopezruiz95} to calculate $C$ is now adapted
to the case of two-dimensional patterns. First, 
we let the system evolve until the asymptotic regime is attained. 
This transient is discarded.
Then, for each time $n$, we map the whole lattice in a binary sequence:
$0$ if $x_i^n<0.5$ and $1$ if $x_i^n>0.5$, for $i=1,\ldots,L$.
This $L$-binary string is analyzed by blocks of $n_o$ bits, where
$n_o$ can be considered the scale of observation.
For this scale, there are $2^{n_o}$ possible states but only some of them are accessible.
These accessible states as well as their probabilities are found in the $L$-binary string.
Next, the magnitudes $H$, $D$, $D_{JS}$, $C$ and $C_{JS}$ are directly calculated for this 
particular time $n$ by applying the formulas (\ref{def-C}),(\ref{eq:c-js}). 
We repeat this process for a set of successive time units $(n,n+1,\cdots,n+m)$. 
The mean values of $H$, $D$, $D_{JS}$, $C$ and $C_{JS}$ for these $m$ time units
are finally obtained and plotted in Figs. \ref{figCML3},\ref{figCML4}.  

Figures \ref{figCML3},\ref{figCML4} show the result for the case of $n_o=10$. 
Let us observe that the highest $C$ and $C_{JS}$ are reached when the dynamics displays 
spatio-temporal intermittency, that is, the {\it most complex patterns} 
are obtained for those values of $p$ that are located
on the borders of the absorbent region $0.258<p<0.335$.
Thus the plot of $C$ and $C_{JS}$ versus $p$ shows two tight peaks around the values
$p=0.256$ and $p=0.34$ (Figs. \ref{figCML3},\ref{figCML4}). Let us remark that  
the LMC complexity $C$ can be neglected far from the absorbent region.
Contrarily to this behavior, the magnitude $C_{JS}$ also shows high peaks
in some other sharp transition of $\beta$ located in the region $0<p<25$, and
an intriguing correlation with the {\it black/white} statistics in the region
$0.4<p<1$. All these facts as well as the stability study of the different dynamical 
regions of system (\ref{eq:xn}) are not the object of the present writing but
they could deserve some attention in a further inspection.  

If the detection of complexity in the two-dimensional case requires to identify 
some sharp change when comparing different patterns, 
those regions in the parameter space
where an abrupt transition happens should be explored 
in order to obtain the most complex patterns.
Smoothness seems not to be at the origin 
of complexity.  As well as a selected few distinct molecules 
among all the possible are in the basis of life \cite{mckay04},
discreteness and its spiky appearance 
could indicate the way towards complexity.
As we show in the next section, the distributions 
with the highest LMC complexity are just those distributions  
with a spiky-like appearance \cite{calbet01}. 
In this line, the striking result here exposed confirms the capability  of the 
LMC-like complexities for signaling a transition to complex behavior when regarding 
two-dimensional patterns \cite{sanchez05,sanchez05-1}. 

\subsection{Complexity versus time: the tetrahedral gas}

As before explained, several definitions of complexity, 
in the general sense of the term, have been presented in the literature.
These can be classified according to their calculation procedure into 
two broad and loosely defined groups. One of these groups is based 
on computational science and consists of all definitions based on
algorithms or automata to derive the complexity. Examples are the 
algorithmic complexity \cite{chaitin66}, the logical depth \cite{bennett85}
and the $\epsilon$-machine complexity \cite{crutchfield89}.
These definitions have been shown to be very useful in describing symbolic
dynamics of chaotic maps, but they have the disadvantage of being
very difficult to calculate. Another broad group consists of those complexities
based on the measure of entropy or entropy rate.
Among these, we may cite the effective measure complexity \cite{grassberger86},
the thermodynamic depth \cite{lloyd88},
the simple measure for complexity \cite{shiner99} and
the metric or K--S entropy rate \cite{kolmogorov58,sinai59}.
These definitions have also been very useful in describing
symbolic dynamics in maps, the simple measure of complexity having been also applied
to some physical situation such as a non-equilibrium Fermi gas \cite{landsberg98}.
They suffer the disadvantage of either being  very difficult
to calculate or having a simple relation to the regular entropy.

Other definition types of complexity have been
introduced. These are based on quantities that can be calculated
directly from the distribution function describing the system.
One of these is based on ``meta-statistics'' \cite{atmanspacher97} 
and the other on the notion of ``disequilibrium'' \cite{lopezruiz95}.
This latter definition has been referred above as the LMC complexity.
These definitions, together with the simple measure for complexity \cite{shiner99},
have the great advantage of allowing easy calculations
within the context of kinetic theory and of permitting their evaluation
in a natural way in terms of statistical mechanics.

As we have shown in the former sections,
the disequilibrium-based complexity is
easy to calculate and shows some interesting properties \cite{lopezruiz95}, 
but suffers from the main drawback of not being very well behaved
as the system size increases, or equivalently, as the distribution
function becomes continuous.
Feldman and Crutchfield \cite{feldman98} tried to solve this problem
by defining another equivalent term for disequilibrium, but
ended up with a complexity that was a trivial function of the entropy.

Whether these definitions of complexity are useful
in non-equilibrium thermodynamics will depend
on how they behave as a function of time.
There is a general belief that, although the second law of thermodynamics
requires average entropy (or disorder) to increase,
this does not in any way forbid local order from arising \cite{gell-mann95}.
The clearest example is seen with life, which can continue
to exist and grow in an isolated system for as long as internal resources last.
In other words, in an isolated system the entropy must increase,
but it should be possible, under certain circumstances, for the complexity to increase.

Here we examine how LMC complexity evolves with time in an isolated system and
we show that it indeed has some interesting properties.
The disequilibrium-based complexity \cite{lopezruiz95} defined in equation (\ref{def-C}) 
actually tends to be maximal as the entropy increases in a Boltzmann integro--differential
equation for a simplified gas.

We proceed to calculate the distributions which maximize and minimize the complexity
and its asymptotic behavior, and also introduce the basic
concepts underlying the time evolution of LMC complexity in Sect. \ref{sec:time}.
Later, in Sects. \ref{sec:tetra} and \ref{sec:evol}, 
by means of numerical computations following
a restricted version of the Boltzmann equation, we apply this to a special
system, which we shall term ``tetrahedral gas''.
Finally, in Sect. \ref{sec:conclu}, the results and conclusions for this system are given, 
together with their possible applications.

\subsubsection{Maximum and minimum complexity}
\label{sec:time}

In this section, we assume that the system can be in one of its $N$ possible
accessible states, $i$. The probability of the system being in state $i$
will be given by the discrete distribution function, $f_i\ge 0$,
with the normalization condition $I \equiv \sum_{i=1}^N f_i = 1$.
The system is defined such that, if isolated, it will reach equilibrium, 
with all the states having equal probability, $f_{\rm e}=\frac{1}{N}$.
Since we are supposing that $H$ is normalized, $0 \leq H \leq 1$,
and $0 \leq D \leq (N-1)/N$, then complexity, $C$, is also normalized, $0 \leq C \leq 1$.

When an isolated system evolves with time, the complexity cannot have any possible 
value in a $C$ versus $H$ map as it can be seen in Fig. \ref{figIntro1}), but it
must stay within certain bounds, $C_{\rm max}$ and $C_{\rm min}$.
These are the maximum and minimum values of $C$ for a given $H$.
Since $C = D \cdot H$, finding the extrema of $C$ for constant $H$
is equivalent to finding the extrema of $D$.

There are two restrictions on $D$:
the normalization, $I$, and the fixed value of the entropy, $H$.
To find these extrema undetermined Lagrange multipliers are used. 
Differentiating expressions of $D$, $I$ and $H$, we obtain
\begin{eqnarray}
\frac{\partial D}{\partial f_j} & = & 2(f_j - f_{\rm e})\,,\\
\frac{\partial I}{\partial f_j} & = & 1\,,\\
\frac{\partial H}{\partial f_j} & = & -\frac{1}{\ln N}\left(\ln f_j + 1\right)\,.
\end{eqnarray}
Defining $\lambda_1$ and $\lambda_2$ as the Lagrange multipliers, we get:
\begin{equation}
2(f_j - f_{\rm e}) + \lambda_1 + \lambda_2(\ln f_j + 1)/\ln N = 0\,.
\end{equation}
Two new parameters, $\alpha$ and $\beta$, which are a
linear combinations of the Lagrange multipliers are defined:
\begin{equation}
f_j + \alpha \ln f_j + \beta = 0\,,
\label{eq:maxmin}
\end{equation}
where the solutions of this equation, $f_j$, are the values that minimize or
maximize the disequilibrium.

In the maximum complexity case
there are two solutions, $f_j$, to Eq.\ (\ref{eq:maxmin})
which are shown in Table \ref{tab:maximum}.
One of these solutions,
$f_{\rm max}$, is given by
\begin{equation}
\label{eq:defhmax}
H = - \frac{1}{\ln N} \left[ f_{\rm max} \ln f_{\rm max} + ( 1 - f_{\rm max} ) \ln
\left( \frac{1 - f_{\rm max}}{N - 1} \right) \right]\,,
\end{equation}
and the other solution by $(1 - f_{\rm max})/(N - 1)$.

The maximum disequilibrium, $D_{\rm max}$, for a fixed $H$ is
\begin{equation}
\label{eq:defdmax}
D_{\rm max} = (f_{\rm max} - f_{\rm e})^2 +
      (N - 1)\left(\frac{1 - f_{\rm max}}{N - 1} - f_{\rm e}\right)^2\,,
\end{equation}
and thus, the maximum complexity, which depends
only on $H$, is
\begin{equation}
\label{eq:defcmax}
C_{\rm max}(H) = D_{\rm max} \cdot H\,.
\end{equation}
The behavior of the maximum value of complexity versus $\ln N$ 
was computed in Ref. \cite{anteneodo96}.

\begin{table}
\caption{Probability values, $f_j$, that give a maximum
of disequilibrium, $D_{\rm max}$, for a given $H$.}
\label{tab:maximum}
\begin{center}
\begin{tabular}{|c|c|c|}
\hline 
Number of states & $ f_j$ & Range of $ f_j$ \\
with $ f_j$      &        &                 \\
\hline
\hline
$ 1$         & $f_{\rm max}$ & $ \frac{1}{N}\ \ldots\ 1$ \\
\hline
$ N - 1$     & $ \frac{1 - f_{\rm max}}{N - 1}$ & $ 0\ \ldots\ \frac{1}{N}$ \\
\hline
\end{tabular}
\end{center}
\end{table}

\begin{table}
\caption{Probability values, $f_j$, that give a minimum
of disequilibrium, $D_{\rm min}$, for a given $H$.}
\label{tab:minimum}
\begin{center}
\begin{tabular}{|c|c|c|}
\hline
Number of states & $ f_j$ & Range of $ f_j$ \\
with $ f_j$      &        &                 \\
\hline
\hline
$ n$         & $ 0$ & $ 0$ \\
\hline
$ 1$         & $ f_{\rm min}$ & $ 0\ \ldots\ \frac{1}{N - n}$ \\
\hline
$ N - n - 1$     &
         $ \frac{1 - f_{\rm min}}{N - n - 1}$ &
         $ \frac{1}{N - n}\ \ldots\ \frac{1}{N - n - 1}$\\
\hline
\end{tabular}
\end{center}
\begin{center}
{$n$ can have the values $0, 1,\ \ldots\, N-2$.}
\end{center}
\end{table}

Equivalently, the values, $f_j$, that give a minimum complexity
are shown in Table \ref{tab:minimum}. One of the solutions,
$f_{\rm min}$, is given by
\begin{equation}
H = - \frac{1}{\ln N} \left[ f_{\rm min} \ln f_{\rm min} + ( 1 - f_{\rm min} ) \ln
     \left( \frac{1 - f_{\rm min}}{N - n - 1} \right) \right]\,,
\end{equation}
where $n$ is the number of states with $f_j = 0$
and takes a value in the range $n = 0, 1,\ \ldots\ ,N - 2$.

\begin{figure}
\centerline{\includegraphics[width=8.5cm,angle=-90]{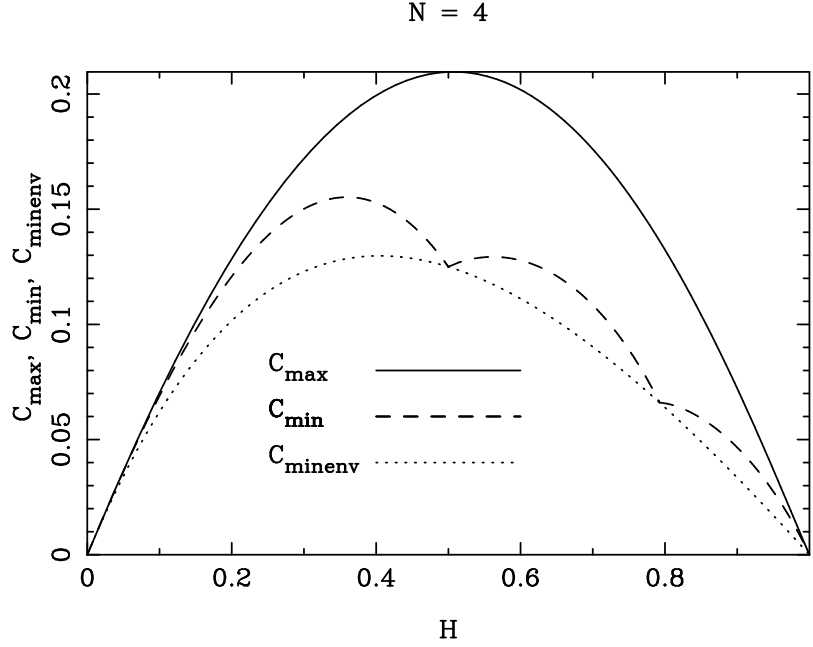}}
\caption{Maximum, minimum, and minimum envelope complexity, $C_{\rm max}$,
$C_{\rm min}$, and $C_{\rm minenv}$ respectively, as a function of
the entropy, $H$, for a system with $N=4$ accessible states.}
\label{fig:c_minmax}
\end{figure}

The resulting minimum disequilibrium, $D_{\rm min}$, for a given $H$ is,
\begin{equation}
D_{\rm min} = (f_{\rm min} - f_{\rm e})^2 + (N - n - 1)
        \left(\frac{1-f_{\rm min}}{N - n - 1} - f_{\rm e}\right)^2 + n f_{\rm e}^2\,.
\end{equation}
Note that in this case $f_j = 0$  is an additional
hidden solution that stems from the positive restriction
in the $f_i$ values. To obtain these solutions explicitly we can define $x_i$ such that
$f_i \equiv {x_i}^2$. These $x_i$ values do not have the restriction of positivity 
imposed to $f_i$ and can take a positive or negative value.
If we repeat the Lagrange multiplier method with these
new variables a new solution arises: $x_j = 0$, or equivalently, $f_j = 0$.

\begin{figure}
\centerline{\includegraphics[width=8.5cm,angle=-90]{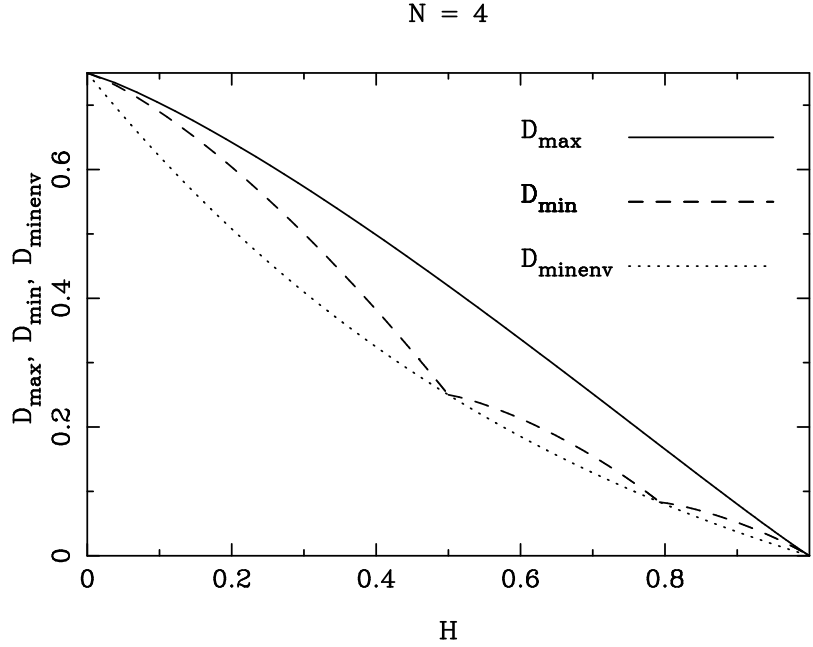}}
\caption{Maximum, minimum, and minimum envelope disequilibrium, $D_{\rm max}$,
$D_{\rm min}$, and $D_{\rm minenv}$ respectively, as a function of
the entropy, $H$, for a system with $N=4$ accessible states.}
\label{fig:d_minmax}
\end{figure}

The resulting minimum complexity, which again only depends on $H$, is
\begin{equation}
\label{defcmin}
C_{\rm min}(H)= D_{\rm min} \cdot H\,.
\end{equation}
As an example, the maximum and minimum of complexity, $C_{\rm max}$ and $C_{\rm min}$,
are plotted as a function of the entropy, $H$, in Fig.~\ref{fig:c_minmax}
for $N=4$. Also, in this figure, it is shown the minimum envelope complexity,
$C_{\rm minenv}=D_{\rm minenv}\cdot H$, where $D_{\rm minenv}$ is defined below.
In Fig.~\ref{fig:d_minmax} the maximum and minimum disequilibrium,
$D_{\rm max}$ and $D_{\rm min}$, versus $H$ are also shown.

As shown in Fig.~\ref{fig:d_minmax} the minimum disequilibrium function is piecewise
defined, having several points where its derivative
is discontinuous. Each of these function pieces corresponds to a different value of
$n$ (Table \ref{tab:minimum}).In some circumstances it might be helpful
to work with the ``envelope'' of the minimum disequilibrium
function. The function, $D_{\rm minenv}$, that traverses
all the discontinuous derivative points in the $D_{\rm min}$ versus $H$ plot is
\begin{equation}
D_{\rm minenv} = e^{- H \ln N} - \frac{1}{N}\,,
\end{equation}
and is also shown in Figure~\ref{fig:d_minmax}. 

When $N$ tends toward infinity the probability, $f_{\rm max}$,
of the dominant state has a linear dependence with the entropy,
\begin{equation}
\lim_{N \rightarrow \infty} f_{\rm max} = 1 - H\,,
\end{equation}
and thus the maximum disequilibrium scales as 
$\lim_{N \rightarrow \infty} D_{\rm max} = (1-H)^2$.
The maximum complexity tends to
\begin{equation}
\label{eq:cmaxlim}
\lim_{N \rightarrow \infty} C_{\rm max} = H \cdot (1-H)^2\,.
\end{equation}
The limit of the minimum disequilibrium and complexity vanishes,
$\lim_{N \rightarrow \infty} D_{\rm minenv} = 0$, and thus
\begin{equation}
\label{eq:cminlim}
\lim_{N \rightarrow \infty} C_{\rm min} = 0\,.
\end{equation}
In general, in the limit $N \rightarrow \infty$,
the complexity is not a trivial function of the entropy,
in the sense that for a given $H$ there exists
a range of complexities between $0$ and $C_{\rm max}$,
given by Eqs. (\ref{eq:cminlim}) and (\ref{eq:cmaxlim}), respectively.

In particular, in this asymptotic limit, the maximum of $C_{\rm max}$ is found when
$H=1/3$, or equivalently $f_{\rm max}=2/3$,
which gives a maximum of the maximum complexity of $C_{\rm max}=4/27$.
This value was numerically calculated in Ref. \cite{anteneodo96}.

\subsubsection{An out equilibrium system: the tetrahedral gas}
\label{sec:tetra}

We present a simplified example of an ideal gas: the tetrahedral
gas.  This system is generated by a simplification of the 
Boltzmann integro--differential equation of an ideal gas. 
We are interested in studying the disequilibrium time evolution.

\begin{figure}
\centerline{\includegraphics[width=5.5cm,angle=-90]{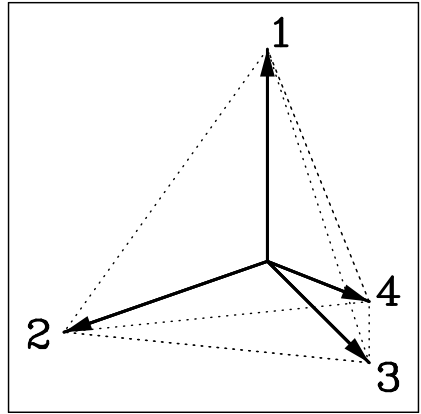}}
\caption{The four possible directions of the velocities of the
tetrahedral gas in space. Positive senses are defined
as emerging from the center point and with integer numbers $1, 2, 3, 4$.}
\label{fig:tetra}
\end{figure}

The Boltzmann integro--differential equation
of an ideal gas with no external forces and no spatial gradients is
\begin{equation}
\label{eq:boltzman}
\frac{\partial f({\bf v};t)}{\partial t} =
        \int d^{3} {\bf v}_{*} \int d \Omega_{\rm c.m.}
        \sigma( {\bf v}_{*} - {\bf v} \rightarrow
        {\bf v}'_{*} - {\bf v}' )
        | {\bf v}_{*} - {\bf v} |  
		\left[ f({\bf v}'_{*};t)
        f({\bf v}';t) - f({\bf v}_{*};t) f({\bf v};t)
        \right]\,,
\end{equation}
where $\sigma$ represents the cross section of a collision
between two particles with initial velocities ${\bf v}$ and
${\bf v}_{*}$ and after the collision with velocities
${\bf v}'$ and ${\bf v}'_{*}$; and $\Omega_{\rm c.m.}$ are
all the possible dispersion angles of the collision as
seen from its center of mass.

In the tetrahedral gas, the particles can travel only in four directions
in three-dimensional space and all have the same absolute velocity. 
These directions are the ones given by joining the center of a tetrahedron
with its corners. The directions can be easily viewed by recalling
the directions given by a methane molecule, or equivalently,
by a caltrop, which is a device with four metal points
so arranged that when any three are on the ground the fourth
projects upward as a hazard to the hooves of horses or to
pneumatic tires (see Fig.~\ref{fig:tetra}).

By definition, the angle that one direction forms with any
other is the same.
It can be shown that the angles between different directions,
$\alpha$, satisfy the relationship $\cos \alpha = - 1/3$,
which gives $\alpha=109.47^{\circ}$. The plane formed by any
two directions is perpendicular to the plane formed by the
remaining two directions.

We assume that the cross-section, $\sigma$, is different from
zero only when the angle between the velocities of the colliding
particles is $109.47^{\circ}$. It is also assumed that this collision
makes the two particles leave in the remaining two directions,
thus again forming an angle of $109.47^{\circ}$.
A consequence of these restrictions is that the
modulus of the velocity is always the same no matter how many collisions
a particle has undergone and they always stay within the directions
of the vertices of the tetrahedron. Furthermore, this type of gas
does not break any law of physics and is perfectly valid, although
hypothetical.

We label the four directions originating from
the center of the caltrop with numbers, ${\bf 1},
{\bf 2}, {\bf 3}, {\bf 4}$ (see Fig.~\ref{fig:tetra}).
The velocity components with the same direction but opposite
sense, or equivalently, directed toward the center of the caltrop,
are labeled with negative numbers ${\bf -1}, {\bf -2}, {\bf  -3}, {\bf -4}$.

In order to formulate the Boltzmann equation for the tetrahedral gas,
and because all directions are equivalent,
we need only study the different collisions that a particle with
one fixed direction can undergo.
In particular if we take a particle with direction ${\bf -1}$
the result of the collision with another particle
with direction ${\bf -2}$ are the same two particles traveling 
in directions ${\bf 3}$ and ${\bf 4}$, that is,
\begin{equation}
({\bf -1}, {\bf -2}) \rightarrow ( {\bf 3}, {\bf 4})\,. 
\end{equation}
With this in mind the last bracket of Eq. (\ref{eq:boltzman}) is,
\begin{equation}
f_{3} f_{4} - f_{-1} f_{-2}\,, 
\end{equation}
where $f_{i}$ denotes the probability of finding
a particle in direction ${\bf i}$.
Note that the dependence on velocity, ${\bf v}$,
of the continuous velocity distribution
function, $f({\bf v};t)$, of Eq. (\ref{eq:boltzman})
is in our case contained in the discrete subindex,
$i$, of the distribution function $f_i$.

We can proceed in the same manner with the other remaining collisions,
\begin{eqnarray}
({\bf -1}, {\bf -3}) &\rightarrow ( {\bf 2}, {\bf 4})\,, \nonumber\\
({\bf -1}, {\bf -4}) &\rightarrow ( {\bf 2}, {\bf 3})\,.
\end{eqnarray}
When a particle with direction ${\bf -1}$ collides with a particle
with direction ${\bf 2}$, they do not form an angle of $109.47^\circ$; i.e., 
they do not collide, they just pass by each other.
This is a consequence of the previous assumption for the tetrahedral gas,
which establishes a null cross section for angles different from $109.47^\circ$.
The same can be said for collisions $({\bf -1}, {\bf 3})$, $({\bf -1}, {\bf 4})$, 
and $({\bf -1}, {\bf 1})$. All these results are summarized in Table \ref{tab:col}.

\begin{table}
\caption{Cross sections, $\sigma$, for a particle in direction ${\bf -1}$ colliding
with particles in the other remaining directions of the tetrahedral gas.}
\label{tab:col}
\begin{center}
\begin{tabular}{|c|c|}
\hline
Collision       &     Cross section \\
 of particles   &     $\sigma$      \\
\hline
$({\bf -1}, {\bf -2}) \rightarrow ( {\bf 3}, {\bf 4})$  &  $1$ \\
$({\bf -1}, {\bf -3}) \rightarrow ( {\bf 2}, {\bf 4})$  &  $1$ \\
$({\bf -1}, {\bf -4}) \rightarrow ( {\bf 2}, {\bf 3})$  &  $1$ \\
Other collisions                                        & $0$ \\
\hline
\end{tabular}
\end{center}
\end{table}

Taking all this into account, Eq. (\ref{eq:boltzman}) for
direction ${\bf -1}$ is reduced to a discrete sum,
\begin{equation}
\frac{d f_{-1}}{d t} =
                       ( f_{3} f_{4} - f_{-1} f_{-2} ) +
                       ( f_{2} f_{4} - f_{-1} f_{-3} ) +
                       ( f_{2} f_{3} - f_{-1} f_{-4} )\,,
\end{equation}
where all other factors have been set to unity for simplicity.

The seven remaining equations for the rest of directions can be easily inferred.
If we now make $f_{i} = f_{-i} ( i=1,2,3,4 )$ initially, this property is conserved in time.
The final four equations defining the evolution of the system are:
\begin{eqnarray}
\label{eq:tetra}
\frac{d f_{1}}{d t} = ( f_{3} f_{4} - f_{1} f_{2} ) +
                                ( f_{2} f_{4} - f_{1} f_{3} ) +
                                ( f_{2} f_{3} - f_{1} f_{4} )\,, \nonumber \\
\frac{d f_{2}}{d t} = ( f_{3} f_{4} - f_{1} f_{2} ) +
                                ( f_{1} f_{4} - f_{2} f_{3} ) +
                                ( f_{1} f_{3} - f_{2} f_{4} )\,, \nonumber \\
\frac{d f_{3}}{d t} = ( f_{2} f_{4} - f_{3} f_{1} ) +
                                ( f_{1} f_{4} - f_{3} f_{2} ) +
                                ( f_{1} f_{2} - f_{3} f_{4} )\,, \nonumber \\
\frac{d f_{4}}{d t} = ( f_{2} f_{3} - f_{4} f_{1} ) +
                                ( f_{1} f_{3} - f_{4} f_{2} ) +
                                ( f_{1} f_{2} - f_{3} f_{4} )\,.
\end{eqnarray}
Note that the ideal gas has been reduced to the tetrahedral gas, which is a four-dimensional
dynamical system. The velocity distribution function, $f_i$, corresponds to a probability 
distribution function with $N=4$ accessible states that evolve in time.

\subsubsection{Evolution of the tetrahedral gas with time}
\label{sec:evol}

To study the time evolution of the complexity, a diagram of $C$ versus time, $t$, 
can be used. But, as we know, the second law of thermodynamics states that
the entropy grows monotonically with time, that is,
\begin{equation}
\frac{d H}{d t} \geq 0.
\end{equation}
This implies that an equivalent way to study the
time evolution of the complexity
can be obtained by plotting $C$ versus $H$.
In this way, the entropy substitutes the
time axis, since the former increases monotonically
with the latter.
The conversion from $C$ vs. $H$ to $C$ vs. $t$ diagrams
is achieved by stretching or shrinking
the entropy axis according to its time evolution.
This method is a key point in all this discussion.
Note that, in any case,
the relationship of $H$ versus $t$ will, in general, not be a simple
one \cite{latora99}.

The tetrahedral gas, Eqs. (\ref{eq:tetra}), reaches equilibrium
when $f_{i} = 1/N$ for $i=1,2,3,4$ and $N=4$.
This stationary state, $d f_i / d t = 0$, represents the equiprobability
towards which the system evolves in time. This is consistent with
the definition of disequilibrium in which we assumed 
that equilibrium was reached at equiprobability, $f_i=f_{\rm e}$, where $D=0$.

\begin{figure}
\centerline{\includegraphics[width=5.5cm,angle=-90]{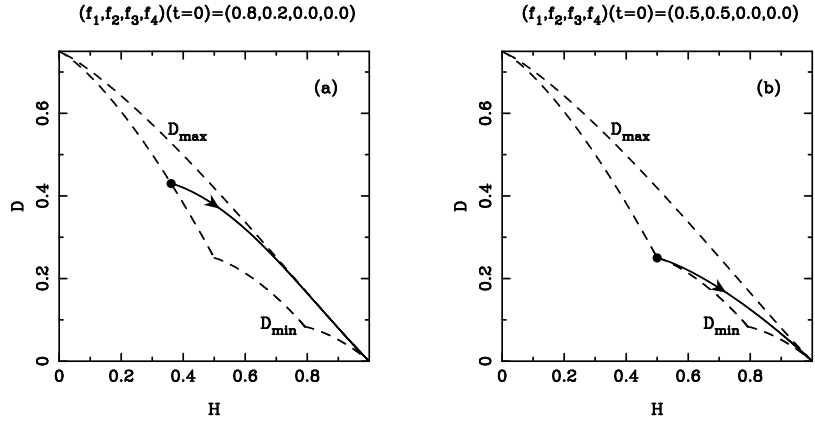}}
\caption{Time evolution of the system in $(H,D)$ phase space
for two different initial conditions at time $t=0$:
(a) $(f_1, f_2, f_3, f_4) = (0.8, 0.2, 0, 0)$
and (b) $(f_1, f_2, f_3, f_4) = (0.5, 0.5, 0, 0)$.
The maximum and minimum disequilibrium are shown by
dashed lines.}
\label{fig:evold}
\end{figure}

As the isolated system evolves it gets closer and closer to equilibrium.
In this sense, one may intuitively think that the disequilibrium
will decrease with time. In fact, it can be analytically shown \cite{calbet01}
that, as the system approaches to equilibrium, $D$ tends to zero monotonically with time:
\begin{equation}
\label{eq:dtlez}
\frac{d D}{d t} \le 0.
\end{equation}
There are even more restrictions on the evolution
of this system. It would be expected that the system approaches equilibrium, 
$D=0$, by following the most direct path.
To verify this, numerical simulations for several initial conditions
have been undertaken. In all of these we observe the additional restriction that $D$
approaches $D_{\rm max}$ on its way to $D=0$. In fact it appears
as an exponential decay of $D$ towards $D_{\rm max}$ in a $D$ versus $H$ plot.
As an example, two of these are shown in Fig.~\ref{fig:evold}, where
Fig.~\ref{fig:evold}(a) shows a really strong tendency towards $D_{\rm max}$.
Contrary to intuition, among all the possible
paths that the system can follow toward equilibrium,
it chooses those closest to $D_{\rm max}$ in particular.

\begin{figure}
\centerline{\includegraphics[width=5.5cm,angle=-90]{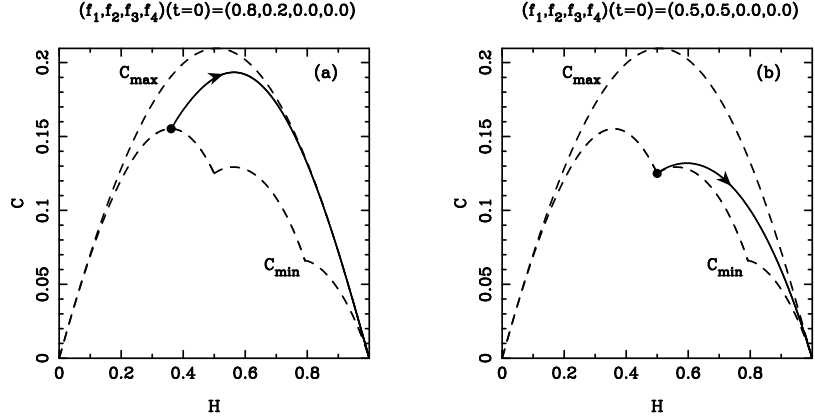}}
\caption{Time evolution of the system in $(H,C)$ phase space
for two different initial conditions at time $t=0$:
(a) $(f_1, f_2, f_3, f_4) = (0.8, 0.2, 0, 0)$
and (b) $(f_1, f_2, f_3, f_4) = (0.5, 0.5, 0, 0)$.
The maximum and minimum complexity are shown by
dashed lines.}
\label{fig:evolc}
\end{figure}

We can also observe this effect in a complexity, $C$, versus $H$ plot.
This is shown for the same two initial conditions in Figure~\ref{fig:evolc}.

\begin{figure}
\centerline{\includegraphics[width=5.5cm,angle=-90]{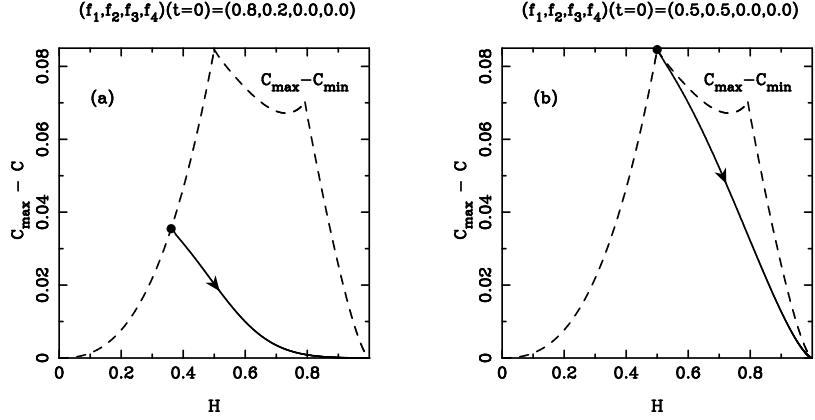}}
\caption{Time evolution of the system in $(H,C_{\rm max}-C)$ phase space
for two different initial conditions at time $t=0$:
(a) $(f_1, f_2, f_3, f_4) = (0.8, 0.2, 0, 0)$
and (b) $(f_1, f_2, f_3, f_4) = (0.5, 0.5, 0, 0)$.
The values $C_{\rm max} - C_{\rm min}$ are shown by dashed lines.}
\label{fig:evolcmax}
\end{figure}

\begin{figure}
\centerline{\includegraphics[width=7.5cm,angle=-90]{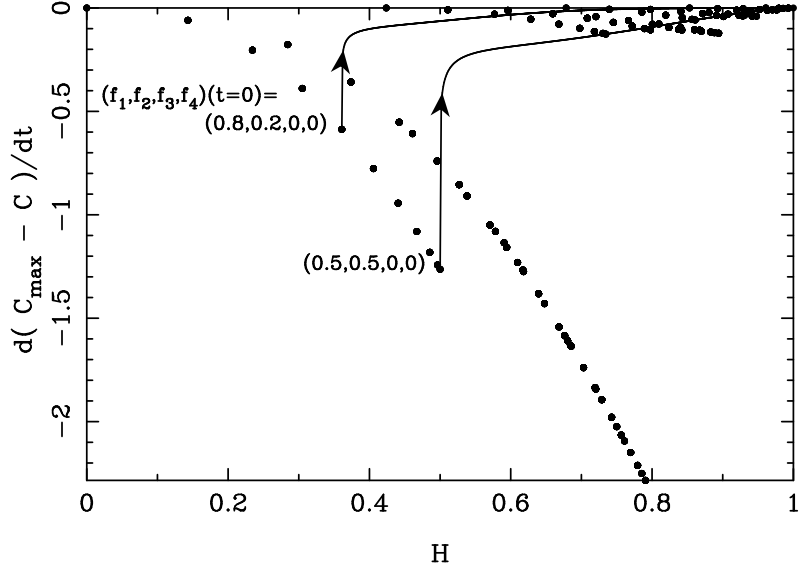}}
\caption{Numerical verification of $d (C_{\rm max} - C)/d t \le 0$.
This time derivative is shown as a function of $H$.
A grid of uniformly spaced, $\Delta f_{i} = 0.5$,
distribution functions, $(f_1, f_2, f_3, f_4)$,
satisfying the normalization condition $I$, have been used.
Two system trajectories for initial
conditions, $t=0$, $(f_1, f_2, f_3, f_4) = (0.8, 0.2, 0, 0)$
and $(f_1, f_2, f_3, f_4) = (0.5, 0.5, 0, 0)$ are also shown 
for illustrative purposes. It can be seen how the above-mentioned 
time derivative always remains negative.}
\label{fig:cmaxmct}
\end{figure}

This additional restriction to the evolution of the system is better viewed by
plotting the difference $C_{\rm max} - C$ versus $H$.
In all the cases analyzed (see two of them in Fig.~\ref{fig:evolcmax})
the following condition is observed:
\begin{equation}
\frac{d (C_{\rm max} - C)}{d t} \le 0\,.
\end{equation}
This has been verified numerically and is illustrated in Figure~\ref{fig:cmaxmct}, 
where this time derivative, which always remains negative, is shown as a function
of $H$ for a grid of uniformly spaced distribution functions,
$(f_1, f_2, f_3, f_4)$, satisfying the normalization condition $I$.
Two system trajectories are also shown for illustrative
purposes. The numerical method used to plot this function is explained in \cite{calbet01}.

We proceed now to show another interesting property of this system.
As shown in Table \ref{tab:maximum}, a collection of maximum complexity
distributions for $N=4$ can take the form
\begin{eqnarray}
\label{eq:coll}
f_1 &=& f_{\rm max} \,,\nonumber \\
f_i &=& \frac{1-f_{\rm max}}{3}\,, i=2,3,4 \,,
\end{eqnarray}
where $f_{\rm max}$ runs from $1/N$ (equiprobability distribution)
to $1$ (``crystal'' distribution). The complexity of this collection
of distributions covers all possible values of $C_{\rm max}$.

There is actually a time evolution of the tetrahedral gas,
or trajectory of the system, formed by this collection of distributions.
Inserting Eqs. (\ref{eq:coll}) in the evolution Eqs. (\ref{eq:tetra}),
it is found that all equations are compatible
with each other and the dynamical equations are reduced to the relation,
\begin{equation}
\frac{d f_{\rm max}}{d t} = \frac{1}{3} ( 4 f_{\rm max}^2 - 5 f_{\rm max} + 1)\,.
\end{equation}
This trajectory is denoted as the {\it maximum complexity path}.

Note that the equiprobability or equilibrium, $f_{\rm max} = 1/4$, 
is a  stable fixed point and the  maximum disequilibrium ``crystal'' distribution,
$f_{\rm max}=1$, is an unstable fixed point.
Thus the maximum complexity path is a heteroclinic
connection between the ``crystal'' and equiprobability distributions.

The maximum complexity path is locally attractive.
Let us assume, for instance, the following perturbed trajectory
\begin{eqnarray}
f_1 &=& f_{\rm max}\,, \nonumber \\
f_2 &=& \frac{1-f_{\rm max}}{3}\,, \nonumber \\
f_3 &=& \frac{1-f_{\rm max}}{3} + \delta\,, \nonumber \\
f_4 &=& \frac{1-f_{\rm max}}{3} - \delta\,,
\end{eqnarray}
whose evolution according to Eqs. (\ref{eq:tetra}) gives the
exponential decay of the perturbation, $\delta$:
\begin{equation}
\frac{d \delta}{d t} \sim - \left( \frac{4 f_{\rm max} + 2}{3} \right) \delta\,,
\end{equation}
showing the attractive nature of these trajectories.

\begin{figure}
\centerline{\includegraphics[width=7.5cm,angle=-90]{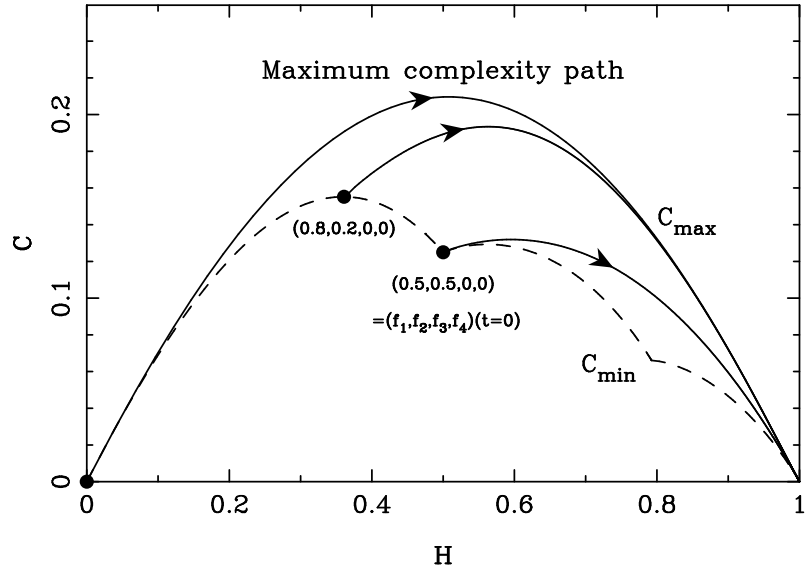}}
\caption{The time evolution of the system for
three different initial conditions,
$t=0$, $(f_1, f_2, f_3, f_4) = (0.8, 0.2, 0, 0)$,
 $(f_1, f_2, f_3, f_4) = (0.5, 0.5, 0, 0)$, and 
 the maximum complexity path are shown.
The minimum complexity is shown by dashed lines.
It can be seen how the system tends to approach the maximum complexity
path as it evolves in time toward equilibrium.}
\label{fig:suma}
\end{figure}

\subsubsection{Conclusions and further remarks}
\label{sec:conclu}

In the former section, the time evolution of the LMC complexity, $C$,
has been studied for a simplified model of an isolated ideal gas: the tetrahedral gas.
In general, the dynamical behavior of this quantity is bounded
between two extremum curves, $C_{\rm max}$ and $C_{\rm min}$,
when observed in a $C$ versus $H$ phase space.
These complexity bounds have been derived and computed.
A continuation of this work applied to the study of complexity in
gases out of equilibrium can be found in Refs. \cite{calbet07,calbet09}.

For the isolated tetrahedral gas two constraints on its dynamics are found.
The first, which is analytically demonstrated, is
that the disequilibrium, $D$, decreases monotonically with
time until it reaches the value $D=0$ for the equilibrium state.
The second is that the maximum complexity paths, $C_{\rm max}$,
are attractive in phase space. In other words,
the complexity of the system tends to equilibrium
always approaching those paths. This has been verified numerically,
that is, the time derivative of the difference between
$C_{\rm max}$ and $C$ is negative.
Fig.~\ref{fig:suma} summarizes the dynamical behavior of 
the tetrahedral gas. The different trajectories starting  with 
arbitrary initial conditions, which represent systems out of
equilibrium, evolve towards equilibrium approaching
the maximum complexity path.

Whether these properties are useful in real
physical systems can need of a further inspection,
particularly the macroscopical nature of the disequilibrium 
in more general systems, such as to the
ideal gas following the complete Boltzmann integro--differential
equation. Another feature that could deserve attention is 
the possibility of approximating the evolution of a real physical
system trajectory to its maximum complexity path.
Note that in general, for a real system, the calculation of the maximum
complexity path will not be an easy task.


\section{The Statistical Complexity in the Continuous Case}

As explained in the former sections,
the LMC statistical measure of complexity \cite{lopezruiz95} identifies the
entropy or information stored in a system and its distance to the
equilibrium probability distribution, the disequilibrium, as the
two ingredients giving the correct asymptotic properties of a
well-behaved measure of complexity. In fact, it vanishes both for
completely ordered and for completely random systems. Besides
giving the main features of an intuitive notion of complexity, it
has been shown that LMC complexity successfully enables us to
discern situations regarded as complex in discrete systems out of
equilibrium: one instance of phase transitions via intermittency 
in coupled logistic maps \cite{sanchez05} or via stochastic 
synchronization in cellular automata \cite{sanchez05-1}, the dynamical
behavior of this quantity in a out-equilibrium gases \cite{calbet01,calbet07,calbet09} 
and other applications in classical statistical mechanics \cite{lopezruiz05,lopezruiz01}.

A possible formula of LMC complexity for continuous systems was
suggested in formula (\ref{def-C-continuo}). Anteneodo and
Plastino \cite{anteneodo96} pointed out some peculiarities concerning
such an extension for continuous probability distributions. 
It is the aim of this section to offer a discussion of the
extension of LMC complexity for continuous systems and to present 
a slightly modified extension \cite{garay02} of expression (\ref{def-C-continuo}) 
that displays interesting and very striking properties.
A further generalization of this work has been done 
in Ref. \cite{romera09,lopezruiz09}.  

In Sect. \ref{sec:h-d} the extension of information and disequilibrium
concepts for the continuous case are discussed. In Sect. \ref{sec:lmc-c} 
the LMC measure of complexity is rewieved and possible extensions for
continuous systems are suggested. We proceed to present some
properties of one of these extensions in Sect. \ref{sec:prop-c}.

\subsection{Entropy/information and disequilibrium}
\label{sec:h-d}

Depending on the necessary conditions to fulfill, the extension of
an established formula from the discrete to the continuous case
always requires a careful study and in many situations some kind
of choice between several possibilities. Next we carry out this
process for the entropy and disequilibrium formulas.

\subsubsection{Entropy or information}
\label{sec:h}

As we know, given a discrete probability distribution $\{p_i\}_{i=1,2,...,N}$
satisfying $p_i\geq 0$ and $\sum_{i=1}^N p_i =1$, the {\it
Boltzmann-Gibss-Shannon formula} \cite{shannon49} that accounts for
the entropy or information, $S$, stored in a system is defined by
\begin{equation}
S(\{p_i\}) = -k \sum_{i=1}^N p_i\log p_i \, , \label{eq:def-h}
\end{equation}
where $k$ is a positive constant. If we identify $H$ with $S$,
then some properties of this quantity
are: (i) {\it positivity}: $H\geq 0$ for any arbitrary set
$\{p_i\}$, (ii) {\it concavity}: $H$ is concave for arbitrary
$\{p_i\}$ and reaches the extremal value for equiprobability
($p_i=1/N$ $\forall i$), (iii) {\it additivity}: $H(A\cup
B)=H(A)+H(B)$ where $A$ and $B$ are two independent systems, and
(iv) {\it continuity}: $H$ is continuous for each of its arguments.
And vice versa, it has been shown that
the only function of $\{p_i\}$ verifying the latter properties is
given by Eq. (\ref{eq:def-h}) \cite{shannon49, khinchin57}.
For an isolated system, the {\it irreversibility} property is 
also verified, that is, the time derivative of $H$ is
positive, $dH/dt\geq 0$, reaching the equality
only for equilibrium. 

Calculation of $H$ for a continuous probability distribution
$p(x)$, with support on $[-L,L]$ and $\int_{-L}^{L}p(x)\,dx = 1$,
can be performed by dividing the interval $[-L,L]$ in small
equal-length pieces $\Delta x = x_i-x_{i-1}$, $i=1,\cdots,n$, with
$x_0=-L$ and $x_n=L$, and by considering the approximated discrete
distribution $\{p_i\} = \{p(\bar{x}_i)\Delta x\}$, $i=1,\cdots,n$,
with $\bar{x}_i$ a point in the segment $[x_{i-1},x_i]$. It gives us
\begin{eqnarray}
H^* & = & H(\{p_i\})\; = \label{eq:def-h*} \\
     & = & -\, k \sum_{i=1}^n
p(\bar{x}_i)\log p(\bar{x}_i)\,\Delta x\, -  \, k \sum_{i=1}^n
p(\bar{x}_i)\log (\Delta x)\,\Delta x \,. \nonumber
\end{eqnarray}
The second adding term of $H^*$ in the expression
(\ref{eq:def-h*}) grows as $\log n$ when $n$ goes to infinity.
Therefore it seems reasonable to take just the first and finite
adding term of $H^*$ as the extension of $H$ to the continuous
case: $H(p(x))$. It characterizes with a finite number the information
contained in a continuous distribution $p(x)$. In the limit
$n\rightarrow\infty$, we obtain
\begin{eqnarray}
H(p(x)) & = & \lim_{n\rightarrow\infty} \left[-k \sum_{i=1}^n
p(\bar{x}_i)\log p(\bar{x}_i)\,\Delta x \right]\; = \nonumber \\
  & = & -k \int_{-L}^{L} p(x)\log p(x)\,dx \,. \label{eq:def-hx}
\end{eqnarray}
If $p(x)\geq 1$ in some region, the entropy defined by Eq.
(\ref{eq:def-hx}) can become negative. Although this situation is
mathematically possible and coherent, it is unfounded from a
physical point of view. See \cite{wehrl78} for a discussion on this
point. Let $f(p,q)$ be a probability distribution in phase space
with coordinates $(p,q)$, $f\geq 0$ and $dp\,dq$ having the
dimension of an action. In this case the volume element is
$dp\,dq/h$ with $h$ the Planck constant. Suppose that $H(f)<0$.
Because of $\int (dp\,dq/h) f = 1$, the extent of the region where
$f>1$ must be smaller than $h$. Hence a negative classical entropy
arises if one tries to localize a particle in phase space in a
region smaller than $h$, that is, if the uncertainty relation is
violated. In consequence, not every classical probability
distribution can be observed in nature. The condition $H(f)=0$
could give us the mininal width that is physically allowed for the
distribution and so the maximal localization of the system under
study. This {\it cutting} property has been used in the calculations 
performed in \cite{lopezruiz01}.

\subsubsection{Disequilibrium}

Given a discrete probability distribution $\{p_i\}_{i=1,2,...,N}$
satisfying $p_i\geq 0$ and $\sum_{i=1}^N p_i =1$, its {\it
Disequilibrium}, $D$, can be defined as 
the quadratic distance of the actual
probability distribution $\{p_i\}$ to equiprobability:
\begin{equation}
D(\{p_i\}) = \sum_{i=1}^N \,\left( p_i - \frac{1}{N} \right)^2 \,.
\label{eq:def-d}
\end{equation}
$D$ is maximal for fully regular systems and vanishes for
completely random ones.

In the continuous case with support on the interval $[-L,L]$, the
rectangular function $p(x)=1/(2L)$, with $-L<x<L$, is the natural
extension of the equiprobability distribution of the discrete
case. The disequilibrium could be defined as
\begin{equation}
D^* = \int_{-L}^L
\left(p(x)-\frac{1}{2L}\right)^2\,dx = \int_{-L}^{L}p^2(x)\,dx - \frac{1}{2L}\,.
\end{equation}
If we redefine $D$ omitting the constant adding term in $D^*$, the
disequilibrium reads now:
\begin{equation}
D(p(x)) = \int_{-L}^L p^2(x)\,dx \,.\label{eq:def-dc}
\end{equation}
$D>0$ for every distribution and it is minimal for the rectangular
function which represents the equipartition. $D$ does also tend
to infinity when the width of $p(x)$ narrows strongly and
becomes extremely peaked.

\subsection{The continuous version $\hat{C}$ of the LMC complexity}
\label{sec:lmc-c}

As shown in the previous sections,
LMC complexity has been successfully calculated in different systems out 
of equilibrium. However, Feldman and Cruchtfield \cite{feldman98} presented as 
a main drawback that $C$ vanishes and it is not an extensive variable for
finite-memory regular Markov chains when the system size
increases. This is not the general behavior of $C$ in the
thermodynamic limit as  it has been suggested by Calbet and
L\'opez-Ruiz \cite{calbet01}. On the one hand, when
$N\rightarrow\infty$ and $k=1/\log N$, LMC complexity is not a
trivial function of the entropy, in the sense that for a given $H$
there exists a range of complexities between $0$ and $C_{max}(H)$,
where $C_{max}$ is given by expression (\ref{eq:cmaxlim}).

Observe that in this case $H$ is normalized, $0<H<1$, because
$k=1/\log N$. On the other hand, non-extensitivity cannot be
considered as an obstacle since it is nowadays well known that
there exists a variety of physical systems for which the classical
statistical mechanics seems to be inadequate and for which an
alternative non-extensive thermodynamics is being hailed as a
possible basis of a theoretical framework appropriate to deal with
them \cite{tsallis98}.

According to the discussion in Section \ref{sec:h-d}, the
expression of $C$ for the case of a continuum number of states,
$x$, with support on the interval $[-L,L]$ and $\int_{-L}^L p(x)\,
dx = 1$, is defined by
\begin{eqnarray}
C(p(x)) & = & H(p(x))\cdot D(p(x))\; = \nonumber \\
    & = & \left( -k \int_{-L}^L p(x)\log p(x)\,dx \right)\cdot
    \left( \int_{-L}^L \,p^2(x)\,dx \right) \, .
    \label{eq:def-cc2}
\end{eqnarray}
Hence, $C$ can become negative. Obviously, $C<0$ implies $H<0$. Although this
situation is coherent from a mathematical point of view, it is not
physically possible. Hence a negative entropy means to localize a
system in phase space into a region smaller than $h$ (Planck
constant) and this would imply to violate the uncertainty
principle (see discussion of Section \ref{sec:h}). Then a
distribution can broaden without any limit but it cannot become
extremely peaked. The condition $H=0$ could indicate the minimal
width that $p(x)$ is allowed to have. Similarly to the discrete
case, $C$ is positive for any situation and vanishes both for an
extreme localization and for the most widely delocalization
embodied by the equiprobability distribution. Thus, LMC complexity
can be straightforwardly calculated for any continuous
distribution by Eq. (\ref{eq:def-cc2}). 
Anyway, the positivity of $C$ for every distribution in the continuous
case can be recovered by taking the exponential of $S$ \cite{dembo91}
and redefining $H$ according to this exponential, i.e. $H=e^{S}$. 
To maintain the same nomenclature than in the precedent text 
we continue to identify $H$ with $S$ and we introduce the symbol $\hat{H}=e^{H}$.
Then the new expression of the statistical measure of complexity  $C$ is identified
as $\hat{C}$ in the rest of this section and is given by \cite{garay02}
\begin{equation}
\hat{C}(p(x)) =  \hat{H}(p(x))\cdot D(p(x)) = e^{H(p(x))}\cdot
D(p(x))\; .  \label{eq:def-cc3}
\end{equation}
In addition to the positivity, $\hat{C}$ encloses other
interesting properties that we describe in the next section.

\subsection{Properties of $\hat{C}$}
\label{sec:prop-c}

The quantity $\hat{C}$ given by  Eq. (\ref{eq:def-cc3}) has been
presented as one of the possible extensions of the LMC complexity
for continuous systems \cite{garay02}. We proceed now to present some of the
properties that characterize such a complexity indicator.

\subsubsection{Invariance under translations and rescaling transformations}

If $p(x)$ is a density function defined on the real axis \R,
$\int_{\R} p(x)\,dx = 1$, and $\alpha>0$ and $\beta$
are two real numbers, we denote by $p_{\alpha,\beta}(x)$ the new
probability distribution obtained by the action of a
$\beta$-translation and an $\alpha$-rescaling transformation on
$p(x)$,
\begin{equation}
p_{\alpha,\beta}(x) = \alpha\, p\,(\alpha\, (x-\beta)) \; .
\label{eq:transf1}
\end{equation}
When $\alpha<1$, $p_{\alpha,\beta}(x)$ broadens whereas if
$\alpha>1$ it becomes more peaked. Observe that
$p_{\alpha,\beta}(x)$ is also a density function. After making the
change of variable $y=\alpha (x-\beta)$ we obtain
\begin{equation}
\int_{\R} p_{\alpha,\beta}(x)\, dx = \int_{\R} \alpha\,
p\,(\alpha\, (x-\beta))\, dx = \int_{\R} p(y)\, dy = 1\; .
\end{equation}
The behaviour of $H$ under the transformation given by Eq.
(\ref{eq:transf1}) is the following:
\begin{eqnarray}
H(p_{\alpha,\beta}) & = & -\int_{\R} p_{\alpha,\beta}(x) \log
p_{\alpha,\beta}(x)\, dx = -\int_{\R} p(y)\log (\alpha p(y))\, dy \nonumber\\
 & = & -\int_{\R} p(y)\log p(y)\, dy - \log \alpha \int_{\R} p(y)\, dy \nonumber\\
 & = & \;\; H(p) - \log \alpha \, .
\end{eqnarray}
Then,
\begin{equation}
\hat{H}(p_{\alpha,\beta}) = e^{H(p_{\alpha,\beta})} =
\frac{\hat{H}(p)}{\alpha} \; .
\end{equation}
It is straightforward to see that $D(p_{\alpha,\beta})=\alpha
D(p)$, and to conclude that
\begin{equation}
\hat{C}(p_{\alpha,\beta}) = \hat{H}(p_{\alpha,\beta})\cdot
D(p_{\alpha,\beta}) = \frac{\hat{H}(p)}{\alpha}\, \alpha D(p) =
\hat{C} (p)\; . \label{eq:transf-c-1}
\end{equation}
Observe that translations and rescaling transformations
keep also the shape of the distributions.
Then it could be reasonable to denominate the invariant quantity
$\hat{C}$ as the {\it shape complexity} of the family formed by 
a distribution $p(x)$ and its transformed $p_{\alpha,\beta}(x)$.
Hence, for instance, the rectangular $\Pi (x)$, the isosceles-triangle
shaped $\Lambda (x)$, the gaussian $\Gamma (x)$, or the
exponential $\Xi (x)$ distributions continue to belong to the
same $\Pi$, $\Lambda$, $\Gamma$ or $\Xi$ family, respectively,
after applying the transformations defined by Eq.
(\ref{eq:transf1}). Calculation of $\hat{C}$ on these distribution
families gives us
\begin{eqnarray}
\hat{C} (\Pi) & = & 1 \\ \hat{C} (\Lambda) & = &
\frac{2}{3}\,\sqrt{e}\approx 1.0991
\\ \hat{C} (\Gamma) & = & \,\sqrt{\frac{e}{2}}\;\approx 1.1658
\\ \hat{C} (\Xi) & = & \;\;\frac{e}{2}\;\;\;\approx 1.3591 \; .
\end{eqnarray}
Remark that the family of rectangular distributions has a 
smaller $\hat{C}$ than the rest of distributions. This fact is 
true for every distribution and it will be proved in Section
\ref{sec:min-c}.

\subsubsection{Invariance under replication}

Lloyd and Pagels \cite{lloyd88} recommend that a complexity measure
should remain essentially unchanged  under replication. We show
now that $\hat{C}$ is replicant invariant, that is, the shape
complexity of $m$ replicas of a given distribution is equal to the
shape complexity of the original one. \par Suppose $p(x)$ a
compactly supported density function, 
$\int_{-\infty}^{\infty} p(x)\, dx = 1$. Take $n$ copies $p_m(x)$,
$m=1,\cdots,n$, of $p(x)$,
\begin{equation}
p_m(x) = \frac{1}{\sqrt{n}}\; p(\sqrt{n}(x-\lambda_m))\, ,\;\;
1\leq m\leq n \, ,
\end{equation}
where the supports of all the $p_m(x)$, centered at $\lambda_m's$
points, $m=1,\cdots,n$, are all disjoint. Observe that
$\int_{-\infty}^{\infty} p_m(x)\, dx = \frac{1}{n}$, what make the
union
\begin{equation}
q(x)=\sum_{i=1}^n p_m (x)
\end{equation}
to be also a normalized probability distribution,
$\int_{-\infty}^{\infty} q(x)\, dx = 1$. For every $p_m(x)$, a
straightforward calculation shows that
\begin{eqnarray}
H(p_m) & = & \frac{1}{n}\, H(p) + \frac{1}{n} \log\sqrt{n} \\
D(p_m) & = & \frac{1}{n\sqrt{n}}\, D(p) \, .
\end{eqnarray}
Taking into account that the $m$ replicas are supported on
disjoint intervals on \R, we obtain
\begin{eqnarray}
H(q) & = & H(p) + \log\sqrt{n}\, , \\
 D(q) & = & \frac{1}{\sqrt{n}}\, D(p) \, .
\end{eqnarray}
Then,
\begin{equation}
\hat{C}(q) = \hat{C} (p) \, ,
\end{equation}
what completes the proof of the replicant invariance of $\hat{C}$.

\subsubsection{Near-continuity}

Continuity is a desirable property of an indicator of complexity.
For a given scale of observation, similar systems should have a
similar complexity. In the continuous case, similarity between
density functions defined on a common support suggests that they
take close values almost everywhere. More strictly speaking, let
$\delta$ be a positive real number. It will be said that two
density functions $f(x)$ and $g(x)$ defined on the interval
$I\in\R$ are {\it $\delta$-neighboring functions} on $I$ if the
Lebesgue measure of the points $x\in I$ verifying $\mid
f(x)-g(x)\mid\geq\delta$ is zero. A real map $T$ defined on
density functions on $I$ will be called {\it near-continuous} if
for any $\epsilon>0$ there exists $\delta(\epsilon)>0$ such that
if $f(x)$ and $g(x)$ are $\delta$-neighboring functions on $I$
then $\mid T(f)-T(g)\mid <\epsilon$.

It can be shown that the information $H$, the disequilibrium $D$
and the shape complexity $\hat{C}$ are near-continuous maps on the
space of density functions defined on a compact support. We must
stress at this point the importance of the compactness condition
of the support in order to have near-continuity. Take, for
instance, the density function defined on the interval $[-1,L]$,
\begin{equation}
g_{\delta,L}(x)= \left\{
\begin{array}{cl}
1-\delta & \mbox{if}\;\; -1\leq x\leq 0 \\
 \frac{\delta}{L} & \mbox{if}\;\;\;\;\;\;\, 0\leq x\leq L \\
 0 & \mbox{otherwise} \; \; ,
\end{array}  \right.
\label{eq:gdl}
\end{equation}
with $0<\delta<1$ and $L>1$. If we calculate $H$ and $D$ for this
distribution we obtain
\begin{eqnarray}
H(g_{\delta,L}) & = & -(1-\delta) \log (1-\delta) - \delta
\log\left(\frac{\delta}{L}\right) \\
 D(g_{\delta,L}) & = & (1-\delta)^2 + \frac{\delta^2}{L} \, .
\end{eqnarray}
Consider also the rectangular density function
\begin{equation}
\chi_{[-1,0]}(x)= \left\{
\begin{array}{cl}
1 & \mbox{if}\;\; -1\leq x\leq 0 \\
 0 &  \mbox{otherwise} \; \; .
\end{array}  \right.
\end{equation}
If $0<\delta<\bar{\delta}<1$, $g_{\delta,L}(x)$ and
$\chi_{[-1,0]}(x)$ are $\bar{\delta}$-neighboring functions. When
$\delta\rightarrow 0$, we have that $\lim_{\delta\rightarrow 0}
g_{\delta,L}(x) = \chi_{[-1,0]}(x)$. In this limit process the
support is maintained and near-continuity manifests itself as
following,
\begin{equation}
\left[\lim_{\delta\rightarrow 0} \hat{C} (g_{\delta,L})\right] =
\hat{C}(\chi_{[-1,0]}) = 1 \, .
\end{equation}
But if we allow the support $L$ to become infinitely large, the
compactness condition is not verified and, although
$\lim_{L\rightarrow\infty} g_{\delta,L}(x)$ and $\chi_{[-1,0]}(x)$
are $\bar{\delta}$-neighboring distributions, we have that
\begin{equation}
\left[\left(\lim_{L\rightarrow\infty} \hat{C}
(g_{\delta,L})\right)\rightarrow \infty \right] \neq
\hat{C}(\chi_{[-1,0]}) = 1 \, .
\end{equation}
Then near-continuity in the map $\hat{C}$ is lost due to the
non-compactness of the support when $L\rightarrow\infty$. This
example suggests that the shape complexity $\hat{C}$ is
near-continuous on compact supports and this property will be
rigorously proved elsewhere.

\subsubsection{The minimal shape complexity}
\label{sec:min-c}

If we calculate $\hat{C}$ on the example given by Eq.
(\ref{eq:gdl}), we can verify that the shape complexity can be as
large as wanted. Take, for instance, $\delta=\frac{1}{2}$. The
measure $\hat{C}$ reads now
\begin{equation}
\hat{C}(g_{\delta=\frac{1}{2},L}) = \frac{1}{2} \sqrt{L}\left( 1 +
\frac{1}{L}\right) \, .
 \label{eq:cd0.5}
\end{equation}
Thus $\hat{C}$ becomes infinitely large after taking the limits
$L\rightarrow 0$ or $L\rightarrow\infty$. Remark that even in the
case $g_{\delta,L}$ has a finite support, $\hat{C}$ is not upper
bounded. The density functions,
$g_{(\delta=\frac{1}{2}),(L\rightarrow 0)}$ and
$g_{(\delta=\frac{1}{2}),(L\rightarrow \infty)}$, of infinitely
increasing complexity have two zones with different probabilities.
In the case $L\rightarrow 0$ there is a narrow zone where
probability rises to infinity and in the case $L\rightarrow\infty$
there exists an increasingly large zone where probability tends to
zero. Both kind of density functions show a similar pattern to
distributions of maximal LMC complexity in the discrete case,
where there is an state of dominating probability and the rest of
states have the same probability.

The minimal $\hat{C}$ given by Eq. (\ref{eq:cd0.5}) is found when
$L=1$, that is, when $g_{\delta,L}$ becomes the rectangular
density function $\chi_{[-1,1]}$. In fact, the value $\hat{C}=1$
is the minimum of possible shape complexities and it is reached
only on the rectangular distributions. We sketch now some steps
that prove this result.

Suppose
\begin{equation}
f = \sum_{k=1}^n \lambda_k \chi_{E_k}
 \label{eq:def-rect}
\end{equation}
to be a density function consisting of several rectangular pieces
$E_k$, $k=1,\cdots,n$, on disjoint intervals. If $\mu_k$ is the
Lebesgue measure of $E_k$, calculation of $\hat{C}$ gives
\begin{equation}
\hat{C}(f) = \prod_{k=1}^n
\left(\lambda_k^{-\lambda_k\mu_k}\right)\cdot\left(\sum_{k=1}^n
\lambda_k^2\mu_k\right) \, .
\end{equation}
Lagrange multipliers method is used to find the real vector
 $(\mu_1,\mu_2,\cdots,\mu_n ;\newline
\lambda_1,\lambda_2,\cdots,\lambda_n)$ that makes extremal the
quantity $\hat{C}(f)$ under the condition $\sum_{k=1}^n
\lambda_k\mu_k = 1$. This is equivalent to studying the extrema of
$\log\hat{C}(f)$. We define the function
$z(\lambda_k,\mu_k)=\log\hat{C}(f)+ \alpha\left(\sum_{k=1}^n
\lambda_k\mu_k-1\right)$, then
\begin{equation}
z(\lambda_k,\mu_k) = -\sum_{k=1}^n \mu_k\lambda_k \log\lambda_k +
\log\left(\sum_{k=1}^n \mu_k\lambda_k^2\right) +
\alpha\left(\sum_{k=1}^n \lambda_k\mu_k-1\right) \, .
\end{equation}
Differentiating this expression and making the result equal to
zero we obtain
\begin{eqnarray}
\frac{\partial z(\lambda_k,\mu_k)}{\partial\lambda_k} & = & -\mu_k
\log\lambda_k - \mu_k + \frac{2\lambda_k\mu_k}{\sum_{j=1}^n
\mu_j\lambda_j^2} + \alpha\mu_k = 0 \label{eq:part1} \\
\frac{\partial z(\lambda_k,\mu_k)}{\partial\mu_k} & = & -\lambda_k
\log\lambda_k + \frac{\lambda_k^2}{\sum_{j=1}^n \mu_j\lambda_j^2}
+ \alpha\lambda_k = 0 \label{eq:part2}
\end{eqnarray}
Dividing Eq. (\ref{eq:part1}) by $\mu_k$ and Eq. (\ref{eq:part2})
by $\lambda_k$ we get
\begin{eqnarray}
\frac{2\lambda_k}{\sum_{j=1}^n \mu_j\lambda_j^2} + \alpha - 1 =
\log\lambda_k \\
 \frac{\lambda_k}{\sum_{j=1}^n \mu_j\lambda_j^2} + \alpha =
 \log\lambda_k\, .
\end{eqnarray}
Solving these two equations for every $\lambda_k$ we have
\begin{equation}
\lambda_k = \sum_{j=1}^n \mu_j\lambda_j^2 \;\;\;\mbox{for all}\;\;
k \, .
\end{equation}
Therefore $f$ is a rectangular function taking the same value
$\lambda$ for every interval $E_k$, that is, $f$ is the
rectangular density function
\begin{equation}
f = \lambda\cdot\chi_L \;\;\;\mbox{with}\;\;
\lambda=\frac{1}{\sum_{i=1}^{n}\mu_i}=\frac{1}{L} \, ,
\end{equation}
where $L$ is the Lebesgue measure of the support.

Then $\hat{C}(f)=1$ is the minimal value for a density function
composed of several rectangular pieces because, as we know for the
example given by Eq. (\ref{eq:cd0.5}), $\hat{C}(f)$ is not upper
bounded for this kind of distributions.

Furthermore, for every compactly supported density function $g$
and for every $\epsilon>0$, it can be shown that near-continuity
of $\hat{C}$ allows to find a $\delta$-neighboring density
function $f$ of the type given by expression (\ref{eq:def-rect})
verifying $\mid\hat{C}(f) - \hat{C}(g)\mid < \epsilon$. The
arbitrariness of the election of $\epsilon$ brings us to conclude
that $\hat{C}(g)\geq 1$ for every probability distribution $g$.
Thus, we can conclude that the minimal value of $\hat{C}$ is $1$
and it is reached only by the rectangular density functions.


\section{Fisher-Shannon Information Product. Some Applications}

\subsection{Fisher-Shannon information: definition and properties}

 The description of electronic properties by means of information
measures was introduced into quantum chemistry by the pioneering works \cite{gadre1,gadre2,gadre3,gadre4,gadre5}. 
Inparticular Shannon entropy \cite{shannon} and Fisher information \cite{Fis}  have attracted special
attention in atomic and molecular physics 
(See e. g. \cite{esquivel96,massen98,massen01,sagar02,nalewajski02,nagy03,massen03,nalewajski03,
romera,parr05,sen05,guevara05,sagar05,romera05,nagy061,nagy062,sagar06,sen06,nagy07,liu07,sen07,patil,
sagar08,nagy08,sanudo2,sanudo3,seo08,sanudo1,nalewajski09}).
It is known that these two information measures give complementary
descriptions of the concentration and uncertainty of the probablility density:
$S_\rho$ ($I_\rho$) can be seen as a global (local) measure of spreading. In
this context, the Fisher-Shannon information  product (that we will define
bellow)  was found as  a link between these information
measures  to improve the
characterzation of a probability density function in terms of information
measures \cite{romera}.

The single-electron density, the basic variable of the  density
functional theory \cite{parr} of  $D$-dimensional many-electron systems is given by 
\begin{equation}
\rho({\bf r})=\int |\Psi({\bf r},{\bf r}_2,...,{\bf r}_N)|^2d^D{\bf r}_2...d^D{\bf r}_N
\end{equation}
where  $\Psi({\bf r}_1,...,{\bf r}_N)$  denotes the normalized wavefunction of the N-electron system 
and  $\rho({\bf r})$ is normalized to unity. The spreading of this quantity is best measured  by the 
Shannon information entropy
\begin{equation}
S_{\rho}=-\int \rho({\bf r}) \ln \rho({\bf r})  d^D{\bf r},
\end{equation} 
or equivalently by the Shannon entropy power \cite{dembo91,shannon}
\begin{equation}
J_{\rho}\equiv \frac{1}{2\pi e}e^{\frac{2}{D}S_{\rho}},
\end{equation}
On the other hand the Fisher information \cite{Fis,dembo91} of $\rho({\bf r})$ is given by
\begin{equation}
I_{\rho}=\int \frac{|\nabla \rho({\bf r})|^2}{\rho({\bf r})}d^D{\bf r}.
\end{equation}
The sharpness, concentration or delocalization of the
electronic cloud is measured by both quantities. 
It is known that  these two information measures give complementary descriptions of the smoothness 
and uncertainty of the electron localization:  $S_{\rho}$  and $I_{\rho}$ are   global and local  
measures of smoothness, respectively \cite{dembo91,gadre1,gadre2,gadre3,gadre4,gadre5,shannon,Fis,romera}. 

For completeness let us point out that the aforementioned information measures, which refer to 
an unity-normalized density $ \rho_1({\bf r})\equiv\rho({\bf r})$, are  related to the corresponding 
measures of the $N$-normalized density $\rho_N({\bf r})$ by 
\begin{equation}
S_{\rho_N}=-N\ln N + N S_{\rho} \quad\quad \mbox{and} \quad\quad I_{\rho_N} = N I_\rho
\end{equation}
for the Shannon and Fisher quantities, respectively.

The information product concept  ${P}_\rho$  was originally  defined  in
\cite{romera} as 
\begin{equation}
{P}_\rho\equiv\frac{1}{D}J_{\rho}I_{\rho},
\end{equation}
and it was applied in the study of electronic properties of quantum systems during last years
(See, eg. \cite{romera,patil,sanudo2,sanudo3,sanudo1,szabo08,montgomery,mitnik}).
Next we will put foward some  mathematical properties which have been
obtained in \cite{romera,romera05,romera01,romera06}
for the Fisher-Shannon information product ${P}_\rho$. 

\subsubsection{Scaling property}

 The Fisher information and the Shannon  entropy power transform as 
\begin{equation}
I_{\rho_{\gamma}}=\gamma^{D-1}I_{\rho};\quad\quad J_{\rho_{\gamma}}=\gamma^{-(D-1)}J_{\rho}
\end{equation} 
under scaling of the probability density $\rho({\bf r})$ by a real scalar factor $\gamma$; i. e. when 
$\rho_{\gamma}({\bf r})=\gamma^D \rho(\gamma{\bf r})$. This indicates that they are homogeneous 
density functionals of degrees $2$ and $-2$, respectively. Consequently, the information product 
${P}_\rho=\frac{1}{D}J_{\rho}I_{\rho}$ is invariant under this scaling transformation, i. e. 
\begin{equation}
{P}_{\rho_{\gamma}}={P}_{\rho}
\end{equation}

\subsubsection{Uncertainty properties}

The Fisher information $I_{\rho}$ and the Shannon entropy power $J_{\rho}$
satisfy the uncertainty relationship  \cite{dembo91}
\begin{equation}
\frac{1}{D}J_{\rho}I_{\rho}\geq 1.
\label{incertidumbre}
\end{equation}
Remark that  when one of the involved quantities decreases near to zero,  the other has to increase 
to a large value. Moreover, it is closely linked to the  uncertainty relation 
  $\langle r^2\rangle \langle p^2\rangle \geq \frac{D^2}{4}$, where $\langle r^2\rangle$ is defined 
  in terms of the charge position density $\rho({\bf r})$ as $\langle r^2\rangle =\int r^2 
  \rho({\bf r})d^D{\bf r}$,  and $\langle p^2\rangle$ is given in terms of the  momentum density 
  $\Pi({\bf p})$ in an analogous way, where $\Pi({\bf p})$ is defined by means of the Fourier transform 
  of $\Psi({\bf r}_1,...,{\bf r}_N)$, $\Phi({\bf p}_1,...,{\bf p}_N)$, as
\begin{equation}
\Pi({\bf p})=\int |\Phi({\bf p},{\bf p}_2,...,{\bf p}_N)|^2 d^D{\bf p}_2...d^D{\bf
  p}_N.
\end{equation}
The Fisher information has been used as a measure of uncertainty in quantum physics 
(See e. g. \cite{romera05,romera06,hall00,hall01,hall02,hall04,luo00,luo03,luo04,petz03,romera99}). 
It has been shown to fulfill the Stam inequalities \cite{stam}
\begin{equation}
I_\rho\le 4\langle p^2\rangle;\quad
I_\pi\le 4\langle r^2\rangle,
\end{equation}
and the Cramer-Rao inequalities \cite{dembo91,romera01,romera99,rao,romera94}
\begin{equation}
I_\rho\ge \frac{D^2}{\langle r^2\rangle};\quad
I_\pi\ge \frac{D^2}{\langle p^2\rangle}
\end{equation}
for the general single-particle systems. The multiplication of each pair of these inequalities produces
\begin{equation}
\frac{D^4}{\langle r^2\rangle\langle p^2\rangle}\le I_\rho I_\pi\le16\langle r^2\rangle\langle p^2\rangle,
\label{bounds}
\end{equation}
valid for ground and excited states of general systems, which shows the close
connection between the Heisenberg-like uncertainty product and the product of
the position and momentum Fisher informations.

Indeed, taken into account 
 $1/D\langle r^2\rangle \geq J_{\rho}$ \cite{BBM} one has that
\begin{equation}
\frac{4}{D^2}\langle p^2\rangle \langle r^2\rangle \geq \frac{1}{D}J_{\rho}I_{\rho}\geq 1\label{union}
\end{equation}
and
\begin{equation}
\frac{4}{D^2}\langle p^2\rangle \langle r^2\rangle \geq \sqrt{{P}_{\rho}{P}_{\pi}}\geq 1.
\end{equation}

It is straightforward to show that the equality limit  of these two
inequalities is reached  for Gaussian densities.

An special case is given by a  single-particle in a central potential. In this
framework an uncertainty Fisher information relation was obtained  in  \cite{romera06}:
\begin{equation}
I_\rho I_\pi\ge 4 D^2 \left[1-\frac{(2l+1)|m|}{2l(l+1)}\right]^2
\label{fis1}
\end{equation}
 and   Fisher information in position space was derived in
\cite{romera05} as 
\begin{equation}
I_\rho=4\langle p^{2}\rangle-2(2l+1)|m|\langle r^{-2} \rangle
\label{fisherpos}
\end{equation}
where $l$ and $m$ are the orbital and magnetic quantum numbers. Taking into
account the duality of the position and momentum spaces as well as the
separability of the wavefunction, one can express the Fisher information of
the momentum distribution density  as
\begin{equation}
I_\pi= 4\langle r^2\rangle-2(2l+1) |m| \langle p^{-2}\rangle.
\label{fishermom}
\end{equation}
On the other hand, the radial expectation values $\langle p^2\rangle$ and
$\langle r^{-2}\rangle$ ($\langle r^2\rangle$ and $\langle p^{-2}\rangle$) are
related \cite{romera05,romera06} by
\begin{eqnarray}
\langle p^2\rangle \ge l(l+1) \langle r^{-2}\rangle,\\
\langle r^2\rangle \ge l(l+1) \langle p^{-2}\rangle,
\end{eqnarray}
and combining above expressions the fisher uncertainty-like relation (\ref{fis1}) is obtained.

\subsubsection{Nonadditivity properties}

The superadditivity of the Fisher information and the subadditivity of the
Shannon information  of a probability density, can be used to prove \cite{romera} that 
\begin{equation}
I_{W}\geq N I_{\rho}
\label{non1} 
\end{equation}
\begin{equation}
S_{W}\leq N S_{\rho}
\label{non}
\end{equation}
where 
\begin{equation}
I_{W}=\int\frac{|\nabla |\Psi({\bf r}_1,...,{\bf
    r}_N)|^2|^2}{|\Psi({\bf r}_1,...,{\bf r}_N)|^2} d{\bf r}_1 ... d {\bf   r}_N
\end{equation} 
and 
\begin{equation}
S_W=\int|\Psi({\bf r}_1,...,{\bf r}_N)|^2\ln|\Psi({\bf r}_1,...,{\bf r}_N)|^2d{\bf r}_1 ... d {\bf r}_N
\end{equation}
for general $N$-fermion systems in three dimensions. The $D$-dimensional
generalization is obvious. We will show the proof below.

Let $\rho({\bf r})$ a probability density on ${\R}^t$, that is, $\rho({\bf r})$ non-negative and 
$\int\rho({\bf r})d{\bf r}=1$.  
We will suppose that Fisher information  and Shannon information  of
$\rho({\bf r})$ exits. Corresponding  to any orthogonal decomposition ${\R}^t=
{\R}^r \oplus {\R}^s$, t=r+s,  the marginal densities are given by:
\begin{equation}
\rho_1({\bf x})=\int_{{\R}^r}\rho({\bf x},{\bf y}) d^r{\bf y}, \quad \rho_2({\bf y})=\int_{{\R}^s}
\rho({\bf x},{\bf y}) d^s{\bf x}
\end{equation}
then \cite{carlen}
\begin{equation}
I_\rho\geq I_{\rho_1}+I_{\rho_2} \quad\quad
\label{19}
\end{equation}
property  which is known as superadditivity of Fisher information, and
\begin{equation}
S_\rho\leq S_{\rho_1}+S_{\rho_2} \quad\quad
\label{20} 
\end{equation}
which is known as subadditivity of Shannon information. Both inequalities saturate 
 when $\rho({\bf x},{\bf y})=\rho_1({\bf x})\rho_2({\bf y})$  \cite{carlen} .

On the other hand, let us consider an  $N$-fermion system and denote the $i$th-electron density by 
\begin{equation}
\rho_i\equiv\rho({\bf r}_i)=\int|\Psi({\bf r}_1,...,{\bf r}_i,...,{\bf r}_N)|^2d{\bf r}_1...
d{\bf r}_{i-1}d{\bf r}_{i+1}...d{\bf r}_N,
\end{equation}
 for  $i=1,...N$. Then, taken into account that  the wavefunction is
 antisymmetric and Eq. (\ref{19}) and (\ref{20}),  the wavefunction Fisher
 information fullfills 
\begin{equation}
I_{W}=\int\frac{|\nabla |\Psi({\bf r}_1,...,{\bf r}_N)|^2|^2}{|\Psi({\bf r}_1,...,{\bf r}_N)|^2} d{\bf r}_1 ... 
d {\bf r}_N\geq\sum_{i=1}^N I_{\rho_i}=N I_\rho,
\label{ab1}
\end{equation}
and  the  wavefunction Shannon information  fulfills:
\begin{equation}
S_W=\int|\Psi({\bf r}_1,...,{\bf r}_N)|^2\ln|\Psi({\bf r}_1,...,{\bf r}_N)|^2d{\bf r}_1 ... d {\bf r}_N\leq 
\sum_{i=1}^N S_{\rho_i}=N S_\rho.
\label{ab2}
\end{equation}
Inequalities (\ref{ab1}) and (\ref{ab2}) are equalities when
$|\Psi({\bf r}_1,...,{\bf r}_N)|^2=\rho({\bf r_1})...\rho({\bf r}_N)$.

These properties have  allowed  us to generalize  the following uncertainty relationships:
\begin{itemize}
\item The Stam's uncertainty relation for wave functions normalized to unity \cite{romera,stam} is generalized 
via the inequality (\ref{ab1}) by
\begin{equation}
N I_\rho\leq I_W\leq 4 N \langle p^2\rangle
\end{equation}
\item The Shannon information uncertainty relation  for wave
  functions normalized to unity \cite{BBM} is generalized via  inequality (\ref{ab2}) by
\begin{equation}
3N(1+\ln \pi)\leq 
-\int|\Psi({\bf r}_1,...,{\bf r}_N)|^2\ln|\Psi({\bf r}_1,...,{\bf
  r}_N)|^2d{\bf r}_1 ... d {\bf r}_N  
\end{equation}
\begin{equation}
-\int|\Phi({\bf p}_1,...{\bf p}_N)|^2\ln|\Phi({\bf p}_1,...,{\bf p}_N)|^2d{\bf p}_1 ... d {\bf p}_N  
\leq N(S_\rho+S_\pi)
\end{equation}
\end{itemize}
where $S_{\rho}(S_\pi)$ denotes the Shannon information  of the single-particle distribution density in 
position (momentum) space.

\subsection{Fisher-Shannon product as an electronic correlation measure}

The Fisher-Shannon information product  was earlier employed \cite{romera}  as a tool for
stuying the electron correlation in atomic systems, in particular in two electron isoelectronic series. 
The application of this indicator to the
electronic shell structure of atoms has received a special attention for
systems running from on-electron atoms to many-electron ones as those
corresponding to the periodic table 
(see, e.g. \cite{sanudo2,sanudo3,sanudo1,szabo08,romera01}).

Many electron systems such as  atoms, molecules and clusters show the electron
correlation phenomenon. This feature has been characterized in terms of the  correlation energy
\cite{Ful}, which gives the difference between the exact non-relativistic
energy and the Hartree-Fock approximation, as well as by some statistical
correlation coefficients  \cite{kul}, which asses radial and angular
correlation in both the position and momentum density distributions. 
Some information-theoretic measures of the electron correlation in many electron
systems have been proposed during last years \cite{romera,grassi,collins,grassi1,mohareji,sagar03,sagar09,
ziesche,huang,gottlieb,juhasz,amovilli}. 
Here we will focus on the Fisher-Shannon Information Product as a measure of
electron correlation. 

The Fisher-Shannon Information Product has been studied in
two types   of two-electron systems \cite{romera}  which differ in the Coulomb- and
oscillator-like form of the electron-nucleus interaction. 
The Hamiltonian of such a system is 
\begin{equation}
H=-\frac{1}{2}\nabla^2_1-\frac{1}{2}\nabla^2_2+V(r_1)+V(r_2)+\frac{1}{|{\bf r}_1-{\bf r}_2|},
\end{equation}
where $V(r_i)$ denotes the electron-nucleus interaction of the $i$th-electron.
 $V(r_i)=Z/r_i$ for He-like ions ($Z$ being the nuclear charge) and
 $V(r_i)=\frac{1}{2} \omega r_i^2$ for the Hooke atoms. The Hooke atom is especially well suited for 
 the understanding of correlation
phenomena because of its amenability to analytical treatment.

\subsubsection{He-like ions}
 
In the bare coulomb field case (BCF), i. e. without coulombic interelectronic interaction in the hamiltonian, 
the ground state wave function of $He(Z)$ is a single Slater determinant and the charge density is a 
hydrogenlike one, so $J_{\rho_Z}=\frac{e}{2\pi^{1/3}}\frac{1}{Z^2}$  and $I_{\rho_Z}=4Z^2$,
so ${P}_{BCF}=K_{BCF}$ with $K_{BCF}\simeq 1.237333$.
To consider the inclusion of electronic interaction we will work with the 204-terms Hylleraas type 
functions of Koga et al. \cite{hyll204} for  the ground states of  H$^{-}$, He, Li$^{+}$, Be$^{2+}$, 
B$^{3+}$, and Ne$^{8+}$ ($Z=1-5$, $10$). 

\begin{figure}[h]
\centerline{\includegraphics[width=12cm]{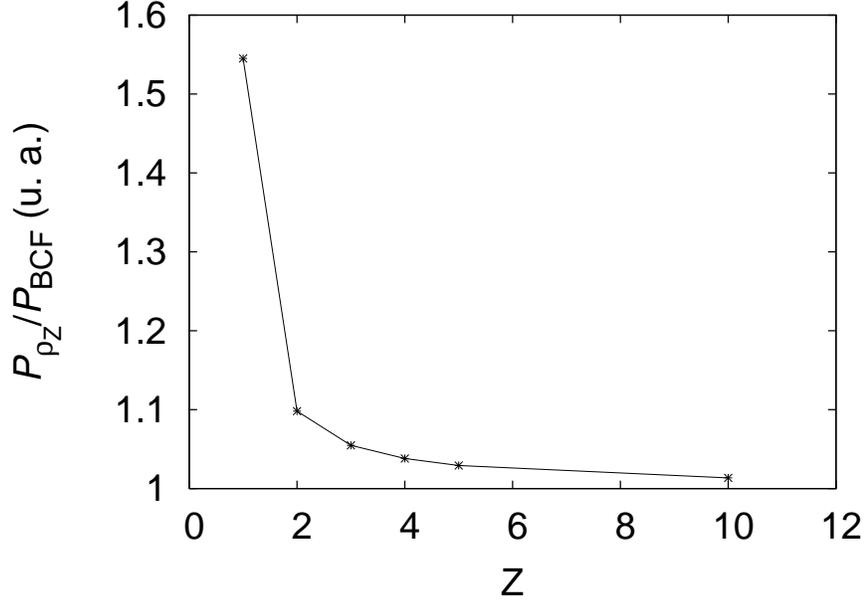}}
\caption{ The ratio  ${P}_{\rho_Z}/{P}_{BCF}$ of the
  information product for the He-like ions and the information product for
  bare two-electron atoms as a function of the nuclear charge $Z$. The points
  correspond to the values of $He(Z)$ ions with $Z=1-5$ and $10$. 
  The solid line  has been drawn only to guide the eye.}
  \label{ER1}
\end{figure}

In Fig. \ref{ER1} we have compared the
dependence of the information product ${P}_{\rho_Z}$  on the nuclear
charge $Z$ for He-like ions with the  bare coulomb field information product. It is apparent the monotonic 
decrease of ${P}_{\rho_Z}$ when $Z$ increased, asymptotically approaching the bare or no-correlation 
value ${P}_{BCF}=K_{BCF}$ and showing that the electron correlation effect gradually decreases 
with respect to the electron-nucleus interaction when the nuclear charge of the system is raised up.

\begin{figure}[h]
\centerline{\includegraphics[width=12cm]{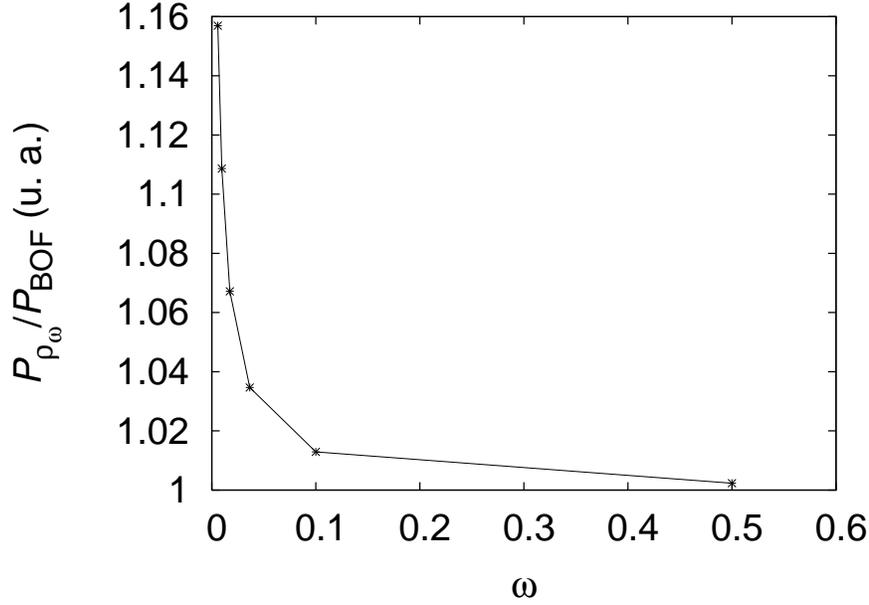}}
\caption{The information product ${P}_{\rho_\omega}/{P}_{BOF}$ 
for  the Hooke atoms with the oscillator strength $\omega=$ 0.5,
  0.1, 0.03653727, 0.01734620, 0.009578420, and 0.005841700 and the
  bare oscilator field information product  ${P}_{BOF}$ . 
  The solid line  has been drawn only to guide the eye.}
  \label{ER2}
\end{figure}

 \subsubsection{Hooke's atoms}  
 
 For the bare oscillator-field case (BOF), it is known
 that $J_{\rho_\omega}=1/(2\omega)$  and $I_{\rho_\omega}=6\omega$,
 so that the information product ${P}_{BOF}=1$. On the other hand the 
 Schr\"odinger equation of the entire Hooke atom can be solved  analytically for an infinite set of 
 oscillator frequencies \cite{Taut}. The use of relative and center of mass coordinates allows the 
 Hamiltonian to be separable so that the total wavefunction for singlet states is given  by $\Psi({\bf r_1},
 \sigma_1,{\bf r_2},\sigma_2)=\xi({\bf R})\Phi({\bf u})\tau(\sigma_1,\sigma_2)$, where $\tau(\sigma_1,\sigma_2)$ 
 is the singlet spin wave function,  $\xi({\bf R})$ and $\Phi({\bf u})$ being the solutions of the Schr\"odinger 
 equations
\begin{equation}
(-\frac{1}{4}\nabla^ 2_R+ \omega R^ 2)\xi({\bf R})=E_R\xi({\bf R}),
\end{equation}
\begin{equation}
(-\nabla_u^2+\frac{1}{4}\omega u^2+\frac{1}{u})\Phi({\bf u})=E_u\Phi({\bf u}),
\end{equation}
respectively, and the total energy $E=E_R+E_u$.

The computed results for the Fisher information and entropy power of these systems are shown in  Fig. \ref{ER2} 
for several  $\omega$ 
 values, (namely,  0.5, 0.1, 0.03653727, 0.01734620, 0.009578420, and 0.005841700). For these particular values 
 the ground state solution can be obtained  \cite{Taut} as 
\begin{equation}
\xi({\bf R})=\left(\frac{2\omega}{\pi}\right)^{3/4}e^{-\omega R^2}\mbox{ and }\Phi({\bf u})=
e^{-\frac{\omega r^2}{4}}Q_n(r)
\end{equation}
where $Q_{n}(r)$ is a polynomial whose coefficients can be determined analytically. 

Cioslowski et al \cite{Har} quantify the domains of the weakly correlated regime of this system which 
corresponds to the values of $\omega$  
greater than $\omega_{c}\simeq 4.011624\times10^{-2}$,  and the strongly correlated 
regime that encompasses the values of $\omega$ smaller than $\omega_c$.

In Fig. \ref{ER2} we have drawn  ${P}_{\rho_\omega}/{P}_{BOF}$ as a function of the oscillator 
electron-nucleus strength $\omega$. 
It is apparent that the value of the electron density functional 
${P}_{\rho_\omega}/{P}_{BOF}$ (dots) is always bigger than unity, when the
electron-electron repulsion becomes very small with respect to the oscillator
electron-nucleus interaction, the  points approach to the value 1, indicating  the
decrease of the  relative importance of electron correlation when the strength
$\omega$ is increased.

\subsection{Fisher information  for a single particle in a central potential}

As another application, let us consider the Fisher information in the position space
(for momentum space is analogous) of a single-particle system in a central potential $V(r)$, defined by
\begin{equation}
I_{\rho}=\int \frac{|\nabla \rho({\bf r})|^2}{\rho({\bf r})}d{\bf r}
\end{equation}
where $\rho({\bf r})=|\psi({\bf r})|^2$ and where $\psi({\bf r})$ is the bound solutions of the Schr\"odinger equation
\begin{equation}
\left[-\frac{1}{2}\nabla^2 + V(r)\right]\psi({\bf r}) = E \psi({\bf r}).
\end{equation}
For bounded states the solution of above equation is given by
\begin{equation}
\psi_{nlm}({\bf r}) = R_{nl}(r) Y_{lm}(\Omega)
\end{equation}
where $R_{nl}(r)$ is the radial part of the function and $Y_{lm}(\Omega)$ is the spherical harmonic of order $l$ 
that is given by
\begin{equation}
Y_{lm}(\Omega)=\frac{1}{\sqrt{2\pi}}e^{im\phi}\Theta_{lm}(\cos{\theta}), \quad (-l\leq m\leq l \quad \mbox{and}
\quad 0\leq\theta\leq\pi, 0\leq\phi\leq 2\pi)
\end{equation}
where $\Theta_{lm}(x)$ are given in terms of the associated Legendre functions of the first kind $P_l^m(x)$:
\begin{equation}
\Theta_{lm}(x)=\sqrt{\frac{2l+1}{2}\frac{(l-m)!}{(l+m)!}} P_l^m(x).
\end{equation}

So the Fisher information for a single particle in a central potential is given by
\begin{equation}
I_{\rho_{nlm}} = 4 \int|\nabla \rho_{nlm}^{1/2}({\bf r})|^2 = \int \left[\Theta_{lm}^2(\theta)
\left(\frac{\partial R_{nl}^2(r)}{\partial r}\right)^2 + \frac{1}{r^2} R_{nl}^2(r)
\left(\frac{\partial \Theta_{lm}(\theta)}{\partial\theta}\right)^2\right]d{\bf r},
\end{equation}
on the other hand the kinetic energy is given by:
\begin{eqnarray}
\langle p^2\rangle_{nlm} & = & \int |\nabla \psi_{nlm}({\bf r})|^2 =
\int \left[ \left(\frac{\partial R_{nl}(r)}{\partial r}\right)^2|Y_{lm}(\Omega)|^2\right]d{\bf r} + \nonumber \\ 
& & \int \left[\frac{1}{r^2} R_{nl}^2(r)\left(\frac{\partial \Theta_{lm}(\theta)}{\partial \theta}\right)^2 + 
\frac{1}{r^2}\frac{1}{\sin^2{\theta}}R_{nl}^2(r)\Theta_{lm}^2(\theta) m^2 \right] d{\bf r}
\end{eqnarray}
thus 
\begin{equation}
I_{\rho_{nlm}} = 4\langle p^2\rangle_{nlm} - 2 \langle r^{-2} \rangle_{nlm} (2l+1)|m|
\end{equation}

\subsubsection{Hydrogen atom}

For this system the potential is $V(r)=-1/r$ and the expectation values $\langle p^2\rangle_{nlm}=
\frac{1}{n^2}$ and $\langle r^{-2}\rangle_{nlm} = \frac{2}{(2l+1) n^3}$ thus
\begin{equation}
I_{\rho_{nlm}}=\frac{4}{n^2}\left(1 - \frac{|m|}{n}\right)
\end{equation}  

\subsubsection{Isotropic harmonic oscillator}

In this case the potential is $V(r)=\frac{1}{2}\omega^2 r^2$ and the expectation values 
$\langle p^2\rangle_{nlm}= \omega (2 n + l + \frac{3}{2})$ and $\langle r^{-2}\rangle_{nlm} = 
\frac{\omega}{(2l+1)}$
\begin{equation}
I_{\rho_{nlm}}=4\omega(2n+l+\frac{3}{2}-|m|)
\end{equation}


\section{Applications to Quantum Systems}

\subsection{Formulas in position and momentum spaces}

Here, we summarize the formulas and the nomenclature that will use
in all this section.

The measure of complexity $C$ has been defined as
\begin{equation}
C = H\cdot D\;,
\end{equation}
where $H$ represents the information content of the system and $D$ gives an idea
of how much concentrated is its spatial distribution. 

The simple exponential Shannon entropy,
in the position and momentum spaces, takes the form, respectively,
\begin{equation}
H_r = e^{S_r}\;, \hspace{1cm}
H_p = e^{S_p}\;,
\end{equation}
where $S_r$ and $S_p$ are the Shannon information entropies,
\begin{equation}
S_r = -\int\rho(\vec{r})\;\log \rho(\vec{r})\; d\vec{r}\;, \hspace{1cm}
S_p = -\int\gamma(\vec{p})\;\log \gamma(\vec{p})\; d\vec{p}\;,
\end{equation}
and $\rho(\vec{r})$ and $\gamma(\vec{p})$ are the densities normalized 
to $1$ of the quantum system in position and momentum spaces, respectively.

The disequilibrium is:
\begin{equation}
D_r = \int\rho^2(\vec{r})\; d\vec{r}\;, \hspace{1cm}
D_p = \int\gamma^2(\vec{p})\; d\vec{p}\;,
\end{equation}
In this manner, the final expressions for $C$ in position and 
momentum spaces are:  
\begin{equation}
C_r = H_r\cdot D_r\;, \hspace{1cm}
C_p = H_p\cdot D_p\;.
\label{eq-Crp}
\end{equation}

Second, the Fisher-Shannon information, $P$, in the position 
and momentum spaces, is given respectively by   
\begin{equation}
P_r = J_r\cdot I_r\;, \hspace{1cm}
P_p = J_p\cdot I_p\;,
\label{eq-Prp}
\end{equation}
where the first factor
\begin{equation}
J_r = {1\over 2\pi e}\;e^{2S_r/3}\;, \hspace{1cm}
J_p = {1\over 2\pi e}\;e^{2S_p/3}\;,
\label{eq-Jrp}
\end{equation}
is a version of the exponential Shannon entropy, 
and the second factor
\begin{equation}
I_r = \int{[\vec{\nabla}\rho(\vec{r})]^2\over \rho(\vec{r})}\; d\vec{r}\;, \hspace{1cm}
I_p = \int{[\vec{\nabla}\gamma(\vec{p})]^2\over \gamma(\vec{p})}\; d\vec{p}\;,
\end{equation}
is the Fisher information measure, that quantifies the narrowness 
of the probability density.

\subsection{The $H$-atom}
\label{SecQ1}

The atom can be considered a complex system. Its structure is determined 
through the well established equations of Quantum Mechanics \cite{landau1981,galindo1991}. 
Depending on the set of quantum numbers defining the state of the atom, different 
conformations are available to it. As a consequence, if the wave function of the atomic state 
is known, the probability densities in the position and the momentum spaces are obtained,
and from them, the different statistical magnitudes such as Shannon and Fisher informations,
different indicators of complexity, etc., can be calculated.

These quantities enlighten new details of the hierarchical organization of the atomic states.
In fact, states with the same energy can display, for instance, different values of complexity.
This is the behavior shown by the simplest atomic system, that is, the hydrogen atom
($H$-atom). Now, we present the calculations for this system \cite{sanudo3}. 

The non-relativistic wave functions of the $H$-atom
in position space ($\vec{r}=(r,\Omega)$, with $r$ the radial distance and
$\Omega$ the solid angle) are:
\begin{equation}
\Psi_{n,l,m}(\vec{r})= R_{n,l}(r)\; Y_{l,m}(\Omega)\;,
\label{eq-Hatom}
\end{equation}
where $R_{n,l}(r)$ is the radial part and $Y_{l,m}(\Omega)$ is the spherical harmonic 
of the atomic state determined by the quantum numbers $(n,l,m)$. The radial part is expressed 
as \cite{galindo1991}
\begin{equation}
R_{n,l}(r)= {2\over n^2} \left[{(n-l-1)!\over (n+l)!}\right]^{1/2}\;
\left({2r\over n}\right)^l\;e^{-{r\over n}}\; L_{n-l-1}^{2l+1}\left({2r\over n}\right)\;,
\end{equation}
being $L_{\alpha}^{\beta}(t)$ the associated Laguerre polynomials.
Atomic units are used here.

The same functions in momentum space ($\vec{p}=(p,\hat{\Omega})$, 
with $p$ the momentum modulus and $\hat{\Omega}$ the solid angle) are:
\begin{equation}
\hat{\Psi}_{n,l,m}(\vec{p})= \hat{R}_{n,l}(p)\; Y_{l,m}(\hat{\Omega})\;,
\label{eq-H1atom}
\end{equation}
where the radial part $\hat{R}_{n,l}(p)$ is now given by the expression \cite{bethe1977}
\begin{equation}
\hat{R}_{n,l}(p)= \left[{2\over\pi}{(n-l-1)!\over (n+l)!}\right]^{1/2}\;
n^2\;2^{2l+2}\;l!\;{n^lp^l\over (n^2p^2+1)^{l+2}}\; 
C_{n-l-1}^{l+1}\left({n^2p^2-1\over n^2p^2+1}\right)\;,
\end{equation}
with $C_{\alpha}^{\beta}(t)$ the Gegenbauer polynomials.

Taking the former expressions, the probability density
in position and momentum spaces,
\begin{equation}
\rho(\vec{r})\;=\;\mid\Psi_{n,l,m}(\vec{r})\mid^2\;, \hspace{1cm}
\gamma(\vec{p})\;=\;\mid\hat{\Psi}_{n,l,m}(\vec{p})\mid^2\;,
\end{equation}
can be explicitly calculated. From these densities, the statistical complexity and 
the Fisher-Shannon information are computed.

\begin{figure}[t]
\centerline{\includegraphics[width=6cm]{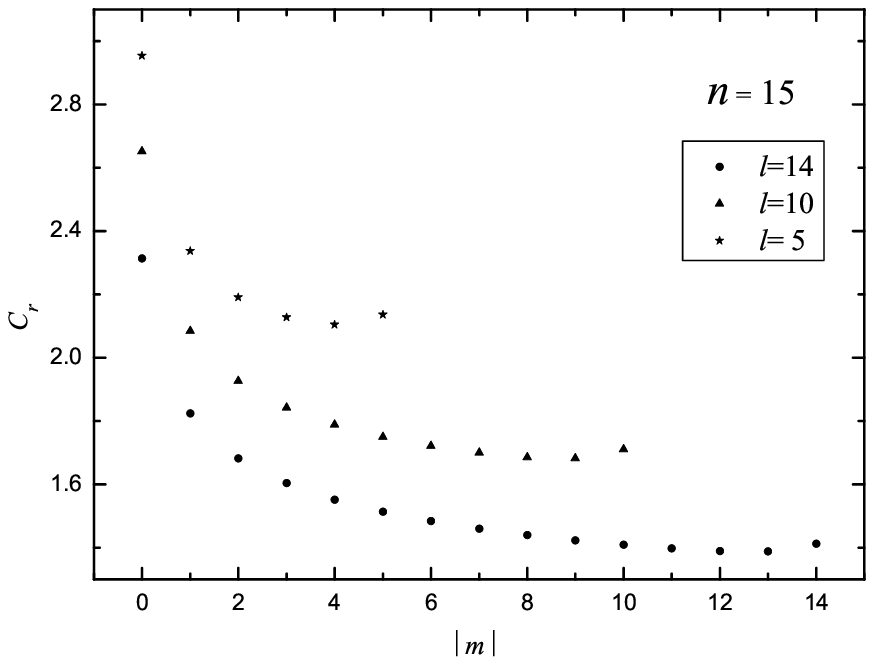}\hskip 5mm\includegraphics[width=6cm]{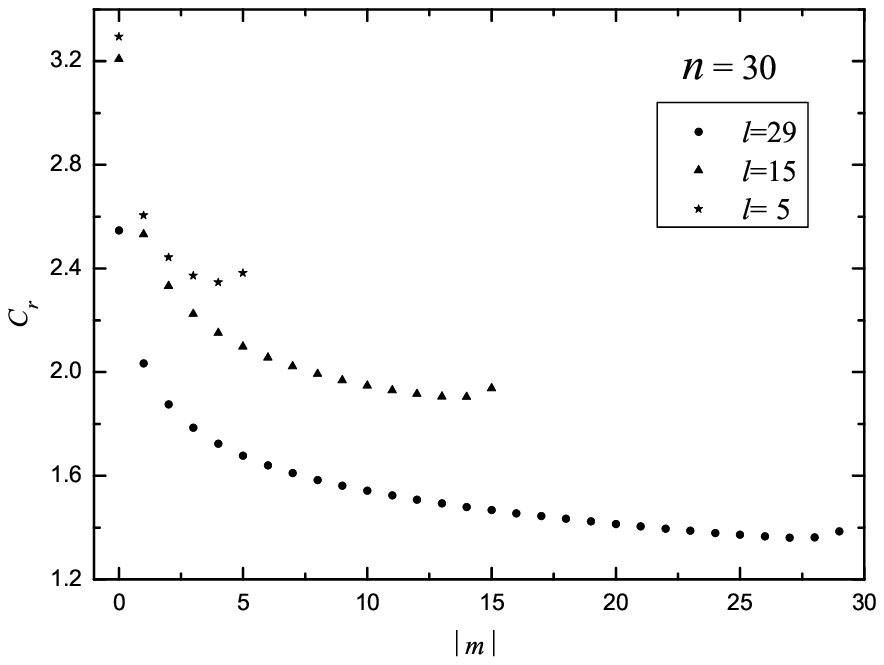}}
\centerline{(a)\hskip 7cm (b)} 
\centerline{\includegraphics[width=6cm]{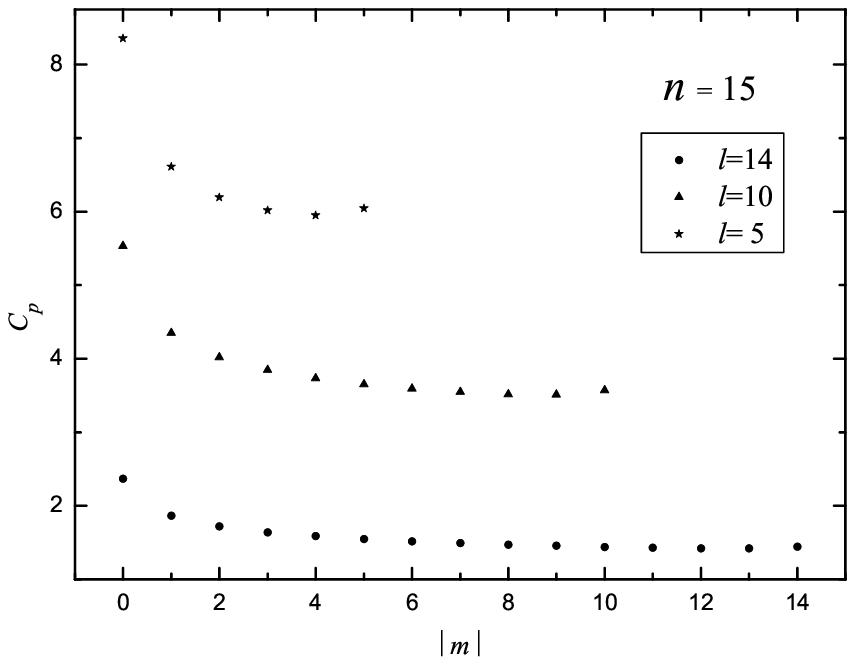}\hskip 5mm\includegraphics[width=6cm]{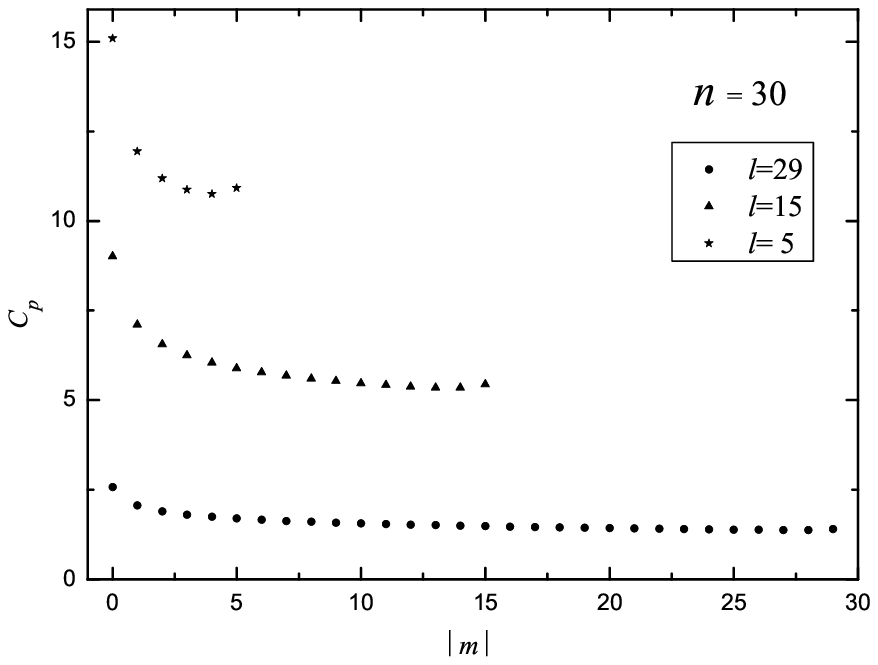}} 
\centerline{(c)\hskip 7cm (d)} 
\caption{Statistical complexity in position space, $C_r$, and momentum space, $C_p$, vs. 
$|m|$ for different $(n,l)$ values in the hydrogen atom. 
$C_r$ for (a) $n=15$ and (b) $n=30$. $C_p$ for (c) $n=15$ and (d) $n=30$. 
All values are in atomic units.}
\label{figQ1}
\end{figure}

$C_r$ and $C_p$ (see expression (\ref{eq-Crp})) are plotted in Fig. \ref{figQ1} 
as function of the modulus of the
third component $m$ of the orbital angular momentum $l$ for different pairs 
of $(n,l)$ values. The range of the quantum numbers is: 
$n\geq 1$, $0\leq l \leq n-1$, and $-l\leq m \leq l$.
Fig. \ref{figQ1}(a) shows $C_r$ for $n=15$ and Fig. \ref{figQ1}(b) shows $C_ r$ for $n=30$.
In both figures, it can be observed that $C_r$ splits in different sets of discrete points.
Each one of these sets is associated to a different $l$ value. 
It is worth to note that the set with the minimum values of $C_r$ corresponds just
to the highest $l$, that is, $l=n-1$. 
The same behavior can be observed in Figs. \ref{figQ1}(c) and \ref{figQ1}(d) for $C_p$.

\begin{figure}[h]
\centerline{\includegraphics[width=6cm]{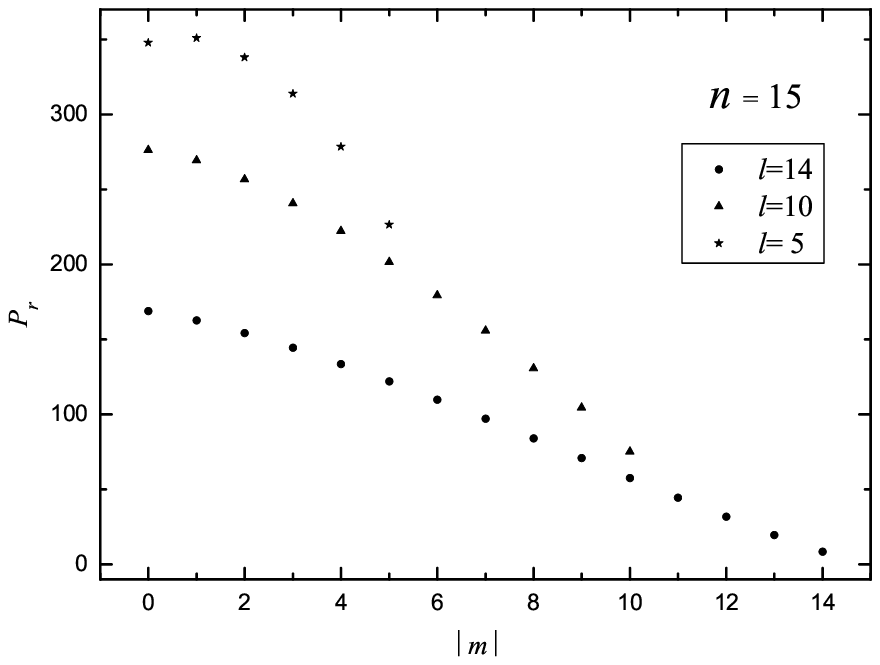}\hskip 5mm\includegraphics[width=6cm]{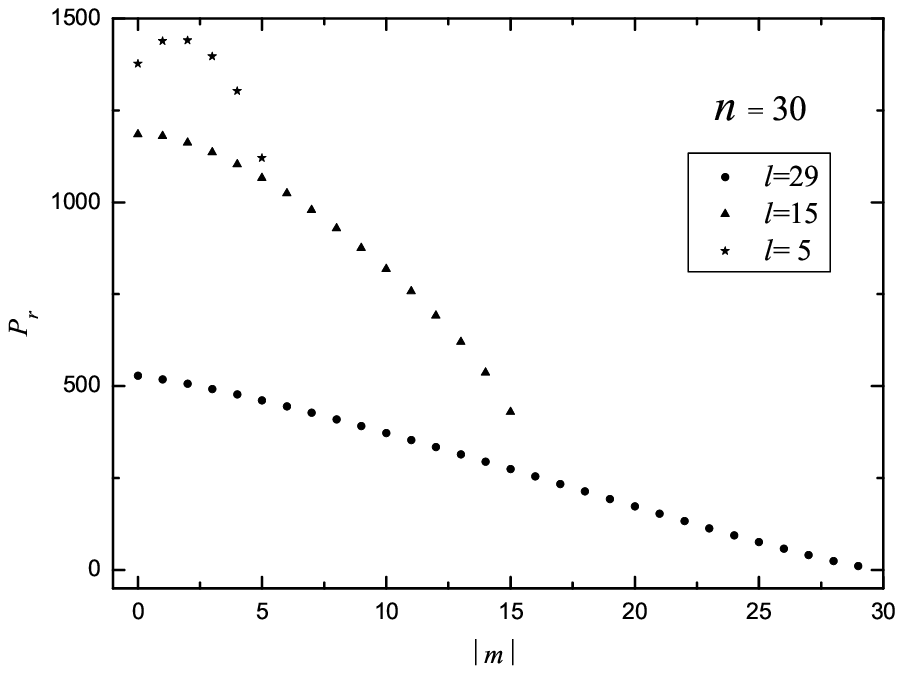}}
\centerline{(a)\hskip 7cm (b)} 
\centerline{\includegraphics[width=6cm]{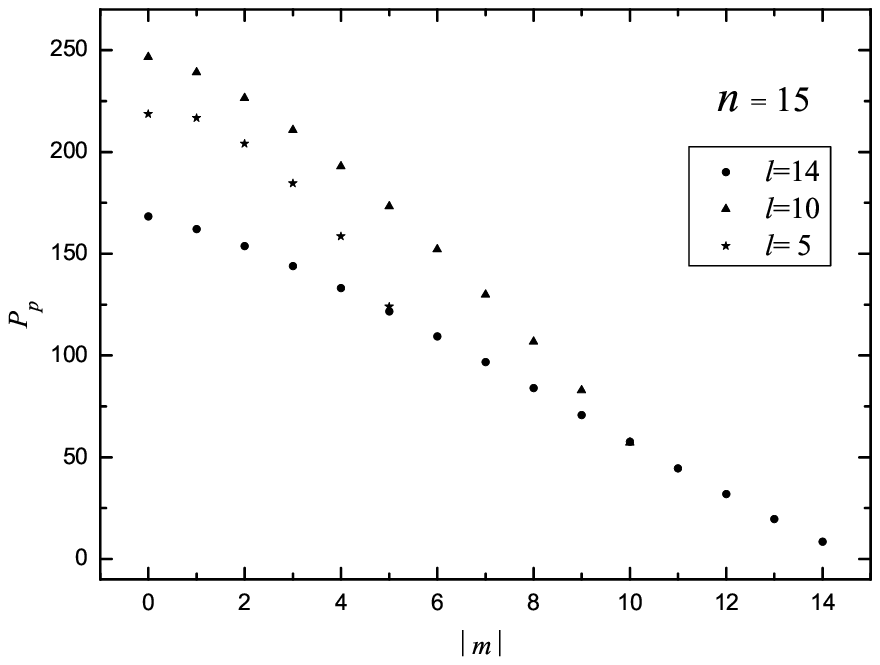}\hskip 5mm\includegraphics[width=6cm]{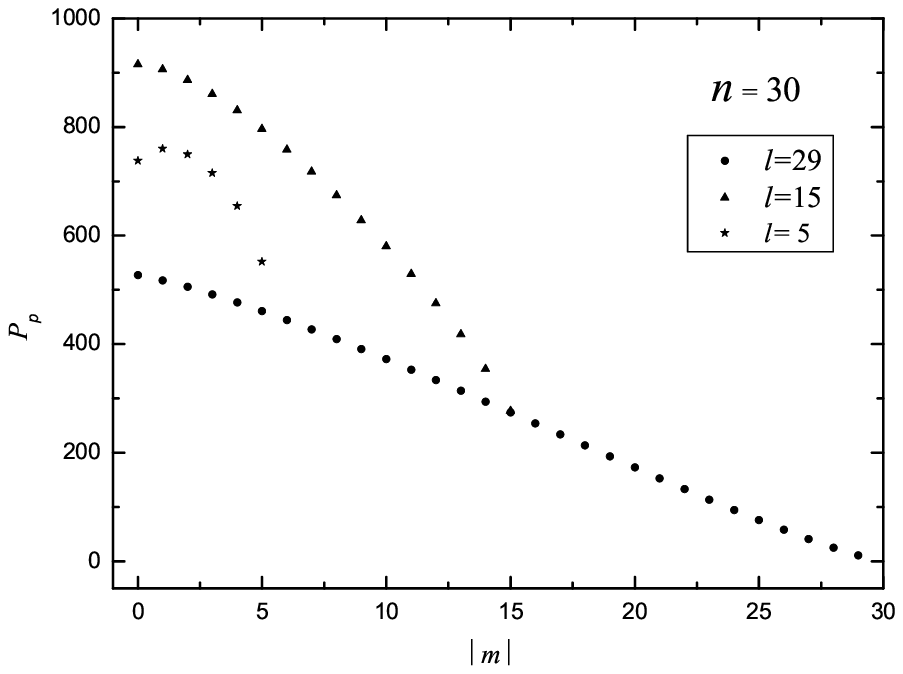}} 
\centerline{(c)\hskip 7cm (d)} 
\caption{Fisher-Shannon information in position space, $P_r$, and momentum space, $P_p$, vs. 
$|m|$ for different $(n,l)$ values in the hydrogen atom. 
$P_r$ for (a) $n=15$ and (b) $n=30$. $P_p$ for (c) $n=15$ and (d) $n=30$. 
All values are in atomic units.} 
\label{figQ2}
\end{figure}

Fig. \ref{figQ2} shows the calculation of $P_r$ and $P_p$ (see expression (\ref{eq-Prp}))
as function of the modulus of the third component $m$ for different pairs of $(n,l)$ values. 
The second factor, $I_r$ or $I_p$, of this indicator can be analytically obtained in both 
spaces (position and momentum). The results are \cite{romera05}:
\begin{equation}
I_r={4\over n^2}\;\left(1-{|m|\over n}\right),
\end{equation}
\begin{equation}
I_p={2n^2}\;\left\{5n^2+1-3l(l+1)-(8n-3(2l+1))\;|m|\right\}\;.
\end{equation}

In Fig. \ref{figQ2}(a), $P_r$ is plotted for $n=15$, and $P_ r$ is plotted for $n=30$ in Fig. \ref{figQ2}(b).
Here $P_r$ also splits in different sets of discrete points, 
showing a behavior parallel to the above signaled for $C$ (Fig. \ref{figQ1}). 
Each one of these sets is also related with a different $l$ value. 
It must be remarked again that the set with the minimum values of $P_r$ corresponds just
to the highest $l$. In Figs. \ref{figQ2}(c) and \ref{figQ2}(d), the same behavior can be observed for $P_p$.

Then, it is put in evidence that, for a fixed level of energy $n$, these statistical magnitudes
take their minimum values for the highest allowed orbital angular momentum, $l=n-1$.
It is worth to remember at this point that the mean radius of an $(n,l=n-1)$ orbital, $<r>_{n,l}$, 
is given by \cite{eisberg1961}  
\begin{equation}
<r>_{n,l=n-1} = n^2\left(1+{1\over 2n}\right),
\end{equation}
that tends, when $n$ is very large, to the radius of the $nth$ energy level, 
$r_{Bohr}=n^2$, of the Bohr atom. The radial part of this particular wave function,
that describes the electron in the $(n,l=n-1)$ orbital, has no nodes. 
In fact, if we take the standard deviation, 
$(\Delta r)=<(r-<r>)^2>^{1/2}$, of this wave function, $(\Delta r)=n\sqrt{2n+1}/2$,
the ratio $(\Delta r)/<r>$ becomes $1/\sqrt{2n}$ for large $n$.
This means that the spatial configuration of this atomic state is 
like a spherical shell that converges to a semiclassical Bohr-like orbit 
when $n$ tends to infinity.  
These highly excited $H$-atoms are referred 
as Rydberg atoms, that have been intensively studied \cite{lebedev1998} for its importance
in areas as astrophysics, plasma physics, quantum optics, etc., and also in studies of 
the classical limit of quantum mechanics \cite{coffey2003}.

We conclude this section by remarking that the minimum values of these statistical measures calculated
from the quantum wave functions of the $H$-atom enhance our intuition by selecting 
just those orbitals that for a large principal quantum number converge to the Bohr-like orbits 
in the pre-quantum image. Therefore, these results show that insights on the structural 
conformation of quantum systems can be inferred from these magnitudes, as it can also be seen
in the next sections.

\subsection{The quantum harmonic oscillator}
\label{SecQ2}

As suggested in the previous section, a variational process on the statistical measures 
calculated in the $H$-atom could select just those orbitals that 
in the pre-quantum image are the Bohr-like orbits.
Now, we show that a similar behavior for the statistical complexity and Fisher-Shannon information
is also found in the case of the isotropic quantum harmonic oscillator \cite{sanudo2}. 

We recall the three-dimensional non-relativistic wave functions of this system 
when the potential energy is written as $V(r)=\lambda^2r^2/2$, with 
$\lambda$ a positive real constant expressing the potential strength. In the same way as
in the $H$-atom (Eq. (\ref{eq-Hatom})), these wave functions in position space ($\vec{r}=(r,\Omega)$, 
with $r$ the radial distance and $\Omega$ the solid angle) are:
\begin{equation}
\Psi_{n,l,m}(\vec{r})= R_{n,l}(r)\; Y_{l,m}(\Omega)\;,
\label{eq:Psi}
\end{equation}
where $R_{n,l}(r)$ is the radial part and $Y_{l,m}(\Omega)$ is the spherical harmonic 
of the quantum state determined by the quantum numbers $(n,l,m)$. 
Atomic units are used here. The radial part is expressed 
as \cite{dehesa1994}
\begin{equation}
R_{n,l}(r)= \left[{2\;n!\;\lambda^{l+3/2}\over \Gamma(n+l+3/2)}\right]^{1/2}\;
r^l\;e^{-{\lambda \over 2}r^2}\; L_{n}^{l+1/2}(\lambda r^2)\;,
\label{eq:R}
\end{equation}
where $L_{\alpha}^{\beta}(t)$ are the associated Laguerre polynomials.
The levels of energy are given by 
\begin{equation}
E_{n,l}=\lambda (2n+l+3/2) = \lambda (e_{n,l}+3/2),
\label{eq:E}
\end{equation}
where $n=0,1,2,\cdots$ and $l=0,1,2,\cdots$.
Let us observe that $e_{n,l}=2n+l$.
Thus, different pairs of $(n,l)$ can give the same $e_{n,l}$,
and then the same energy $E_{n,l}$.

The wave functions in momentum space ($\vec{p}=(p,\hat{\Omega})$, 
with $p$ the momentum modulus and $\hat{\Omega}$ the solid angle) 
present the same form as in the $H$-atom (Eq. \ref{eq-H1atom}):
\begin{equation}
\hat{\Psi}_{n,l,m}(\vec{p})= \hat{R}_{n,l}(p)\; Y_{l,m}(\hat{\Omega})\;,
\label{eq:Psih}
\end{equation}
where the radial part $\hat{R}_{n,l}(p)$ is now given by the expression \cite{dehesa1994}
\begin{equation}
\hat{R}_{n,l}(p)= \left[{2\;n!\;\lambda^{-l-3/2}\over \Gamma(n+l+3/2)}\right]^{1/2}\;
p^l\;e^{-{p^2\over 2\lambda}}\; L_{n}^{l+1/2}({p^2/\lambda})\;.
\label{eq:Rh}
\end{equation}

Taking the former expressions, the probability density
in position and momentum spaces,
\begin{equation}
\rho_{\lambda}(\vec{r})\;=\;\mid\Psi_{n,l,m}(\vec{r})\mid^2\;, \hspace{1cm}
\gamma_{\lambda}(\vec{p})\;=\;\mid\hat{\Psi}_{n,l,m}(\vec{p})\mid^2\;,
\label{eq:rho}
\end{equation}
can be explicitly calculated. From these densities, 
the statistical complexity (see expression (\ref{eq-Crp}))
and the Fisher-Shannon information (see expression (\ref{eq-Prp})) are computed.
It is shown in section \ref{subsec-A} that these quantities are independent of $\lambda$, 
the potential strength, and also that they are the same in both position and momentum
spaces, i.e. $C_r=C_p$ and $P_r=P_p$. 

\begin{figure}[t]
\centerline{\includegraphics[width=6cm]{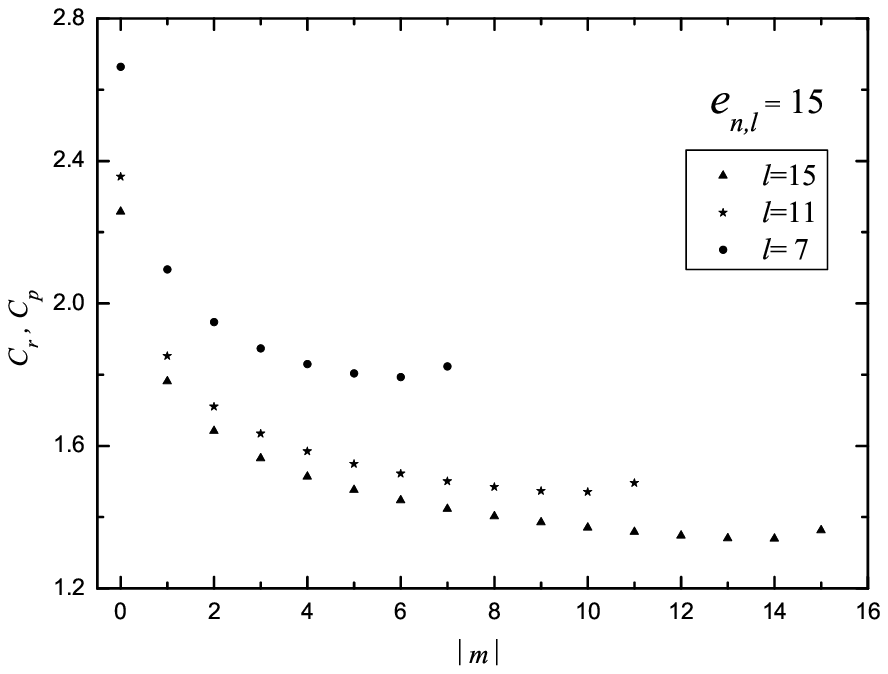}\hskip 5mm\includegraphics[width=6cm]{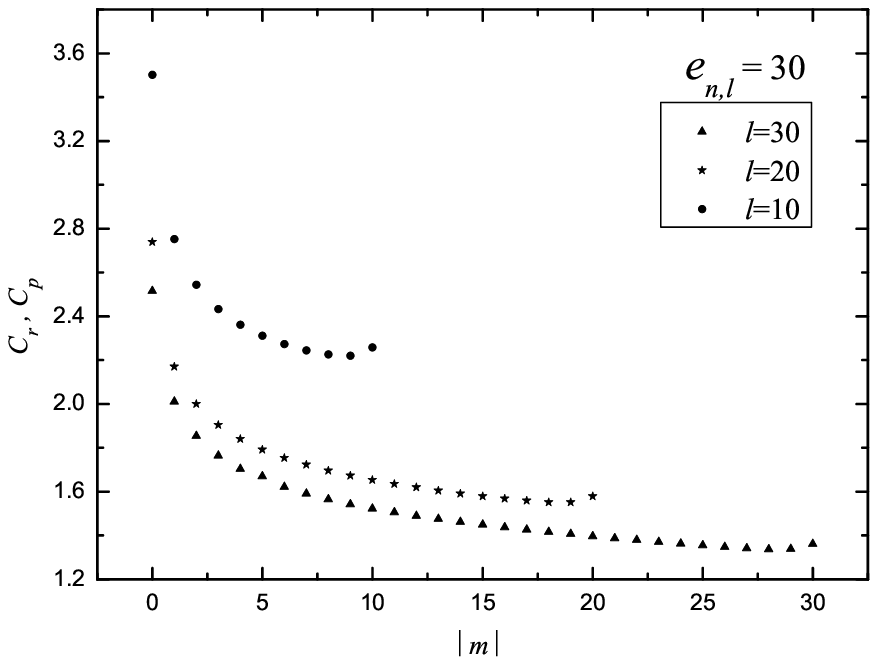}}
\centerline{(a)\hskip 7cm (b)} 
\caption{Statistical complexity in position space, $C_r$, and momentum space, $C_p$, vs. 
$|m|$ for different energy $e_{n,l}$-values in the quantum isotropic harmonic oscillator 
for (a) $e_{n,l}=15$ and (b) $e_{n,l}=30$. Recall that $C_r=C_p$. All values are in atomic units.}
\label{figQ3}
\end{figure}

In Fig. \ref{figQ3}, $C_r$ (or $C_p$) is plotted as function of the modulus of the
third component $m$, $-l\leq m \leq l$, of the orbital angular momentum $l$ for different $l$ values
with a fixed energy. That is, according to expression (\ref{eq:E}), 
the quantity $e_{n,l}=2n+l$ is constant in each figure.
Fig. \ref{figQ3}(a) shows $C_r$ for $e_{n,l}=15$ and Fig. \ref{figQ3}(b) shows $C_ r$ for $e_{n,l}=30$.
In both figures, it can be observed that $C_r$ splits in different sets of discrete points.
Each one of these sets is associated to a different $l$ value. 
It is worth noting that the set with the minimum values of $C_r$ corresponds just
to the highest $l$, that is, $l=15$ in Fig. \ref{figQ3}(a) and $l=30$ in Fig. \ref{figQ3}(b). 

\begin{figure}[h]
\centerline{\includegraphics[width=6cm]{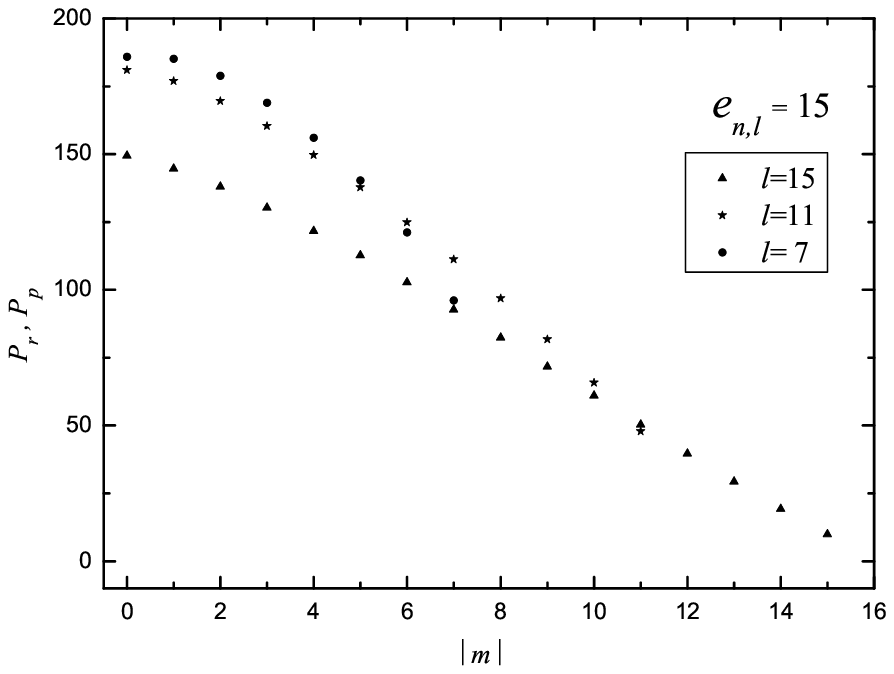}\hskip 5mm\includegraphics[width=6cm]{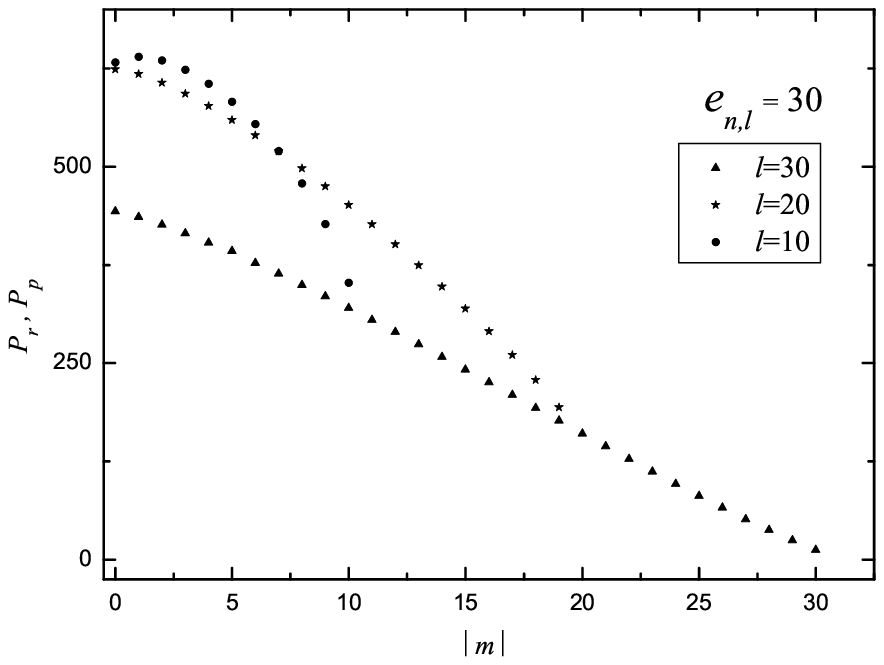}}
\centerline{(a)\hskip 7cm (b)} 
\caption{Fisher-Shannon information in position space, $P_r$, and momentum space, $P_p$, vs. 
$|m|$ for different energy $e_{n,l}$-values in the quantum isotropic harmonic oscillator. 
for (a) $e_{n,l}=15$ and (b) $e_{n,l}=30$. Recall that $P_r=P_p$. 
All values are in atomic units.} 
\label{figQ4}
\end{figure}

Fig. \ref{figQ4} shows $P$ as function of the modulus of the
third component $m$ for different pairs of $(e_{n,l}=2n+l,l)$ values. 
The second factor, $I_r$ or $I_p$, of this indicator can be analytically obtained in both 
spaces (position and momentum) \cite{romera05}:
\begin{eqnarray}
\hspace{2cm} I_r & = & 4\,(2n+l+3/2-|m|)\,\lambda \, , \label{eq:Ir}\\
\hspace{2cm} I_p & = & 4\,(2n+l+3/2-|m|)\,\lambda^{-1}. \label{eq:Ip}
\end{eqnarray}
Let us note that $I_r$ and $I_p$ depend on $\lambda$,
although the final result for $P_r$ and $P_p$ are 
non $\lambda$-dependent (see section \ref{subsec-A}).
In Fig. \ref{figQ4}(a), $P_r$ (or $P_p$) is plotted for $e_{n,l}=15$, 
and $P_ r$ is plotted for $e_{n,l}=30$ in Fig. \ref{figQ4}(b).
Here, $P_r$ also splits in different sets 
of discrete points, showing a behavior similar to that of $C$ in Fig. \ref{figQ3}. 
Each one of these sets is related with a different $l$ value, 
and the set with the minimum values of $P_r$ also corresponds just
to the highest $l$, that is, $l=15$ and $l=30$, respectively.

As in the $H$-atom, we also see here that, for a fixed level of energy, let us say $e_{n,l}=2n+l$, 
these statistical quantities take their minimum values for the highest allowed 
orbital angular momentum, $l=e_{n,l}$. 
It is worth remembering at this point that the radial part of this particular wave function,
that describes the quantum system in the $(n=0,l=e_{n,l})$ orbital, has no nodes. This means
that the spatial configuration of this state is, in some way, a spherical-like shell. 
In section \ref{subsec-B}, the mean radius of this shell, $<r>_{n,l,m}$, is found for the 
case $(n=0,l=e_{n,l},m)$. This is: 
\begin{equation}
<r>_{n=0,l=e_{n,l},m}\equiv <r>_{n=0,l=e_{n,l}} \simeq 
\,\sqrt{\lambda^{-1}\,(e_{n,l}+1)}\left(1+\Theta({e_{n,l}^{-1}})\right),
\label{eq:re}
\end{equation}
that tends, when $e_{n,l}\gg 1$, to the radius of the $Nth$ energy level, 
$r_{N}=\sqrt{\lambda^{-1}\,(N+1)}$, taking $N=e_{n,l}$ in the Bohr-like picture 
of the harmonic oscillator (see section \ref{subsec-B}). 

Then, we can remark again that the minimum values of the statistical measures calculated
from the wave functions of the quantum isotropic harmonic oscillator
also select just those orbitals that in the pre-quantum image are the Bohr-like orbits.

\subsubsection{Invariance of $C$ and $P$ under rescaling transformations}
\label{subsec-A} 

Here, it is shown that the statistical complexities $C_r$ and $C_p$ are equal 
and independent of the strength potential, $\lambda$,
for the case of the quantum isotropic harmonic oscillator.
Also, the same behavior is displayed by $P_r$ and $P_p$.

For a fixed set of quantum numbers, $(n,l,m)$,
let us define the normalized probability density $\hat{\rho}(\vec{t})$:
\begin{equation}
\hat{\rho}(\vec{t})= {2\;n!\over \Gamma(n+l+3/2)}\;
t^{2l}\;e^{-{t^2}}\; \left[L_{n}^{l+1/2}({t^2})\right]^2\;|Y_{l,m}(\Omega)|^2.
\end{equation}
From expressions (\ref{eq:Psi}), (\ref{eq:R}) and (\ref{eq:rho}),
it can be obtained that
\begin{equation}
\rho_{\lambda}(\vec{r})=\lambda^{3/2}\;\hat{\rho}(\lambda^{1/2}\vec{r}),
\end{equation}
where $\rho_{\lambda}$ is the normalized probability density of expression (\ref{eq:rho}). 
Now, it is straightforward to find that
\begin{equation}
H_r(\rho_{\lambda})= \lambda^{-3/2}\;H(\hat{\rho})\,,
\end{equation}
and that 
\begin{equation}
D_r(\rho_{\lambda})={\lambda^{3/2}\;D(\hat{\rho})}.
\end{equation}
Then,
\begin{equation}
C_r(\rho_{\lambda})=C(\hat{\rho}),
\end{equation} 
and the non $\lambda$-dependence of $C_r$ is shown.

To show that $C_r$ and $C_p$ are equal, let us note that,
from expressions (\ref{eq:Psih}), (\ref{eq:Rh}) and (\ref{eq:rho}),
the normalized probability density $\gamma_{\lambda}(\vec{p})$ for the same
set of quantum numbers $(n,l,m)$ can be written as
\begin{equation}
\gamma_{\lambda}(\vec{p})=\lambda^{-3/2}\;\hat{\rho}(\lambda^{-1/2}\vec{p}).
\end{equation} 
Now, it is found that
\begin{equation}
H_p(\gamma_{\lambda})=\lambda^{3/2}\;H(\hat{\rho})\,,
\end{equation}
and that 
\begin{equation}
D_p(\gamma_{\lambda})={\lambda^{-3/2}\;D(\hat{\rho})}.
\end{equation}
Then,
\begin{equation}
C_p(\gamma_{\lambda})=C(\hat{\rho}),
\end{equation} 
and the equality of $C_r$ and $C_p$, and their non $\lambda$-dependence
are shown.

Similarly, from expressions (\ref{eq-Prp}), (\ref{eq-Jrp}), (\ref{eq:Ir}) 
and (\ref{eq:Ip}), it can be found that $P_r=P_p$, and that these quantities
are also non $\lambda$-dependent.

\subsubsection{Bohr-like orbits in the quantum isotropic harmonic oscillator} 
\label{subsec-B}

Here, the mean radius of the orbital with the lowest complexity is calculated 
as function of the energy. Also, the radii of the orbits in the Bohr picture are 
obtained.

The general expression of the mean radius of a state represented by the
wave function $\Psi_{n,l,m}$ is given by 
\begin{equation}
<r>_{n,l,m}\equiv <r>_{n,l}={n!\over \Gamma(n+l+3/2)}\;{1\over \lambda^{1/2}}\;
\int_0^{\infty} t^{l+1}\;e^{-{t}}\; \left[L_{n}^{l+1/2}({t})\right]^2\;dt.
\label{eq:r}
\end{equation}
For the case of minimum complexity (see Fig. \ref{figQ3} or \ref{figQ4}),
the state has the quantum numbers $(n=0,l=e_{n,l})$. The last expression (\ref{eq:r}) becomes:
\begin{equation}
<r>_{n=0,l=e_{n,l}}={(e_{n,l}+1)!\over \Gamma(e_{n,l}+3/2)\lambda^{1/2}}\;,
\end{equation}
that, in the limit $e_{n,l}\gg 1$, simplifies to expression (\ref{eq:re}):
\begin{equation}
<r>_{n=0,l=e_{n,l}\gg 1}\simeq 
\,\sqrt{\lambda^{-1}\,(e_{n,l}+1)}\left(1+\Theta({e_{n,l}^{-1}})\right).
\label{eq:rn0}
\end{equation}

Now we obtain the radius of an orbit in the Bohr-like image of the 
isotropic harmonic oscillator. Let us recall that this image establishes the quantization 
of the energy through the quantization of the classical orbital angular momentum.
So, the energy $E$ of a particle of mass $m$ moving with velocity $v$ on a circular 
orbit of radius $r$ under the harmonic potential $V(r)=m\lambda^2r^2/2$ is:
\begin{equation}
E={1\over 2}m\lambda^2r^2+{1\over 2}mv^2.
\label{eq:E1}
\end{equation}
The circular orbit is maintained by the central force through the equation:
\begin{equation}
{mv^2\over r}=m\lambda^2r.
\end{equation}
The angular momentum takes discrete values according to the condition
\begin{equation}
mvr=(N+1)\hbar \hspace{1cm} (N=0,1,2,\ldots).
\label{eq:mv}
\end{equation}
Combining the last three equations (\ref{eq:E1})-(\ref{eq:mv}), 
and taking atomic units, $m=\hbar=1$, the radius $r_N$ of a Bohr-like orbit
for this system is obtained
\begin{equation}
r_N=\sqrt{\lambda^{-1}(N+1)} \hspace{1cm} (N=0,1,2,\ldots).
\end{equation}
Let us observe that this expression coincides with the quantum mechanical radius 
given by expression (\ref{eq:rn0}) when $e_{n,l}=N$ for $N\gg 1$.

\subsection{The square well}

Statistical complexity has been calculated in different atomic systems,
such as in the $H$ atom (Sec. \ref{SecQ1}) and in the quantum harmonic 
oscillator (Sec. \ref{SecQ2}).
The behavior of this statistical magnitude in comparison with that
of the energy displays some differences. 
Among other applications, the energy has a clear physical meaning \cite{landau1981} 
and it can be used to find the equilibrium states of a system.
In the same way, it has also been shown that the complexity can give 
some insight about the equilibrium configuration in the ground state of 
the $H_2^+$ molecule \cite{montgomery}. 
In this case, Montgomery and Sen have reported that the minimum of the statistical complexity
as a function of the internuclear distance for this molecule is an 
accurate result comparable with that obtained with the minimization of the energy.
This fact could suggest that energy and complexity are two magnitudes strongly correlated
for any quantum system. But this is not the general case. See, for example, the behavior of
both magnitudes in the previous sections for the H-atom and for the quantum isotropic harmonic 
oscillator. In both systems, the degeneration of the energy 
is split by the statistical complexity, in such a way that the minimum of complexity 
for each level of energy is taken on the wave function with the maximum orbital angular momentum.
Therefore, energy and complexity are two independent variables.

In this section, we wonder if there exists a quantum system where degeneration 
of the complexity can be split by the energy. 
The answer will be affirmative \cite{lopez2009}. We show it in two steps.
First, a new type of invariance by replication for the statistical complexity is established, 
and, second, it is seen that the energy eingestates of the 
quantum infinite square well fulfill the requirements of this kind of invariance. 
From there, it is revealed that the degeneration of complexity in this quantum system 
is broken by the energy.  
  
Different types of replication can be defined on a given probability density.
One of them was established in Ref. \cite{garay02}.
Here, a similar kind of replication is presented, in such a manner that
the complexity $C$ of $m$ replicas of a given distribution is equal to the
complexity of the original one. 
Thus, if $\R$ represents the support of the density function $p(x)$,
with $\int_{\R} p(x)\,dx = 1$, take $n$ copies $p_m(x)$,
$m=1,\cdots,n$, of $p(x)$,
\begin{equation}
p_m(x) = \; p(n (x-\lambda_m))\, ,\;\;
1\leq m\leq n \, ,
\label{eqrep}
\end{equation}
where the supports of all the $p_m(x)$, centered at $\lambda_m's$
points, $m=1,\cdots,n$, are all disjoint. Observe that
$\int_{\R} p_m(x)\, dx = \frac{1}{n}$, what makes the replicas union
\begin{equation}
q_n(x)=\sum_{i=1}^n p_m (x)
\label{eqqn}
\end{equation}
to be also a normalized probability distribution,
$\int_{\R} q_n(x)\, dx = 1$. For every $p_m(x)$, a
straightforward calculation shows that the Shannon entropy is 
\begin{equation}
S(p_m) =  \frac{1}{n}\, S(p) \, ,
\end{equation}
and the disequilibrium is
\begin{equation}
D(p_m) =  \frac{1}{n}\, D(p) \, .
\end{equation}

Taking into account that the $m$ replicas are supported on
disjoint intervals on \R, we obtain
\begin{equation}
S(q_n)  =  S(p) \, , \\
\end{equation}
\begin{equation}
D(q_n)  =  \, D(p) \, .
\end{equation}
Then, the statistical complexity ($C=e^S\cdot D$) is
\begin{equation}
C(q_n) = C (p) \, ,
\end{equation}
and this type of invariance by replication for $C$ is shown.

Let us see now that the probability density of the eigenstates of the energy
in the quantum infinite square well display this type of invariance.
The wave functions representing these states for a particle in a box, that is 
confined in the one-dimensional interval $[0,L]$, are given by \cite{cohen1977}
\begin{equation}
\varphi_k(x)=\sqrt{2\over L}\,\sin\left(k\pi x\over L\right),\,\, k=1,2,\ldots
\end{equation}
Taking $p(x)$ as the probability density 
of the fundamental state $(k=1)$,
\begin{equation}
p(x)=\mid \varphi_1(x)\mid^2, 
\end{equation}
the probability density of the $kth$ excited state,
\begin{equation}
q_k(x)=\mid \varphi_k(x)\mid^2, 
\end{equation}
can be interpreted as the union of $k$ replicas of the fundamental state density, 
$p(x)$, in the $k$ disjoint intervals 
$[{(m-1)L/ k}, {mL/ k}]$, with $m=1,2,\ldots,k$. That is, we find expression (\ref{eqqn}), 
$q_k(x)=\sum_{i=1}^k p_m (x)$, with 
\begin{equation}
p_m(x)={2\over L}\,\sin^2\left({k\pi x\over L}-\pi(m-1)\right),\,\, m=1,2,\ldots,k,
\end{equation}
where in this case the $\lambda_m's$ of expression (\ref{eqrep})
are taken as $(m-1)L/k$. Therefore, we conclude that the complexity is degenerated
for all the energy eigenstates of the quantum infinite square well. Its value can be
exactly calculated. Considering that $L$ is the natural length unit in this problem, 
we obtain
\begin{equation}
C(p)=C(q_k)={3\over e}=1.1036...
\end{equation}
In the general case of a particle in a $d$-dimensional box of width $L$ in each dimension,
it can also be verified that complexity is degenerated for all its energy eigenstates with a 
constant value given by $C=(3/e)^d$. 

Here we have shown that, in the same way that the complexity breaks the energy degeneration
in the H-atom and in the quantum isotropic harmonic oscillator, the contrary behavior is also possible. 
In particular, the complexity is constant
for the whole energy spectrum of the $d$-dimensional quantum infinite square well. 
This result is due to the same functional 
form displayed by all the energy eigenstates of this system. Therefore, 
it suggests that the study of the statistical complexity in a quantum system 
allows to infer some properties on its structural conformation.

\subsection{The periodic table}

The use of these statistical magnitudes to study the electronic structure of atoms 
is another interesting application 
\cite{gadre4,sen07,panos2005,panos2007,borgoo2007,angulo2008,romera2008,borgoo2008}. 
The basic ingredient to calculate these statistical indicators is the electron probability density,
$\rho(\vec{r})$, that can be obtained from the numerically derived Hartree-Fock atomic wave function
in the non-relativistic case \cite{panos2005,panos2007}, 
and from the Dirac-Fock atomic wave function in the relativistic case \cite{borgoo2007}.
The behavior of these statistical quantifiers with the atomic number $Z$ has revealed a
connection with physical measures, such as the ionization potential and the static dipole 
polarizability \cite{sen07}. All of them, theoretical and physical magnitudes,
are capable of unveiling the shell structure of atoms, specifically the closure of shells 
in the noble gases. Also, 
it has been observed that statistical complexity fluctuates around an average 
value that is non-decreasing as the atomic number $Z$ increases in the non-relativistic 
case \cite{panos2007,borgoo2007}.
This average value becomes increasing in the relativistic case \cite{borgoo2007}.
This trend has also been confirmed when the atomic electron density is obtained
with a different approach \cite{sanudo2008-}. In another context where the main interactions
have a gravitational origin, as it is the case of a white dwarf,  
it has also been observed that complexity grows as a function of the
star mass, from the low-mass non-relativistic case to the extreme relativistic limit.
In particular, complexity for the white dwarf reaches a maximum finite value in 
the Chandrasekhar limit as it was calculated by Sa\~nudo and L\'opez-Ruiz \cite{sanudo2009}. 

\begin{figure}[h]  
\centerline{\includegraphics[width=12cm]{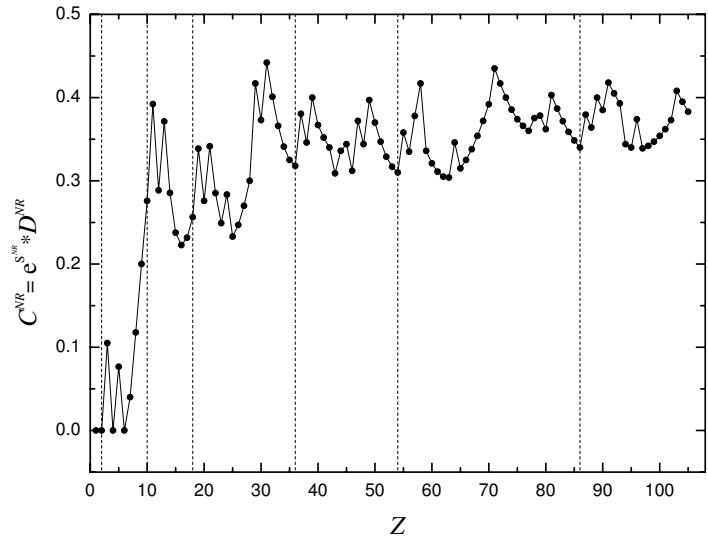}}  
\caption{Statistical complexity, $C$, vs. $Z$ in the non relativistic case ($C^{NR}$). 
The dashed lines indicate the position of noble gases. For details, see the text.}  
\label{figQ5}  
\end{figure}  

\begin{figure}[h]  
\centerline{\includegraphics[width=12cm]{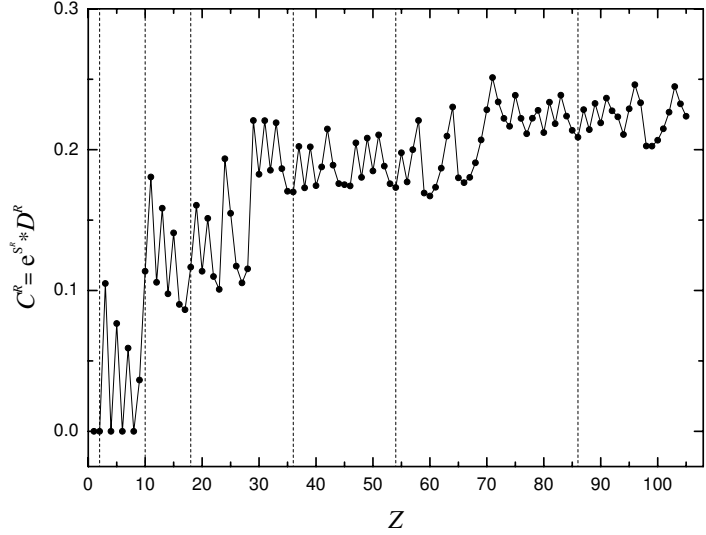}}     
\caption{Statistical complexity, $C$, vs. $Z$ in the relativistic case ($C^R$). 
The comments given in Fig. \ref{figQ5} are also valid here.}  
\label{figQ6}  
\end{figure}  

An alternative method to calculate the statistical magnitudes can be used when the atom
is seen as a discrete hierarchical organization. The atomic shell structure can also be captured
by the fractional occupation probabilities of electrons in the different atomic orbitals.
This set of probabilities is here employed to evaluate all these quantifiers
for the non-relativistic ($NR$) and relativistic ($R$) cases.
In the $NR$ case, a non-decreasing trend in complexity as $Z$ increases is obtained
and also the closure of shells for some noble gases is observed \cite{sanudo1, panos2009}. 
 
For the $NR$ case, each electron shell of the atom is given by $(nl)^w$ \cite{bransden2003}, 
where $n$ denotes the principal quantum number, $l$ the orbital angular momentum $(0\leq l\leq n-1)$ 
and $w$ is the number of electrons in the shell $(0\leq w\leq 2(2l+1))$. 
For the $R$ case, due to the spin-orbit interaction, 
each shell is split, in general, in two shells \cite{cowan1981}: 
$(nlj_-)^{w_-}$, $(nlj_+)^{w_+}$, where $j_{\pm}=l\pm 1/2$ 
(for $l = 0$ only one value of $j$ is possible, $j=j_+=1/2$) and
$0\leq w_{\pm}\leq 2j_{\pm}+1$. 
As an example, we explicitly give the electron configuration of  
$Ar(Z=18)$ in both cases,
\begin{equation}
 Ar(NR)  :  (1s)^2(2s)^2(2p)^6(3s)^2(3p)^6, 
\end{equation}
\begin{equation}
 Ar(R)  :  (1s1/2)^2(2s1/2)^2(2p1/2)^2(2p3/2)^4(3s1/2)^2(3p1/2)^2(3p3/2)^4. 
\end{equation}

For each atom, a fractional occupation probability distribution of electrons 
in atomic orbitals $\{p_k\}$, $k = 1,2,\ldots,\Pi$,  being $\Pi$ the number of shells of the atom,
can be defined. This normalized probability distribution $\{p_k\}$ $(\sum p_k=1)$ 
is easily calculated by dividing the superscripts $w_{\pm}$ 
(number of electrons in each shell) by $Z$, the total number of electrons in neutral atoms, 
which is the case we are considering here. The order of shell filling
dictated by nature \cite{bransden2003} has been chosen. Then, from this probability distribution,
the different statistical magnitudes (Shannon entropy, disequilibrium, statistical complexity and
Fisher-Shannon entropy) is calculated.

In order to calculate the statistical complexity $C=H\cdot D$, with $H=e^S$,
we use the discrete versions of the Shannon entropy $S$ and disequilibrium $D$:
\begin{eqnarray}
 S & = & -\sum_{k=1}^{\Pi}p_k\log p_k \; , \label{eq-Sdiscrete}\\
 D & = & \;\;\sum_{k=1}^{\Pi}(p_k-1/\Pi)^2 \; . \label{eq-Ddiscrete}
\end{eqnarray}
To compute the Fisher-Shannon information, $P=J\cdot I$, with $J=\frac{1}{2\pi e}\,e^{2S/3}$,
the discrete version of $I$ is defined as \cite{sanudo1,panos2009} 
\begin{equation}
I = \sum_{k=1}^{\Pi} {(p_{k+1}-p_k)^2\over p_k}\;, \label{eq-Idiscrete}
\end{equation}
where $p_{\Pi+1}=0$ is taken.

\begin{figure}[h]  
\centerline{\includegraphics[width=12cm]{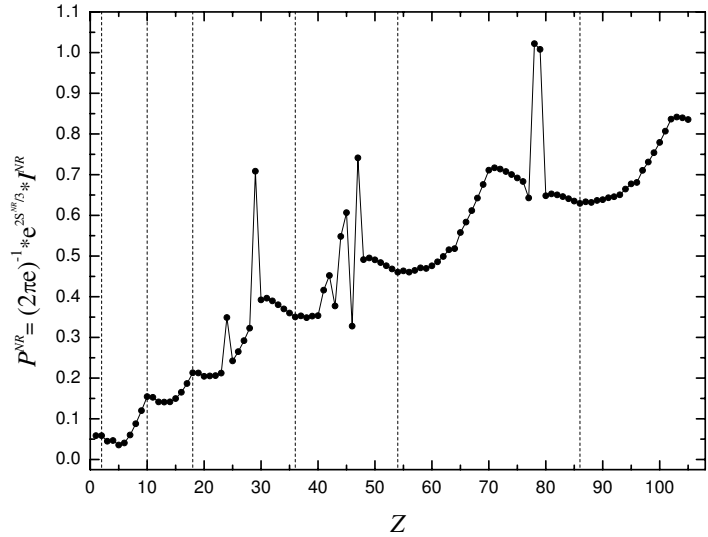}}     
\caption{Fisher-Shannon entropy, $P$, vs. $Z$, in the non relativistic case ($P^{NR}$). 
The dashed lines indicate the position of noble gases. For details, see the text.}  
\label{figQ7}  
\end{figure}  

\begin{figure}[h]  
\centerline{\includegraphics[width=12cm]{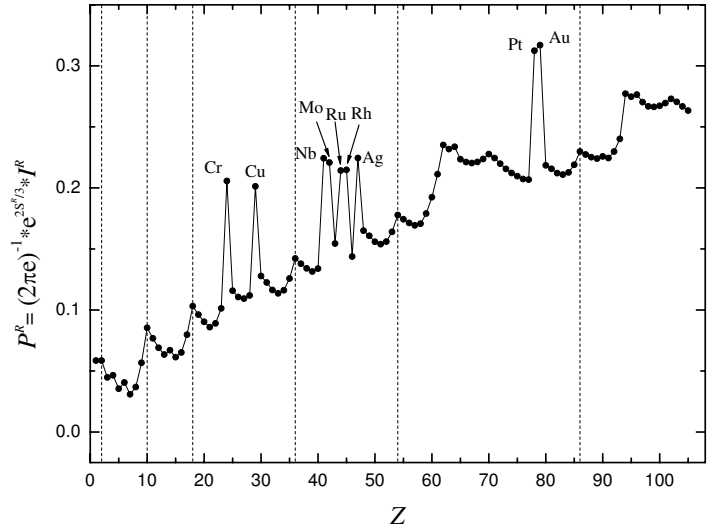}}     
\caption{Fisher-Shannon entropy, $P$, vs. $Z$, in the relativistic case ($P^R$). 
The comments given in Fig. \ref{figQ7} are also valid here.}  
\label{figQ8}  
\end{figure}  

The statistical complexity, $C$, as a function of the atomic number, $Z$, 
for the $NR$ and $R$ cases for neutral atoms is given in Figs. \ref{figQ5} and \ref{figQ6}, 
respectively. 
It is observed in both figures that this magnitude fluctuates around an increasing average value 
with $Z$. This increasing trend recovers the behavior obtained
by using the continuous quantum-mechanical wave functions \cite{panos2007,borgoo2007}.
A shell-like structure is also unveiled in this approach by looking at the minimum values of $C$
taken on the noble gases positions (the dashed lines in the figures) with the 
exception of $Ne(Z=10)$ and $Ar(Z=18)$. This behavior can be interpreted as special arrangements
in the atomic configuration for the noble gas cases out of the general increasing trend
of $C$ with $Z$. 

The Fisher-Shannon entropy, $P$, as a function of $Z$, for the $NR$ and $R$ cases in neutral 
atoms is given in Figs. \ref{figQ7} and \ref{figQ8}, respectively. 
The shell structure is again displayed in the special atomic arrangements, 
particularly in the $R$ case (Fig. \ref{figQ8})
where  $P$ takes local maxima for all the noble gases 
(see the dashed lines on $Z = 2, 10, 18, 36, 54, 86$).
The irregular filling (i.f.) of $s$ and $d$  shells \cite{bransden2003} is also detected by peaks
in the magnitude $P$, mainly in the $R$ case. In particular, see the elements $Cr$ and $Cu$  
(i.f. of $4s$ and $3d$ shells); $Nb$, $Mo$, $Ru$, $Rh$, and $Ag$ (i.f. of $5s$ and $4d$ shells); 
and finally $Pt$ and $Au$ (i.f. of $6s$ and $5d$ shells). 
$Pd$ also has an irregular filling, but $P$ does not display a peak on it
because the shell filling in this case does not follow the same procedure 
as the before elements (the $5s$ shell is empty and the $5d$ is full). 
Finally, the increasing trend of $P$ with $Z$ is clearly observed.

Then, it is found that $P$, the Fisher-Shannon entropy, 
in the relativistic case (Fig. \ref{figQ8}) reflects in a clearer way the increasing trend with $Z$,
the shell structure in noble gases, and the irregular shell filling of some specific elements.
The same method that uses the fractional occupation probability distribution is applied
in the next section to another many particle system, the atomic nucleus, 
that has also been described by a shell model.

\subsection{Magic numbers in nuclei}

Nucleus is another interesting quantum system that can be described by a shell model \cite{krane1988}.
In this picture, just as electrons in atoms,
nucleons in nuclei fill in the nuclear shells by following a determined hierarchy.
Hence, the fractional occupation probabilities of nucleons in the different nuclear orbitals
can capture the nuclear shell structure. This set of probabilities, as explained in the above section,
can be used to evaluate the statistical quantifiers for nuclei as a function of the number of nucleons. 
In this section, by following this method, the calculation of statistical
complexity and Fisher-Shannon information for nuclei is presented \cite{lopez2009+}. 

The nuclear shell model is developed by choosing an intermediate three-dimensional potential, 
between an infinite well and the harmonic oscillator, in which nucleons evolve under the 
Schr\"odinger equation with an additional spin-orbit interaction \cite{krane1988}.
In this model, each nuclear shell is given by $(nlj)^w$, 
where $l$ denotes the orbital angular momentum ($l=0,1,2,\ldots$), 
$n$ counts the number of levels with that $l$ value,
$j$ can take the values $l+1/2$ and $l-1/2$
(for $l = 0$ only one value of $j$ is possible, $j=1/2$), 
and $w$ is the number of one-type of nucleons (protons or neutrons)
in the shell $(0\leq w\leq 2j+1)$. 

\begin{figure}[h]  
\centerline{\includegraphics[width=12cm]{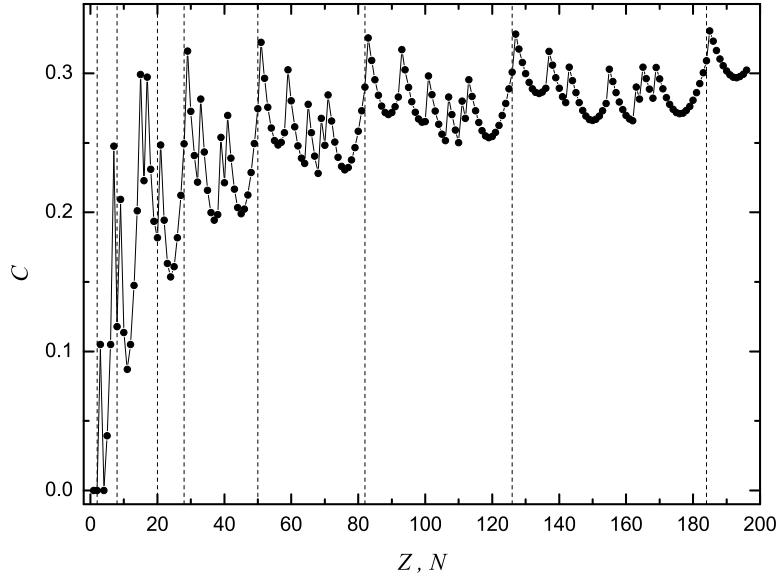}}  
\caption{Statistical complexity, $C$, vs. number of nucleons, $Z$ or $N$.
The dashed lines indicate the positions of magic numbers
$\{2,8,20,28,50,82,126,184\}$. For details, see the text.}  
\label{figQ9}  
\end{figure}  

As an example, we explicitly give the shell configuration of a nucleus formed
by $Z=20$ protons or by $N=20$ neutrons. In both cases, it is obtained \cite{krane1988}:
 \begin{equation}
 \left\{\begin{array}{c}
 (Z=20) \\
 (N=20)
 \end{array} \right\}
 :  (1s1/2)^2(1p3/2)^4(1p1/2)^2(1d5/2)^6(2s1/2)^2(1d3/2)^4. 
 \end{equation}

When one-type of nucleons (protons or neutrons) in the nucleus is considered, 
a fractional occupation probability distribution of this type of nucleons 
in nuclear orbitals $\{p_k\}$, $k = 1,2,\ldots,\Pi$,  being $\Pi$ the 
number of shells for this type of nucleons, can be defined in the same way 
as it has been done for electronic calculations in the atom in the previous section. 
This normalized probability distribution $\{p_k\}$ $(\sum p_k=1)$ is easily found by dividing 
the superscripts $w$ by the total of the corresponding type of nucleons ($Z$ or $N$). 
Then, from this probability distribution, the different statistical magnitudes 
(Shannon entropy, disequilibrium, statistical complexity and
Fisher-Shannon entropy) by following expressions (\ref{eq-Sdiscrete}-\ref{eq-Idiscrete})
are obtained.

\begin{figure}[h]  
\centerline{\includegraphics[width=12cm]{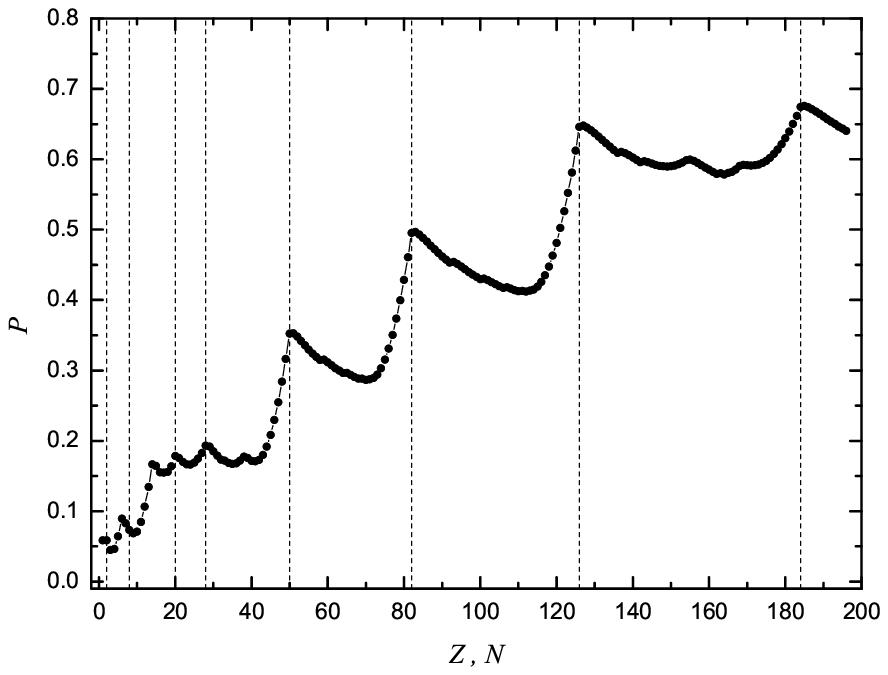}}     
\caption{Fisher-Shannon entropy, $P$, vs. the number of nucleons, $Z$ or $N$.
The dashed lines indicate the positions of magic numbers
$\{2,8,20,28,50,82,126,184\}$. For details, see the text.}  
\label{figQ10}  
\end{figure}  

The statistical complexity, $C$, of nuclei as a function of the number of nucleons, 
$Z$ or $N$, is given in Fig. \ref{figQ9}. 
Here we can observe that this magnitude fluctuates around an increasing average value 
with $Z$ or $N$. This trend is also found for the electronic structure of atoms (see previous section), 
reinforcing the idea that, in general, complexity increases with the number of units forming a system.
However, the shell model supposes that the system encounters certain ordered rearrangements 
for some specific number of units (electrons or nucleons).
This shell-like structure is also unveiled by $C$ in this approach to nuclei.
In this case, the extremal values of $C$ are not taken just on the closed shells as 
happens in the noble gases positions for atoms, if not that they appear to be in the
positions one unit less than the closed shells.
  
The Fisher-Shannon entropy, $P$, of nuclei as a function of $Z$ or $N$ is given in Fig. \ref{figQ10}. 
It presents an increasing trend with $Z$ or $N$. The spiky behavior of $C$ provoked by 
the nuclear shell structure becomes smoother for $P$, that presents
peaks (changes in the sign of the derivative) only at a few $Z$ or $N$, 
concretely at the numbers $2,6,14,20,28,50,82,126,184$. Strikingly, the sequence of
magic numbers is $\{2,8,20,28,50,82,126,184\}$ (represented as dashed vertical lines
in the figures). Only the peaks at $6$ and $14$ disagree with the sequence of magic numbers,
what could be justified by saying that statistical indicators work better for high numbers.
But in this case, it should be observed that the carbon nucleus, $C_{Z=6}^{N=6}$, and the 
silicon nucleus, $Si_{Z=14}^{N=14}$, apart from their great importance in nature and industry, 
they are the stable isotopes with the greatest abundance in the corresponding isotopic series, 
$98.9\%$ and $92.2\%$, respectively.

Then, the increasing trend of these statistical magnitudes with $Z$ or $N$,
and the reflect of the shell structure in the spiky behavior of their plots are found
when using for their calculation the fractional occupation probability distribution 
of nucleons, $Z$ or $N$.
It is worth to note that the relevant peaks in the Fisher-Shannon information
are revealed to be just the series of magic numbers in nuclei.
This fact indicates again that these statistical indicators are able to enlighten 
some structural aspects of quantum many-body systems. 

\begin{acknowledgement}
R.L-R. thanks Prof. K.D. Sen for his invitation to prepare and to present this chapter
in the book "Statistical Complexity" published by Springer in 2011 (ISBN 978-90-481-3889-0). 
\end{acknowledgement}
%

%
%

\end{document}